\documentclass[11pt,a4paper]{article}

\usepackage{jheppub}

\usepackage{epsfig,epsf}
\usepackage{amsmath}
\usepackage{amsthm}
\usepackage{amsfonts}
\usepackage{amssymb}
\usepackage{dsfont}

\usepackage{multirow}

\usepackage{slashed}

\usepackage[active]{srcltx}
\usepackage{psfrag}

\def\II{\hbox{{1}\kern-.25em\hbox{l}}}

\newcommand{\convent}{\rho}
\newcommand \vev [1] {\langle{#1}\rangle}

\title{Operator product expansion in QCD in off-forward kinematics:
       Separation of kinematic and dynamical contributions}

\author[a]{V. M. Braun}
\author[a,b]{and A. N. Manashov}

\affiliation[a]{
   Institut f\"ur Theoretische Physik, Universit\"at
   Regensburg \\ D-93040 Regensburg, Germany}
\affiliation[b]{
   Department of Theoretical Physics,  St.-Petersburg State
   University \\
   199034, St.-Petersburg, Russia}
\emailAdd{vladimir.braun@physik.uni-r.de}

\emailAdd{~~~~~~~alexander.manashov@physik.uni-r.de}

\abstract{
We develop a general approach to the calculation of target mass
and finite $t=(p'-p)^2$ corrections
in hard processes which can be studied in the
framework of the operator product expansion and involve momentum transfer from the initial to the
final hadron state.
Such corrections, which are usually referred to as kinematic,
can be defined as contributions of operators of all twists that can be
reduced to total derivatives of the
leading twist operators.
As the principal result, we provide a set of projection operators that
pick up the ``kinematic'' part of an arbitrary flavor-nonsinglet twist-four operator in QCD.
A complete expression is derived for the
time-ordered product of two electromagnetic currents that
includes all kinematic corrections to twist-four accuracy.
The results are immediately applicable to the studies of
deeply-virtual Compton scattering, transition $\gamma^*\to M\gamma$
form factors and related processes.
As a byproduct of this study, we find a series of
``genuine'' twist-four flavor-nonsinglet quark-antiquark-gluon operators
which have the same anomalous dimensions as the leading twist
quark-antiquark operators.
       }

\keywords{OPE, conformal symmetry, DVCS}

\begin{document}

\maketitle

\section{Introduction}

As well known, target mass corrections $\sim (m^2/Q^2)^k$ to the structure functions
in deep-inelastic lepton-nucleon scattering (DIS) can be calculated exactly in terms
of the leading-twist parton densities. On a technical level, these corrections arise
because of the subtractions that are needed to form traceless operators, as demonstrated
in the pioneering paper by Nachtmann~\cite{Nachtmann:1973mr}. The Nachtmann power corrections
have been studied in much detail (see e.g.~\cite{Blumlein:1998nv}) and are routinely taken into
account in the analysis of experimental data.
Another classical result is due to Wandzura and Wilczek~\cite{Wandzura:1977qf}
who have shown that the twist-three structure function $g_2(x,Q^2)$ for massive targets with spin~1/2
receives a contribution related to the leading-twist
structure function $g_1(x,Q^2)$. This contrbution appears to be
numerically dominant for the nucleon, see
e.g.~\cite{Accardi:2009au} for a recent analysis. The Wandzura-Wilczek relation
follows from Lorentz invariance and can be understood as spin rotation
in the target rest frame~\cite{Ball:1998sk,Anikin:2001ge}.

The target mass effects are usually referred to as kinematic corrections since they can be
expressed in terms of leading-twist parton distributions and do not involve ``genuine''
nonperturbative effects due to quark-gluon correlations. In this paper we develop an
approach to the calculation of kinematic corrections to a larger class of hard
processes that can be studied in the framework of the operator product expansion (OPE) and
at the same time involve  momentum transfer from the initial to the final hadron state.
This, more general, kinematics is relevant for \emph{exclusive} scattering processes which
involve off-forward matrix elements of the type
$$\langle p'|\mathcal{O}|p\rangle\,,\qquad p^2=p'^2=m^2\,,\qquad t= (p'-p)^2 $$
and  form-factors at large momentum transfers (and weak decays of heavy mesons/baryons)
that involve hadron-to-vacuum matrix elements
$$\langle 0|\mathcal{O}|p\rangle\,, \qquad p^2 =m^2\,,$$
e.g. $\gamma^*\to \eta\gamma$, $B\to K^*\ell\bar\nu_\ell$ etc.

The forthcoming studies of hard exclusive scattering processes at the future 12~GeV facility
at Jefferson Lab provide the main motivation for our study.
It is generally accepted that such reactions allow one to access a three-dimensional
picture of the proton in longitudinal and transverse plane~\cite{Burkardt:2002hr},
encoded in generalized parton distributions
(GPDs)~\cite{Diehl:2003ny,Belitsky:2005qn}. One of the principal reactions in this context
is Compton scattering with one real and one highly-virtual photon (DVCS) which has
received a lot of attention. The QCD description of DVCS is based on the OPE
 of the time-ordered product of two electromagnetic currents where the GPDs
appear as operator matrix elements and the coefficient functions can be calculated
perturbatively. In order to access the transverse proton structure one is interested
in the dependence of the amplitude on the momentum transfer to the target
$t=(p'-p)^2$ in a sufficiently broad range. Since the available photon
virtualities $Q^2$ are not very large, corrections of the type $\sim t/Q^2$ which are
formally twist-four effects, can have significant impact and should be taken into account.

The problem is well known and its importance for phenomenology
has been  acknowledged by many
authors~\cite{Belitsky:2005qn,Anikin:2000em,Blumlein:2000cx,Kivel:2000rb,Radyushkin:2000ap,Belitsky:2000vx,Belitsky:2001hz,Belitsky:2010jw,Geyer:2004bx,Blumlein:2006ia,Blumlein:2008di}.
At first sight, such corrections are seemingly disconnected
from nonperturbative effects (e.g. one may consider a theoretical
limit $\Lambda_{\rm QCD}^2 \ll t \ll Q^2$). Yet their separation from generic twist-four
corrections $\sim \Lambda_{\rm QCD}^2/Q^2$ proves to be surprisingly difficult.

The kinematic corrections $\sim t/Q^2$ and $\sim m^2/Q^2$
induced by the subtraction of traces in the leading-twist operators,
which are a direct generalization of the Nachtmann's corrections in DIS~\cite{Nachtmann:1973mr}
were considered already in Refs.~\cite{Blumlein:2000cx,Belitsky:2001hz,Belitsky:2010jw,Geyer:2004bx,Blumlein:2006ia,Blumlein:2008di}. These results are, however, incomplete, because
$\sim t/Q^2$ corrections in hard exclusive processes
(and for spin-1/2 targets also $\sim m^2/Q^2$ corrections)
also receive contributions from higher-twist  operators that can be reduced
to total derivatives of the twist-two ones.
 Indeed, let $\mathcal{O}_{\mu_1\ldots\mu_n}$ be a multiplicatively renormalizable
(conformal) local twist-two operator,
symmetrized and traceless over all indices. Then the operators
\begin{equation}
  \mathcal{O}_1 \sim \partial^2 \mathcal{O}_{\mu_1\ldots\mu_n}\,, \qquad
  \mathcal{O}_2 \sim \partial^{\mu_1}\mathcal{O}_{\mu_1\ldots\mu_n}
\label{eq:O1O2}
\end{equation}
are, on the one hand, twist-four, and on the other hand their matrix elements
are obviously given by the reduced matrix elements of the twist-two operators, times the
momentum transfer squared (and for spin-1/2 hadrons also target mass corrections).
Thus, the contributions of both operators in Eq.~(\ref{eq:O1O2}) must be taken into account.
Moreover, as we will see below, the distinction between the kinematic corrections
due to contributions of leading-twist and higher-twist operators is not Lorentz invariant
and has no physical meaning; they must always be summed up.

Again, at first sight, taking into account the operators that are shown
schematically in  Eq.~(\ref{eq:O1O2}) should not be difficult, at least to the leading
order in strong coupling. The problem arises because $\mathcal{O}_2$ has very peculiar
properties that can be traced to the  Ferrara-Grillo-Parisi-Gatto (FGPG) theorem
\cite{Ferrara:1972xq}): divergence of a conformal operator vanishes in a free theory. As a
consequence, using QCD equations of motion (EOM) the operator $\mathcal{O}_2$ can be
expressed as a sum of contributions of quark-antiquark-gluon operators. The simplest
example of such relation
is~\cite{Kolesnichenko:1984dj,Braun:2004vf,Anikin:2004vc,Anikin:2004ja}
\begin{equation}
  \partial^\mu O_{\mu\nu} = 2i\bar q g F_{\nu\mu}\gamma^\mu q\,,
\label{eq:puzzle}
\end{equation}
where
$
  O_{\mu\nu} = (1/2)[\bar q \gamma_\mu\!
  \stackrel{\leftrightarrow}{D}_\nu \!q
  + (\mu\leftrightarrow\nu)]
$
is the quark part of the energy-momentum tensor.
{}For simplicity we consider massless quarks. The operator on the r.h.s.
of Eq.~(\ref{eq:puzzle}) involves the gluon field strength and,  naively,
one would expect hadronic matrix elements of this operator to be
of the order of $\Lambda_{\rm QCD}^2$, which is wrong.
More complicated examples involving leading-twist operators with two derivatives
can be found in \cite{Balitsky:1989ry,Ball:1998ff}, e.g.
\begin{eqnarray}
\frac{4}{5} \partial^\mu \mathcal{O}_{\mu\alpha\beta} & = & -12 i
 \bar q \gamma^\rho \left\{gF_{\rho\beta}\!
\stackrel{\rightarrow}{D}_\alpha
- \stackrel{\leftarrow}{D}_\alpha\! gF_{\rho\beta} +
  (\alpha\leftrightarrow \beta) \right\} q
-4\partial^\rho \bar q (\gamma_\beta g\widetilde{F}_{\alpha\rho} +
\gamma_{\alpha} g\widetilde{F}_{\beta\rho} ) \gamma_5 q
\nonumber\\
& & {} - \frac{8}{3}\,
\partial_\beta \bar q \gamma^\sigma g\widetilde{F}_{\sigma\alpha} \gamma_5
q - \frac{8}{3}\,\partial_\alpha \bar q \gamma^\sigma
g\widetilde{F}_{\sigma\beta} \gamma_5 q
+\frac{28}{3}\, g_{\alpha\beta} \partial_\rho \bar q \gamma^\sigma
g\widetilde{F}_{\sigma\rho} q\,,
\label{eq:Omualphabeta}
\end{eqnarray}
where
$$\mathcal{O}_{\mu\alpha\beta} =\mathrm{Sym}_{\mu\alpha\beta}\left[
\frac{15}{2}\bar q \gamma_\mu\!\stackrel{\leftrightarrow}{D}_{\alpha}
\stackrel{\leftrightarrow}{D}_{\beta}  q
-\frac{3}{2} \partial_{\alpha} \partial_{\beta}\bar q \gamma_\mu
q \right]  - \mathrm{traces}
$$
and $\mathrm{Sym}_{\mu\alpha\beta}$ stands for 
symmetrization in the Lorentz indices.
The general structure of such relations is, schematically
\begin{equation}
 (\partial\mathcal{O})_{N} = \sum_k a^{(N)}_{k}\,{G}_{N,k}\,,
\label{eq:partialO}
\end{equation}
where ${G}_{Nk}$ are twist-four quark-antiquark-gluon
(and more complicated) operators and $a^{(N)}_{k}$ the numerical coefficients.
The subscript $N$ stands for the number of
derivatives in $\mathcal{O}_N$ and the summation
goes over all contributing operators, with and without total derivatives
(so that $k$ is a certain multi-index).
The same operators, ${G}_{Nk}$, also appear in the OPE for
the current product of interest at the twist-four level:
\begin{equation}
 T\{j(x)j(0)\}^{t=4} = \sum_{N,k} c_{N,k}(x)\,{G}_{N,k}\,.
\label{eq:Tproduct}
\end{equation}
A separation of ``kinematic'' and ``dynamical'' contributions implies
rewriting this expansion in such a way that the contribution
of a particular combination appearing in (\ref{eq:partialO}) is
separated from the remaining twist-four contributions.
The ``kinematic'' approximation would correspond to taking into account
this term only, and neglecting contributions of ``genuine'' quark-gluon operators.
The main problem is to make this separation consistently, such
that ``genuine'' quark-gluon operators are in some well-defined sense ``orthogonal'' to the kinematic
contribution.

The guiding principle is that the separation of kinematic
effects is only meaningful if they have autonomous scale dependence.
The twist-four operators in Eq.~(\ref{eq:O1O2}) are special
in that they obviously have the  same anomalous dimensions as their ``parent'' twist-two
operators. Hence we can reformulate the problem in the following way:

All existing twist-four operators with the same quantum numbers and of the same dimension
mix with each other and satisfy a renormalization group (RG) equation which can be solved,
at least in principle. Note that we have to include operators with total derivatives,
so that this is a large matrix equation, in general.
Let $\mathcal{G}_{N,k}$ be the set of multiplicatively
renormalizable twist-four operators which we can express in terms of linear combinations of
the original operators, i.e.
\begin{equation}
  \mathcal{G}_{N,k} = \sum_{k'} \psi^{(N)}_{k,k'}\, {G}_{N,k'}\,.
\end{equation}
Relation (\ref{eq:partialO}) tells us that one of the solutions
of the RG equation is known {\it without calculation}. Namely, there exists
a twist-four operator with the anomalous dimension equal to the anomalous
dimension of the leading twist operator, and Eq.~(\ref{eq:partialO})
gives the corresponding eigenvector.
(For simplicity we ignore the contributions of $\partial^2\mathcal{O}_N$
operators in this discussion; they do not pose a ``problem'' and can
be taken into account relatively simply.)

Let us assume that this special solution corresponds to $k=0$, so that
$\mathcal{G}_{N,k=0} \equiv (\partial\mathcal{O})_{N}$ and
$\psi^{(N)}_{k=0,k'}= a_{k'}$. Inverting the
matrix of coefficients $\psi^{(N)}_{k,k'}$ we can expand an
arbitrary twist-four operator in terms of the multiplicatively renormalizable
ones
\begin{equation}
 G_{N,k} = \phi^{(N)}_{k,0} (\partial\mathcal{O})_{N} + \sum_{k'\not=0}\phi^{(N)}_{k,k'}\,
\mathcal{G}_{N,k'}\,.
\label{eq:eee}
\end{equation}
Inserting this expansion in Eq.~(\ref{eq:Tproduct}) one obtains
\begin{equation}
  T\{j(x)j(0)\}^{\rm tw-4} = \sum_{N,k} c_{N,k}(x)\phi^{(N)}_{k,0}\,(\partial\mathcal{O})_{N}
 + \ldots\,,
\label{eq:solve}
\end{equation}
where the term in $(\partial\mathcal{O})_{N}$ defines the kinematic
correction which we are looking for,
and the ellipses stand for the contributions of ``genuine'' twist-four
operators.
The problem with this (formal) solution is that finding the coefficients
$\phi^{(N)}_{k,0}$ in general requires knowledge of the full
matrix $\psi^{(N)}_{k,k'}$, in other words the explicit solution
of the twist-four RG equations, which is not available.

In this work we show how to overcome this difficulty.
Our main result is that the explicit solution is \emph{not needed};
it can be avoided by making use of the symmetries of the RG equation
that are rooted in conformal symmetry of the QCD Lagrangian~\cite{Braun:2003rp}.

Twist-four operators in QCD can be divided in two classes:
quasipartonic~\cite{Bukhvostov:1985rn}, that only involve ``plus''
components of the fields, and non-quasipartonic which also include
``minus'' light-cone projections. Our starting observation is that
quasipartonic operators are irrelevant for the present discussion
since they have autonomous evolution (to the one-loop accuracy).
Hence terms in $(\partial\mathcal O)_N$ do not appear in the re-expansion
of quasipartonic operators in multiplicatively renormalizable operators, Eq.~(\ref{eq:eee}):
the corresponding coefficients
$\phi^{(N)}_{k,0}$ vanish. As the result, the kinematic power correction
$\sim (\partial\mathcal O)_N$ is entirely due to contributions of
non-quasipartonic operators.

Renormalization of twist-four non-quasipartonic operators was
studied systematically in~\cite{Braun:2008ia,Braun:2009vc}.
The main result is that with the appropriate choice of the operator basis,
the corresponding RG equations can be written in terms
of several $SL(2)$-invariant kernels. Using this technique,
we are able to prove that
the anomalous dimension matrix for non-quasipartonic operators is
hermitian with respect to a certain scalar product, which implies that
different eigenvectors are mutually orthogonal, i.e.
\begin{equation}
 \sum_k \mu^{(N)}_k \psi^{(N)}_{l,k} \psi^{(N)}_{m,k} \sim \delta_{lm}\,,
\label{eq:muNk}
\end{equation}
where $\mu^{(N)}_k$ is the corresponding (nontrivial) measure.

The knowledge of $\mu^{(N)}_k$ is sufficient:
using this orthogonality relation and the expression (\ref{eq:partialO})
for the relevant eigenvector, one obtains, for the non-quasipartonic
operators
\begin{equation}
 \phi^{(N)}_{k,0} = a^{(N)}_{k} ||a^{(N)}||^{-2}\,,
\label{eq:coc}
\end{equation}
where $||a^{(N)}||^2 =  \sum_k \mu^{(N)}_k (a^{(N)}_{k})^2$.
Inserting this expression in (\ref{eq:solve}) we end up with
the desired separation of kinematic effects.

The actual derivation proves to be rather involved. It is done using the two-component
spinor formalism in intermediate steps and requires some specific techniques
of the $SL(2)$ representation theory. Some of our results were reported earlier
in~\cite{Braun:2011zr}.

The presentation is organized as follows.
Section~2 mainly serves to introduce the necessary formalism and
notation. Following Refs.~\cite{Braun:2008ia,Braun:2009vc} we
define a complete basis of non-quasipartonic twist-four operators
which correspond to $G_{Nk}$ in Eqs.~(\ref{eq:partialO}),~(\ref{eq:Tproduct})
and discuss symmetries of the corresponding RG equations. A new result in
this section is the hermiticity of the anomalous dimension matrix with
respect to a certain scalar product, cf.~Eq.~(\ref{eq:muNk}).

Section~3 is the central one.
Here we derive the expansion for divergence of the leading
twist operators, $(\partial\mathcal{O})_N$,
in terms of non-quasipartonic quark-gluon operators in the conformal basis
(i.e. the precise version of Eq.~(\ref{eq:partialO})), and use hermiticity
of the anomalous dimension matrix to reconstruct the coefficient in front
of $(\partial\mathcal{O})_N$ in the inverse expansion of an arbitrary
quark-gluon operator in terms of the multiplicatively
renormalizable twist-four operators, the analogue of Eq.~(\ref{eq:coc}).
As a byproduct of this study, we find a series of
``genuine'' twist-four flavor-nonsinglet quark-antiquark-gluon operators
which do not reduce to total derivatives and have the same anomalous
dimensions as the leading twist quark-antiquark operators. For reader's
convenience, we present a summary of the ``kinematic projections'' for
arbitrary twist-four flavor-nonsinglet quark-gluon operators in Section~4.

The next three Sections are devoted to the application of this formalism
to the OPE for the time-ordered product of two electromagnetic currents,
which is the most interesting case  phenomenologically.
In Section 5 we demonstrate how the twist expansion
can be done conveniently in the spinor formalism, Section 6 contains the
explicit construction of the kinematic contributions
and in Section~7 we give the final result for the $T$-product in two
different representations for the crucial twist-four part.

Section~8 contains a short discussion of typical matrix elements that one
encounters in
physics applications. One issue that we want to emphasize here is that
the distinction between the kinematic  corrections of Nachtmann's type,
i.e. due to contributions of
leading-twist~\cite{Radyushkin:2000ap,Belitsky:2001hz,Belitsky:2000vx,Belitsky:2010jw,%
Geyer:2004bx,Blumlein:2006ia,Blumlein:2008di}, and of higher-twist
operators calculated
in this work is not invariant under translations along the line connecting
the electromagnetic currents in the $T$-product. As we show on a simple
example, translation invariance is only restored in the sum of all
contributions.
We also use this example to illustrate application of the leading-twist
projection operators and elucidate on the general structure of
higher-twist matrix elements that contribute to
hard exclusive scattering from a proton target, e.g. DVCS.

The final Section~9 is reserved for a summary and conclusions. The paper
also contains several Appendices where we explain some technical issues
and details of the derivation.

\section{Conformal operator basis}
\subsection{Spinor formalism}
The conformal symmetry which plays a crucial role in the further discussion becomes more
transparent in the spinor formalism. The latter also presents a convenient framework for
twist separation in comparison with the conventional vector formalism. Unfortunately,
there are no standard notations (conventions) which vary from paper to
paper. In this work we follow  the notations  adopted in
Refs.~\cite{Braun:2008ia,Braun:2009vc} which are summarized below.

In the spinor formalism each covariant four-vector $x_\mu$ is mapped to a hermitian
matrix~$x$:
 $$
x_{\alpha\dot\alpha}=x_\mu (\sigma^{\mu})_{\alpha\dot\alpha}\,,
\qquad
\bar x^{\dot\alpha\alpha}=x_\mu (\bar\sigma^{\mu})^{\dot\alpha\alpha}\,,
$$
where $\sigma^\mu=(\II,\vec{\sigma})$, $\bar\sigma^\mu=(\II,-\vec{\sigma})$ and
$\vec{\sigma}$ are the usual Pauli matrices.
The Dirac matrices in the spinor representation read
\begin{align}
\gamma^{\mu}=\begin{pmatrix}0&[\sigma^\mu]_{\alpha\dot\beta}\\
                            [\bar\sigma^{\mu}]^{\dot\alpha\beta}&0 \end{pmatrix},&&
\sigma^{\mu\nu}=\begin{pmatrix}{[\sigma^{\mu\nu}]_{\alpha}}^{\beta}&0\\
                           0& {[\bar\sigma^{\mu\nu}]^{\dot\alpha}}_{\dot\beta} \end{pmatrix},
&&
\gamma_5=\begin{pmatrix}-\delta_{\alpha}^\beta&0\\
                           0&\delta^{\dot\alpha}_{\dot\beta}  \end{pmatrix}\,.
\end{align}
Here $\sigma^{\mu\nu}=\frac{i}2[\gamma^\mu,\gamma^\nu]$,
$\gamma_5=i\gamma^0\gamma^1\gamma^2\gamma^3$ and
\begin{align}
{(\sigma^{\mu\nu})_{\alpha}}^{\beta}=
\frac{i}2{\left[\sigma^{\mu}\bar\sigma^\nu-\sigma^{\nu}\bar\sigma^\mu\right
]_{\alpha}}^\beta\,, &&
{(\bar\sigma^{\mu\nu})^{\dot\alpha}}_{\dot\beta}=
\frac{i}2{\left[\bar\sigma^{\mu}\sigma^\nu-\bar\sigma^{\nu}\sigma^\mu\right
]^{\dot\alpha}}_{\dot\beta}\,.
\end{align}
The Dirac (quark) spinor $q$ and the conjugated spinor $\bar q$ are written as
\begin{align}
q=\begin{pmatrix}\psi_\alpha\\ \bar\chi^{\dot\beta}\end{pmatrix},
&& \bar q=(\chi^\beta,\bar\psi_{\dot\alpha})\,,
\end{align}
where $\psi_{\alpha}$, $\bar \chi^{\dot\beta}$ are two-component Weyl spinors,
$\bar\psi_{\dot\alpha}=(\psi_{\alpha})^\dagger$,
$\chi^{\alpha}=(\bar\chi^{\dot\alpha})^\dagger$. We accept the following rule for raising
and lowering of spinor indices (cf.~\cite{Sohnius})
\begin{align}\label{raise}
u^\alpha=\epsilon^{\alpha\beta}u_\beta\,,&& u_\alpha=u^\beta\epsilon_{\beta\alpha}\,,&&
\bar u^{\dot\alpha}=\bar u_{\dot\beta}\epsilon^{\dot\beta\dot\alpha}\,,&&
\bar u_{\dot\alpha}=\epsilon_{\dot\alpha\dot\beta}\bar u^{\dot\beta}\,,
\end{align}
where the antisymmetric Levi-Civita tensor $\epsilon$ is defined  as follows
\begin{align}\label{eps-n}
\epsilon_{12}=\epsilon^{12}=-\epsilon_{\dot1\dot2}=-\epsilon^{\dot1\dot2}=1.
\end{align}
 Note  that for this definition
${\epsilon_\alpha}^\beta=-{\epsilon^\beta}_\alpha=\delta^\beta_\alpha$ and
${\epsilon^{\dot\alpha}}_{\dot\beta}=-{\epsilon_{\dot\beta}}^{\dot\alpha}=
\delta_{\dot\beta}^{\dot\alpha}$ and
$(\epsilon^{\alpha\beta})^*=\epsilon^{\dot\beta\dot\alpha}$. For arbitrary Weyl spinors we
define the invariant products as
\begin{align}\label{}
(uv)=u^\alpha v_\alpha = - u_\alpha v^\alpha\,, &&
(\bar u\bar v)=\bar u_{\dot\alpha} \bar v^{\dot\alpha} = -\bar u^{\dot\alpha} \bar v_{\dot\alpha}\,.
\end{align}
The gluon strength tensor $F_{\mu\nu}$ and its dual $\widetilde F^{\mu\nu}=\frac12
\epsilon^{\mu\nu\rho\sigma} F_{\rho\sigma}$ can be decomposed as
\begin{align}
F_{\alpha\beta,\dot\alpha\dot\beta}=&\sigma^\mu_{\alpha\dot\alpha}\sigma^\nu_{\beta\dot\beta}
F_{\mu\nu}=
2\left(\epsilon_{\dot\alpha\dot\beta} f_{\alpha\beta}-
\epsilon_{\alpha\beta} \bar f_{\dot\alpha\dot\beta}
\right)\,,
\notag\\
i {\widetilde F}_{\alpha\beta,\dot\alpha\dot\beta}
=&i\sigma^\mu_{\alpha\dot\alpha}\sigma^\nu_{\beta\dot\beta}
\widetilde F_{\mu\nu}=
2(\epsilon_{\dot\alpha\dot\beta}
f_{\alpha\beta}+
\epsilon_{\alpha\beta}\bar f_{\dot\alpha\dot\beta})\,.
\end{align}
Here $f_{\alpha\beta}$ and $\bar f_{\dot\alpha\dot\beta}$ are chiral and anti-chiral
symmetric tensors, $f^*=\bar f$, which belong to $(1,0)$ and $(0,1)$ representations of
the Lorenz group, respectively.

The explicit expression for the matrix $x_{\alpha\dot\alpha}$ reads
\begin{align}\label{xd}
x=\begin{pmatrix}
x_0+x_3 & x_1-ix_2\\
x_1+ix_2 & x_0-x_3
\end{pmatrix} \equiv x_{\mu}\sigma^\mu\,
\end{align}
so that one immediately recognizes the two diagonal entries as the light-cone ``plus'' and ``minus''
components of a four-vector $x_{\mu}$~\cite{Kogut:1969xa},
whereas the two off-diagonal entries correspond to the (complex) coordinates in the transverse plane.
The spinor representation for four-vectors is in this sense similar to introduction of the familiar
light-cone coordinates. In order to retain the Lorentz covariance in the light-cone formalism
one usually introduces two auxiliary light-like vectors
$$
  n^2 =0\,,\qquad \tilde n^2 =0\,,\qquad (n\tilde n)\slashed{=} 0\,,
$$
so that arbitrary four-vector can be expanded as
$$
  (n \tilde n)\, x_\mu = x_+\, \tilde n_\mu + x_-\, n_\mu + (n \tilde n) \,\vec{x}_\perp
$$
where $x_+ = x_\mu n^\mu$ and $x_- = x_\mu \tilde n^\mu$.
The same  decomposition can be made quite elegantly in the spinor formalism
by observing that a light-like vector can always be written as a product of spinors:
\begin{equation}
  n_{\alpha\dot\alpha}=\lambda_{\alpha}\bar\lambda_{\dot\alpha}\,, \qquad
  \tilde n_{\alpha\dot\alpha}=\mu_\alpha\bar\mu_{\dot\alpha}\,,
\label{2:nlambda}
\end{equation}
where $\bar\lambda=\lambda^\dagger$,  $\bar\mu=\mu^\dagger$.
The basis vectors in the plane transverse to $n,\tilde n$ can be chosen as
$\mu_{\alpha}\bar\lambda_{\dot\alpha}$ and $\lambda_{\alpha}\bar\mu_{\dot\alpha}$.
Thus
\begin{equation}
(\mu\lambda)(\bar\lambda\bar\mu)\,x_{\alpha\dot\alpha}=
x_{++}\,\mu_{\alpha} \bar\mu_{\dot\alpha} + x_{--}\,\lambda_{\alpha}\bar\lambda_{\dot\alpha}
+x_{-+}\,\lambda_{\alpha}\bar\mu_{\dot\alpha}+x_{+-}\,\mu_{\alpha}\bar\lambda_{\dot\alpha}\,,
\end{equation}
where
\begin{equation}
 x_{++} = \lambda^\alpha x_{\alpha\dot\alpha}\bar\lambda^{\dot\alpha}\,,\qquad
 x_{+-} = \lambda^\alpha x_{\alpha\dot\alpha}\bar\mu^{\dot\alpha}\,,\qquad
 x_{-+} = \mu^\alpha x_{\alpha\dot\alpha} \bar\lambda^{\dot\alpha}\,,\qquad
 x_{--} = \mu^\alpha x_{\alpha\dot\alpha} \bar\mu^{\dot\alpha}\,.
\end{equation}
so that $x_{++}$ and $x_{--}$ correspond to the ``plus'' and ``minus'' coordinates, respectively,
whereas $x_{+-}$ and $x_{-+}$ are the two (holomorphic and anti-holomorphic) coordinates in the
transverse plane. Note that in difference to Ref.~\cite{Braun:2008ia} we keep the
normalization of the auxiliary spinors $\lambda^\alpha$, $\mu^\alpha$ and
hence the product $n_\mu\tilde n^\mu  = (1/2)(\mu\lambda)(\bar\lambda\bar\mu)$ arbitrary.
This freedom will offer some advantages at the cost of slightly more cumbersome notation.

Similarly, for quark and gluon field operators we define
\footnote{This ``plus-minus'' notation is, unfortunately, somewhat ambiguous
because e.g. $x_{++}=\lambda^\alpha x_{\alpha\dot\alpha}\bar\lambda^{\dot\alpha}$ but
$f_{++}=\lambda^\alpha\lambda^\beta f_{\alpha\beta}$. A more consistent but less intuitive
``lambda-mu'' notation~\cite{Braun:2008ia,Braun:2009vc}
is to write $x_{++} \equiv x_{\lambda\bar\lambda}$,  $x_{+-} \equiv x_{\lambda\bar\mu}$,
 $f_{++}\equiv f_{\lambda\lambda}$, $f_{+-}\equiv f_{\lambda\mu}$ etc.}
\begin{align}\label{2:plusf}
\psi_+=\lambda^\alpha\psi_\alpha\,,&&\chi_+=\lambda^\alpha\chi_\alpha\,, &&
f_{++}=\lambda^\alpha\lambda^\beta f_{\alpha\beta}\,,
\nonumber\\
\bar\psi_+ = \bar\psi_{\dot\alpha}\bar\lambda^{\dot\alpha}\,,&&
\bar \chi_+ = \bar\chi_{\dot\alpha}\bar\lambda^{\dot\alpha}\,, &&
\bar f_{++}= \bar f_{\dot\alpha\dot\beta} \bar\lambda^{\dot\alpha}\bar\lambda^{\dot\beta}\,,
\nonumber\\
 \psi_-=\mu^\alpha \psi_\alpha\,,&&
 \bar \psi_-= \bar\psi_{\dot\alpha}\bar\mu^{\dot\alpha}\,, &&f_{+-}=\lambda^\alpha\mu^\beta
f_{\alpha\beta}\,,
\end{align}
etc., so that, for example
\begin{equation}
 (\lambda\mu)\, \psi_\alpha = \mu_\alpha \psi_+ - \lambda_\alpha \psi_-
\end{equation}
The ``plus'' field components correspond to ``good'' components that present independent degrees
of freedom in the light-cone quantization~\cite{Kogut:1969xa}. For example, $f_{++}$ and $\bar f_{++}$
gluon operators can be expanded in terms of creation and annihilation operators for gluons
with definite helicity, cf.~\cite{Braun:2011aw}.
Further details and useful identities involving the algebra of $\sigma_\mu$ matrices
can be found in~\cite{Braun:2008ia,Braun:2009vc}.

Finally we note that in a vector theory like QCD contributions of left-handed and
right-handed quarks, $\psi_\alpha$ and $\bar\chi^{\dot\alpha}$, always enter in parallel.
In the subsequent sections we mostly consider operators built of left-handed quarks;
the remaining ones can be added in the final answers using general symmetry considerations.

\subsection{Conformal symmetry and the $SU(1,1)$ scalar product}

Despite being broken by quantum corrections, conformal symmetry of the QCD Lagrangian has important
consequences and in particular appears to be very useful in the analysis of scale dependence
of physical observables, see Ref.~\cite{Braun:2003rp} for a review. In the applications to
hard processes the so-called collinear subgroup $SL(2,\mathbb{R})$ of the conformal group
plays a special role. It corresponds to fractional-linear transformations of the coordinate
along a certain light-like direction which is fixed by the kinematics of the process:
\begin{equation}
 x_\mu = z n_\mu\,,\qquad z\to z'= \frac{az+b}{cz+d}\,,
\end{equation}
where $a,b,c,d$ are real numbers, $ad-bc =1$. In particular we will be dealing with quark and
gluon fields ``living'' on the light cone:
\begin{equation}
 q(z)\equiv q(zn)\,,\qquad F_{\mu\nu}(z)\equiv F_{\mu\nu}(zn)\,.
\end{equation}
Representations of the $SL(2,\mathbb{R})$ group $T^j$ are labeled by conformal spin $j$
which can take integer and/or half-integer values.
For a generic field (or function of the coordinate) the corresponding transformation law is
\begin{align}\label{SL2R}
\varphi(z)\to T^j\varphi(z) = \frac{1}{(cz+d)^{2j}}\varphi\left(\frac{az+b}{cz+d}\right)\,,
\end{align}
The generators of the infinitesimal transformations corresponding
to~(\ref{SL2R}) take the form
\begin{align}\label{generators}
S_+=z^2\partial_z+2jz\,, && S_0=z\partial_z+j\,, && S_-=-\partial_z\,
\end{align}
and satisfy the usual $SL(2)$ algebra
\begin{align}\label{generators1}
[S_+,S_-] = 2 S_0\,, && [S_0,S_\pm] = \pm S_\pm\,.
\end{align}
The quadratic Casimir operator for functions of one variable is just a number:
\begin{equation}
  S^2 = S_0^2 -S_0 + S_+S_-\,,\qquad [S^2,S_i]=0\,,\qquad S^2  = j(j-1)\,.
\end{equation}
The separation of quark and gluon fields in ``plus'' and ``minus'' components simultaneously
serves to separate the terms that transform according to different representations of the
collinear conformal group.
One can show \cite{Braun:2003rp,Braun:2008ia} that ``minus'' projections of the quark fields
$\psi_-,\bar\chi_-$ correspond to $j=1/2$, ``plus'' projections of the quark fields $\psi_+,\bar\chi_+$
together with $f_{+-}$, $\bar f_{+-}$ gluons transform according to  $j=1$,
whereas $f_{++}, \bar f_{++}$ have spin $j=3/2$.

Multiplicatively renormalizable quark-antiquark operators of leading twist can be constructed
(see e.g.~\cite{Braun:2003rp,Braun:2008ia}) as highest weights of  irreducible components in the direct product
$T^{j=1}\otimes T^{j=1}$ corresponding to the total conformal spin $j=N+2$ where $N$ is the
total number of derivatives. E.g. for the left-handed quarks
\begin{eqnarray}
\mathcal{O}_N(y) &=& (-\partial_+)^N \bar\psi_+(y) C_N^{3/2}
\left(\frac{\stackrel{\rightarrow}{D}_+-\stackrel{\leftarrow}{D}_+}
           {\stackrel{\rightarrow}{D}_++\stackrel{\leftarrow}{D}_+} \right)\psi_+(y)
\label{ON}
\end{eqnarray}
where $C_N^{3/2}(x)$ is the Gegenbauer polynomial, $D_\mu = \partial_\mu -i g A_\mu$
is the covariant derivative and
$\partial_+ = \partial_\mu n^\mu = (1/2)\lambda^\alpha\partial_{\alpha\dot\alpha}\bar\lambda^{\dot\alpha}$.
For simplicity we tacitly assume flavor nonsinglet operators.

The generators of conformal transformations acting on 
products
of fields at different light-cone positions (light-ray operators) are
given by the sum of the
generators acting on the field coordinates, e.g.
\begin{eqnarray}
  S^{(j_1,j_2)}_{+} &=& S^{(j_1)}_{+} + S^{(j_2)}_{+} = z_1^2\partial_{z_1}+z_2^2\partial_{z_2}+2j_1 z_1++2j_2 z_2\,,
\nonumber\\
  S^{(j_1,j_2,j_3)}_{+} &=&
 z_1^2\partial_{z_1}+z_2^2\partial_{z_2}+z_3^2\partial_{z_3}+2j_1 z_1++2j_2 z_2+2j_3 z_3\,,
\end{eqnarray}
etc. For brevity we will sometimes omit the conformal spin indices if their values and also the
number of the fields are clear from the context.

Light-ray operators
can be viewed as generating functions for local operators of a given twist. Let
\begin{equation}
  O_{++}(z_1,z_2) = \bar \psi_+(z_1)[z_1,z_2]\psi_+(z_2)\,,
\label{O++}
\end{equation}
where
\begin{equation}
  [z_1,z_2] = \mathrm{Pexp}\Big\{\frac12 igz_{12}\int_{0}^{1} \!d u\,A_{++}(z_{21}^u)\Big\}
\end{equation}
is the light-like Wilson line. Here and below we use the following shorthand notations:
\begin{equation}
 z_{12} = z_1 -z_2\,, \qquad z_{21}^u = \bar u z_2 + u z_1\,,\qquad \bar u = 1-u\,.
\end{equation}
The conformal operator (\ref{ON}) can be written in terms of $O_{++}(z_1,z_2)$ as
\begin{equation}\label{ONz12}
 \mathcal{O}_N(y) = (\partial_{z_1}+\partial_{z_2})^N C_N^{3/2}
\left(\frac{\partial_{z_1}-\partial_{z_2}}{\partial_{z_1}+\partial_{z_2}}\right)
O_{++}(z_1n+y,z_2n+y)\Big|_{z_i=0}\,.
\end{equation}
Alternatively, one can expand the light-ray operator (\ref{O++}) in terms of conformal operators:
\begin{eqnarray}\label{Oconf-ex}
 O_{++}(z_1,z_2) &=& \sum_N \varkappa_N z_{12}^N \int_0^1\!du\, (u\bar u)^{N+1}\, \mathcal{O}_N(z_{21}^u)
 \nonumber\\&=&
\sum_N \varkappa_N z_{12}^N \sum_k\frac{1}{k!}\int_0^1\!du\, (u\bar u)^{N+1}\,(z_{21}^u)^k  \mathcal{O}_{Nk}(0)\,,
\end{eqnarray}
where
\begin{equation}
 \varkappa_N = \frac{2(2N+3)}{(N+1)!}\,
\end{equation}
and
\begin{equation}
  \mathcal{O}_{Nk} = \partial_+^k \mathcal{O}_{N}\,.
\label{eq:ONk}
\end{equation}
The operators $\mathcal{O}_{Nk}$, $k=0,1,2,\ldots$
form what is usually referred to as conformal tower~\cite{Braun:2003rp} and the lowest
dimension operator in a tower, $\mathcal{O}_{N0} \equiv \mathcal{O}_{N}$, is called a conformal operator.
A formal definition is that a conformal operator is annihilated by the special
conformal transformations, $\delta\mathcal{O}_{N}=[\tilde n_\mu K^\mu,\mathcal{O}_{N} ]=0$ (see below).

Note that the interpretation (or definition) of a light-ray operator as the generating
function means that one has effectively to deal with polynomials in light-cone coordinates $z_i$.
It turns out to be convenient to consider $z_i$ as complex variables and exploit the
equivalence of the $SL(2,R)$ and $SU(1,1)$ groups.
A $SU(1,1)$ transformation is defined by the same formula~(\ref{SL2R}),
where $a,b,c,d$ are now complex numbers such that $d=\bar a$ and $c=\bar b$:
\begin{equation}
  z'= \frac{az+b}{\bar b z+\bar a}\,.
\end{equation}
The operator $T^j$, Eq.~(\ref{SL2R}), is a unitary operator with respect to the following scalar product:
\begin{equation}
\langle\phi,\psi\rangle_j =
\frac{2j-1}{\pi} \int\limits_{|z|<1}\!\!d^2z\,
(1-|z|^2)^{2j-2} \bar\phi(z)\psi(z) \equiv
\int\limits_{|z|<1}\!\!\mathcal{D}_jz\,\bar\phi(z)\psi(z)\,,
\qquad 
\end{equation}
where the integration goes over the interior of the unit circle
(i.e. over the unit disc) in the complex plane.
In what follows we will often drop the subscript $j$ in the notation
for the scalar product $\langle\ldots\rangle_j$ and the measure $\int\!\mathcal{D}_jz$
if this cannot yield confusion.
The $SU(1,1)$-invariance implies that
\begin{eqnarray}
  \langle T^j \phi,T^j \psi\rangle &=& \langle\phi,\psi\rangle
\end{eqnarray}
which can be checked by explicit calculation using that
\begin{eqnarray}
  d^2z' &=& \frac{d^2z}{(\bar bz + \bar a)^2 (b\bar z + a)^2}\,.
\end{eqnarray}
It is easy to show that for simple powers
\begin{equation}
 \langle z^n,z^{n'}\rangle =\delta_{nn'}||z^n||^2\,,\qquad ||z^n||^2 = \frac{\Gamma(2j)\,n!}{\Gamma(n+2j)}\,.
\end{equation}
The $\mathrm{SU}(1,1)$ generators are defined as in
Eq.~(\ref{generators}) and have the following hermiticity properties
(with respect to the above scalar product):
\begin{equation}\label{hermiticity}
 S_0^\dagger = S_0\,,\qquad (S_+)^\dagger = - S_-\,.
\end{equation}
Many of the specific techniques that make this realization useful for our present purposes
are based on the following representation of the unit operator (reproducing kernel)
\begin{eqnarray}\label{repr}
\phi(z) &=& \int_{|w|<1}\!\!\!\!\mathcal{D}_{j}w\,\mathcal{K}_j(z,\bar w)
\phi(w)\,,
\end{eqnarray}
where
\begin{align}\label{}
\mathcal{K}_j(z,\bar w)=\frac{1}{(1-z\bar w)^{2j}}\,
\end{align}
which will be used in what follows.

This construction is easily generalized for functions of several variables.
The generators of conformal transformations are defined as the sums of the generators
in Eq.~(\ref{generators}) with spins $j_1,\ldots,j_n$ which act on the corresponding
variables $z_1,\ldots, z_n$. The scalar product
is also modified in an appropriate way,
e.g. for two variables it takes the form
\begin{equation}\label{multisc}
\langle\phi,\psi\rangle_{j_1,j_2} =
\int\limits_{|z_1|<1}\!\!\mathcal{D}_{j_1}z_1\int\limits_{|z_2|<1}\!\!\mathcal{D}_{j_2}z_2\,\bar\phi(z_1,z_2)\psi(z_1,z_2)\,.
\end{equation}
One can show that
\begin{eqnarray}\label{eq:z12n}
||z_{12}^n||^2_{j_1j_2}&=&n!\frac{\Gamma(2j_1)\Gamma(2j_2)}{\Gamma(n+2j_1)\Gamma(n+2j_2)}
\frac{\Gamma(2n+2j_1+2j_2-1)}{\Gamma(n+2j_1+2j_2-1)}\,.
\end{eqnarray}

With the help of the $SU(1,1)$ scalar product one can project the conformal operator
(\ref{ON}) from the light-ray operator $O_{++}(z_1,z_2)$ as follows~\cite{Belitsky:2005gr}
\begin{equation}
\mathcal{O}_{N} = \convent_N \big\langle z_{12}^N,O_{++}(z_1,z_2)\big\rangle_{11}
=\convent_N 
\iint\limits_{|z_i|<1} \mathcal{D}_1z_1 \mathcal{D}_1{z_2} \, \bar z_{12}^N\,
O_{++}(z_1,z_2)\,,
\label{eq:Copera}
\end{equation}
where
\begin{align}
\convent_N =\frac12 (N+1)(N+2)!\,.
\label{eq:rhoN}
\end{align}
This representation can be proven in the following way.
We remind that the operator is called conformal if it is annihilated by the generator of special
conformal transformations, $\delta_K\mathcal{O}_{N}=[\tilde n_\mu K^\mu,
\mathcal{O}_{N} ]=0$.  The action of the quantum operator $\tilde n_\mu K^\mu$ on the
quantum fields can be replaced by the differential operator acting on the field
coordinates, in particular
\begin{align}
\delta_K O_{++}(z_1,z_2)=2(n\bar n)\, S_+^{(1,1)}O_{++}(z_1,z_2)\,,
\end{align}
where $$S_+^{(1,1)}=S^+_{z_1}+S^+_{z_2} = z_1^2\partial_{z_1}+z_2^2\partial_{z_2}+ 2 z_1 + 2 z_2.$$
Applying the special conformal transformation to Eq.~(\ref{eq:Copera}) one obtains
\begin{eqnarray}
 \delta_K\mathcal{O}_{N}&=&
\convent_N  \iint\limits_{|z_i|<1} \mathcal{D}z_1 \mathcal{D}{z_2}\, \bar z_{12}^N\, \delta_K O_{++}(z_1,z_2)
 \sim  \iint\limits_{|z_i|<1} \mathcal{D}z_1 \mathcal{D}{z_2}\, \bar z_{12}^N\, S^{(1,1)}_+
O_{++}(z_1,z_2)
\nonumber\\
&=& - \iint\limits_{|z_i|<1} \mathcal{D}z_1 \mathcal{D}{z_2} \Big(S_-^{(1,1)}z_{12}^N\Big)^*
O_{++}(z_1,z_2) = 0\,,
\end{eqnarray}
as expected. The result is zero because $S_-^{(1,1)} = -(\partial_{z_1}+\partial_{z_2})$,
and thus $S_-^{(1,1)}(z_1-z_2)^N=0$. It follows that the l.h.s. of Eq.~(\ref{eq:Copera})
is indeed a conformal operator, up to a (convention-dependent) coefficient which can be
calculated explicitly. A useful formula is
\begin{eqnarray}
\frac{\rho_N}{\pi^2}\iint\limits_{|z_i|<1} d^2 z_1 d^2z_2 \, (\bar z_1-\bar z_2)^N e^{ip_1z_1+ip_2z_2}
&=&i^N (p_1+p_2)^N C^{3/2}_N\left(\frac{p_1-p_2}{p_1+p_2}\right)\,.
\end{eqnarray}

In the same way one finds that
\begin{equation}
\mathcal{O}_{Nk} = \convent_N \big\langle \Psi^{t=2}_{Nk}(z_1,z_2),O_{++}(z_1,z_2)\big\rangle_{11}
=\convent_N
\iint\limits_{|z_i|<1} \mathcal{D}_1z_1 \mathcal{D}_1{z_2} \, \Psi^{t=2}_{Nk}(\bar z_1,\bar z_2)\,
O_{++}(z_1,z_2)\,,
\label{eq:Cotower}
\end{equation}
where
\begin{equation}
   \Psi^{t=2}_{Nk}(z_1,z_2) = (S^{(1,1)}_+)^k z_{12}^N
\label{eq:psit2}
\end{equation}
can be thought of as a ``wave function''. The superscript $t=2$ serves to remind that we are dealing
here with twist-two operators.

The functions $\Psi^{t=2}_{Nk}(z_1,z_2)$ form an orthogonal basis
\begin{equation}
  \big\langle \Psi^{t=2}_{Nk},\Psi^{t=2}_{N'k'}\big\rangle = \delta_{NN'}\delta_{kk'}||\Psi^{t=2}_{Nk}||^2\,,
\end{equation}
so that one can represent the nonlocal operator in the form
\begin{align}\label{expansion++}
O_{++}(z_1,z_2)=\sum_{N=0}^\infty\sum_{k=0}^\infty\convent_N^{-1}||\Psi^{t=2}_{Nk}||^{-2}\Psi^{t=2}_{Nk}(z_1,z_2)\mathcal{O}_{Nk}\,.
\end{align}
The norm $||\Psi^{t=2}_{Nk}||^2$ can be calculated recurrently
$$
||\Psi^{t=2}_{Nk}||^2=\big\langle S_+^kz_{12}^N,S_+^kz_{12}^N\big\rangle = - \big\langle
S_-S_+^kz_{12}^N,S_+^{k-1}z_{12}^N\big\rangle=k(2N+k+3)||\Psi^{t=2}_{Nk-1}||^2.
$$
One obtains
\begin{equation}
 ||\Psi^{t=2}_{Nk}||^2 = ||z^N_{12}||_{11}^2\, p_{Nk}^{-1}\,,\qquad  p_{Nk} = \frac1{k!}\frac{\Gamma(2N+4)}{\Gamma(2N+4+k)}
\label{eq:pNk}
\end{equation}
so that finally
\begin{equation}\label{ONk}
O_{++}(z_1,z_2)=\sum_{N=0}^\infty\sum_{k=0}^\infty \omega_{Nk}\Psi^{t=2}_{Nk}(z_1,z_2)\mathcal{O}_{Nk}(0)\,,
\end{equation}
where
\begin{equation}\label{omegaNk}
  \omega_{Nk} = \varkappa_N \frac{1}{k!}\frac{\Gamma(N+2)\Gamma(N+2)}{\Gamma(2N+4+k)}\,.
\end{equation}
Explicit expression for the ``wave functions'' $\Psi^{t=2}_{Nk}(z_1,z_2)$ can be obtained
observing that $S_+$ is the generator of special conformal transformations so that
$\exp[aS_+]$ corresponds to a finite conformal transformation $z\to z/(1-az)$.
Therefore
\begin{eqnarray}
   \exp[aS_+^{(j_1,j_2)}] z_{12}^N &=& \frac{z_{12}^N}{(1-az_1)^{N+2j_1} (1-az_2)^{N+2j_2}}
\nonumber\\
  &=& \frac{\Gamma[2N+2j_1+2j_2]}{\Gamma[N+2j_1]\Gamma[N+2j_2]}
  \int_0^1\!dt\, \frac{z_{12}^N\, t^{N+2j_1-1}\bar t^{N+2j_2-1}}{(1-az_{21}^t)^{2N+2j_1+2j_2}},
\end{eqnarray}
where $j_1=j_2=1$ for the case at hand.
In order to find $\Psi^{t=2}_{Nk}(z_1,z_2)$ one has to differentiate this expression
$k$ times in $a$ and set $a=0$.
This gives:
\begin{eqnarray}
  (S_+^{(j_1,j_2)})^k z_{12}^N = z_{12}^N \frac{\Gamma[2N+2j_1+2j_2+k]}{\Gamma[N+2j_1]\Gamma[N+2j_2]}
  \int_0^1\!dt\, t^{N+2j_1-1}\bar t^{N+2j_2-1}\, (z_{21}^t)^k.
\label{eq:itrafo}
\end{eqnarray}
Using this representation the expansion~(\ref{ONk}) can easily be recast in
the form~(\ref{Oconf-ex}). What has to be
stressed here is that the functions which project the local operators from the nonlocal
one are the same functions that appear in the expansion of the nonlocal operator over the
conformal operators and its descendants (conformal tower).
This is a general property which holds for
arbitrary operators~\cite{Braun:2008ia,Derkachov:2010zz}.

%
\subsection{Conformal non-quasipartonic operators}
%

All existing (gauge-invariant) composite operators in QCD can be divided in two
classes: quasipartonic~\cite{Bukhvostov:1985rn} and non-quasipartonic.
By definition, operators built of ``plus'' components of the fields and arbitrary amount of
``plus'' covariant derivatives are called quasipartonic. The light-ray operator defined
in Eq.~(\ref{O++}) is the generating function for local quasipartonic operators of the leading twist.
The complete operator basis for the leading twist-two and also subleading twist-three operators can
always be chosen to contain quasipartonic operators only. The subset of quasipartonic operators of a
given twist is special in that it is closed under renormalization~\cite{Bukhvostov:1985rn} at one-loop.
Moreover, since building blocks of quasipartonic operators
transform as the primary fields under $SL(2,\mathbb{R})$ transformations, the corresponding RG equations
are explicitly $SL(2,\mathbb{R})-$invariant~\cite{Bukhvostov:1985rn,Braun:2008ia}.
Examples of twist-four quasipartonic operators are $\bar \psi_+\psi_+\bar\psi_+\psi_+$,
$\bar\psi_+ f_{++}\bar f_{++} \psi_+$, etc.

All other operators are naturally called non-quasipartonic, e.g. they can include ``minus'' fields
components and ``minus'' or transverse derivatives. Non-quasipartonic operators cannot be avoided
starting with twist-four. The complete operator basis for
twist-four operators, including non-quasipartonic ones, was constructed in \cite{Braun:2008ia}.

The main problem in dealing with non-quasipartonic operators is due to transverse derivatives
which, generally, do not have ``good'' conformal properties. Note that in the spinor formalism
we distinguish two transverse derivatives: $D_{+-}$ and $D_{-+}$. One can show~\cite{Braun:2008ia}
that e.g. $D_{+-}\psi_+$ transforms as a primary field with conformal spin $j=3/2$, whereas
$D_{-+}\psi_+$ does not transform homogeneously, it is a ``bad'' object as far as the $SL(2)$ symmetry
is concerned. The procedure suggested in Ref.~\cite{Braun:2008ia} is to keep the ``good'' derivatives
and get rid of the ``bad'' ones using equations of motion $\bar D^{\dot\alpha\alpha} \psi_\alpha =0$ etc.
The possibility to make this separation is the crucial advantage of the spinor formalism in the
present context.

In this work we will be interested in operators with quantum numbers such that they
can contribute to the time-ordered product of two electromagnetic currents.
With the further restriction to flavor-nonsinglet contributions,
we are left with three (non-quasipartonic) operators of twist-four that involve
the self-dual gluon field $f$,
\begin{align}
Q_1(z_1,z_2,z_3)=&(\bar\lambda\bar\mu)\bar\psi_+(z_1)f_{+-}(z_2)\psi_+(z_3)\,,
&& T^{j=1}\otimes T^{j=1}\otimes T^{j=1}\,,
\notag\\
Q_2(z_1,z_2,z_3)=&(\bar\lambda\bar\mu)\bar\psi_+(z_1)f_{++}(z_2)\psi_-(z_3)\,,
&& T^{j=1}\otimes T^{j=3/2}\otimes T^{j=1/2}\,,
\notag\\
Q_3(z_1,z_2,z_3)=&\frac12(\bar\lambda\bar\mu)[D_{-+}\bar\psi_+](z_1)f_{++}(z_2)\psi_+(z_3)\,,
&& T^{j=3/2}\otimes T^{j=3/2}\otimes T^{j=1}\,,
\label{eq:Q}
\end{align}
and  three operators involving $\bar f$:
\begin{align}
\bar Q_1(z_1,z_2,z_3)=&(\mu\lambda)\bar\psi_+(z_1)\bar f_{+-}(z_2)\psi_+(z_3)\,,
&& T^{j=1}\otimes T^{j=1}\otimes T^{j=1}\,,
\notag\\
\bar Q_2(z_1,z_2,z_3)=&(\mu\lambda)\bar\psi_-(z_1)\bar f_{++}(z_2)\psi_+(z_3)\,,
&& T^{j=1/2}\otimes T^{j=3/2}\otimes T^{j=1}\,,
\notag\\
\bar Q_3(z_1,z_2,z_3)=&\frac12(\mu\lambda)\bar\psi_+(z_1)\bar f_{++}(z_2)[D_{+-}\psi_+](z_3)\,,
&& T^{j=1}\otimes T^{j=3/2}\otimes T^{j=3/2}\,.
\label{eq:barQ}
\end{align}
The rationale for including the factors $(\mu\lambda)$ and $(\bar\lambda\bar\mu)$ in the definition of the
operators is that, in this form, they contain equal amount of auxiliary spinors with and without
a ``bar'' (i.e. equal amount of
$\lambda$ and $\bar \lambda$ and separately $\mu$ and $\bar\mu$) and therefore can be rewritten
in terms of auxiliary vectors $n$, $\tilde n$, e.g.
\begin{align}
Q_1(n,\tilde n; z_1,z_2,z_3) = \bar\psi(z_1n)\bar n f(z_2n) \tilde n \bar n \psi(z_3n)\,.
\end{align}
Note that
$$(\mu\lambda)(\bar\lambda\bar\mu) = 2(n\cdot \tilde n).$$
Operators in usual vector notation can easily be converted to this basis, e.g.
\begin{eqnarray}
(n\tilde n)
\bar q_L(z_1)\big[F_{+\mu}(z_2)+ i \widetilde F_{+\mu}(z_2)\big]\gamma^\mu q_L(z_3)
 = Q_2(z_1,z_2,z_3) - Q_1(z_1,z_2,z_3)\,,
\nonumber\\
(n\tilde n)
\bar q_L(z_1)\big[F_{+\mu}(z_2) - i \widetilde F_{+\mu}(z_2)\big]\gamma^\mu q_L(z_3)
 = \bar Q_2(z_1,z_2,z_3) - \bar Q_1(z_1,z_2,z_3)\,,
\end{eqnarray}
where $q_L = \frac12(1-\gamma_5)q$.

The operators with and without a ``bar'' are related by hermitian conjugation:
\begin{equation}
 \bar Q_i(z_1,z_2,z_3)=(Q_i(z_3,z_2,z_1))^\dagger
\label{eq:herm}
\end{equation}
and transform as direct products of primary fields under $SL(2,\mathbb{R})$ transformations,
as indicated.
We stress that the operators that include a transverse derivative, $Q_3$ and $\bar Q_3$,
must be included in the basis in order to ensure simple realization of the conformal symmetry,
although they may be dispensed off at the later stage (see below).

The following combinations
\begin{align}\label{Qpm}
Q^\pm_k(z_1,z_2,z_3)=Q_k(z_1,z_2,z_3)\pm \bar Q_k(z_3,z_2,z_1)
\end{align}
transform in a different way under the combined charge conjugation and parity
transformations and therefore cannot mix under renormalization.
By a direct calculation one finds
\begin{equation}\label{QpmPC}
Q^\pm_k(n,\tilde n; z_1,z_2,z_3)\xrightarrow{\rm CP} \pm Q^\pm_k(\tilde n, n; z_1,z_2,z_3)\,,
\end{equation}
where we have displayed explicitly the dependence on the auxiliary vectors $n$
and $\tilde n$. Note that the auxiliary vectors are interchanged $n\leftrightarrow \tilde n$
by tbe CP transformation. For comparison, the leading-twist light-ray operator~(\ref{O++})
transforms as
\begin{align}
O_{++}(n,z_1,z_2)\xrightarrow{\rm CP} - O_{++}(\tilde n,z_2,z_1)\,,
\end{align}
which implies that for conformal operators ~(\ref{eq:Copera})
\begin{align}\label{QNPC}
\mathcal{O}_N(n)\xrightarrow{\rm CP} (-1)^{N+1}\mathcal{O}_N(\tilde n)\,.
\end{align}
Note that interchanging the auxiliary vectors does not have any physical significance
as they only serve to simplify the leading-twist projection for local operators.

To save space, in the following discussion we will sometimes use vector notation:
\begin{equation}
  \overrightarrow{Q}(z_1,z_2,z_3) =
\begin{pmatrix}
{Q}_1(z_1,z_2,z_3)\\
{Q}_2(z_1,z_2,z_3)\\
{Q}_3(z_1,z_2,z_3)
\end{pmatrix}
\end{equation}
and similar for $\bar Q_i$.

\subsection{Renormalization and hermiticity}

Let  $\overrightarrow{P}(z_1,z_2,z_3,z_4)$ be the complete set of quasipartonic twist-four operators
of the type  $\bar \psi_+\psi_+\bar\psi_+\psi_+$, $\bar\psi_+ f_{++}\bar f_{++} \psi_+$, etc. Explicit
expressions will not be needed in what follows. The complete RG equation (to the one-loop accuracy)
has a block-triangular form
\begin{align}\label{RGQP}
\left(\mu\frac{\partial}{\partial\mu}+\beta(\alpha_s)\frac{\partial}{\partial\alpha_s}\right)
\begin{pmatrix}
\overrightarrow{Q}\\
\overrightarrow{\bar Q}\\
\overrightarrow{P}
\end{pmatrix}=-\frac{\alpha_s}{2\pi}
\begin{pmatrix}
\mathbb{H}_{QQ}& 0 &\mathbb{H}_{QP}\\
0 & \mathbb{H}_{\bar Q\bar Q}&\mathbb{H}_{\bar QP}\\
0 & 0 & \mathbb{H}_{PP}
\end{pmatrix}
\begin{pmatrix}
\overrightarrow{Q}\\
\overrightarrow{\bar Q}\\
\overrightarrow{P}
\end{pmatrix}\,.
\end{align}
so that the quasipartonic operators have autonomous evolution, whereas the non-quasi\-partonic
ones,
$\overrightarrow{Q}$ and $\overrightarrow{\bar Q}$, do not mix with each other
but can mix with the quasipartonic operators.
Due to this structure, the RG equation (\ref{RGQP}) actually decouples in two equations
which contain $Q$ and $\bar Q$ operators:
\begin{eqnarray}\label{RGQP1}
\left(\mu\frac{\partial}{\partial\mu}+\beta(\alpha_s)\frac{\partial}{\partial\alpha_s}\right)
\begin{pmatrix}
\overrightarrow{Q}\\
\overrightarrow{P}
\end{pmatrix}&=&-\frac{\alpha_s}{2\pi}
\begin{pmatrix}
\mathbb{H}_{QQ}& \mathbb{H}_{QP}\\ 0 & \mathbb{H}_{PP}
\end{pmatrix}
\begin{pmatrix}
\overrightarrow{Q}\\
\overrightarrow{P}
\end{pmatrix}\,,
\nonumber\\
\left(\mu\frac{\partial}{\partial\mu}+\beta(\alpha_s)\frac{\partial}{\partial\alpha_s}\right)
\begin{pmatrix}
\overrightarrow{\bar Q}\\
\overrightarrow{P}
\end{pmatrix}&=&-\frac{\alpha_s}{2\pi}
\begin{pmatrix}
 \mathbb{H}_{\bar Q\bar Q}&\mathbb{H}_{\bar QP}\\
 0 & \mathbb{H}_{PP}
\end{pmatrix}
\begin{pmatrix}
\overrightarrow{\bar Q}\\
\overrightarrow{P}
\end{pmatrix}\,.
\end{eqnarray}
The operators $Q_i$ and $\bar Q_i$ are hermitian conjugate to one another, cf.~(\ref{eq:herm}),
so that the two RG equations in (\ref{RGQP1}) are equivalent and it is sufficient to consider
one of them.
This implies in particular that the spectrum
of one-loop anomalous dimensions of non-quasipartonic operators is double-degenerate: there are two operators
with different CP-parity for each anomalous dimension.

The ``Hamiltonians'' $\mathbb{H}$ are known
explicitly~\cite{Braun:2009vc} and can be written in terms of $SL(2)$ invariant
integral operators (kernels). ``Diagonal'' Hamiltonians $\mathbb{H}_{QQ}$, $\mathbb{H}_{\bar Q\bar Q}$
and $\mathbb{H}_{PP}$
have a two-particle structure, i.e. each contributing kernel
involves light-cone coordinates of two field operators only, whereas the mixing
ones,  $H_{QP}$ and $H_{\bar QP}$, involve $2\to 3$ transitions.

The light-ray operators $\overrightarrow{P}$, $\overrightarrow{Q}$, $\overrightarrow{\bar Q}$ can be expanded in
multiplicatively renormalizable local operators, cf. Eq.~(\ref{expansion++}).
Thanks to the block-triangular structure of the mixing matrix, expansion of
$\overrightarrow{P}$ only involves quasipartonic local operators, which we denote by $\mathcal{P}_{Np}$,
whereas the expansion of $\overrightarrow{Q}$ ($\overrightarrow{\bar Q}$) involves both
$\mathcal{P}_{Np}$ and multiplicatively
renormalizable non-quasipartonic operators, which we denote by $\mathcal{Q}_{Np}$ ($\bar{\mathcal{Q}}_{Np}$):
\begin{eqnarray}
 \overrightarrow{P} &=& \sum_{Np} a_{Np}(z_1,z_2,z_3,z_4)\mathcal{P}_{Np}\,,
\nonumber\\
 \overrightarrow{Q} &=& \sum_{Np} b_{Np}(z_1,z_2,z_3)\mathcal{Q}_{Np}
 + \sum_{Np} c_{Np}(z_1,z_2,z_3)\mathcal{P}_{Np}\,,
\nonumber\\
 \overrightarrow{\bar Q} &=& \sum_{Np} \bar b_{Np}(z_1,z_2,z_3){\mathcal{Q}}_{Np}
 + \sum_{Np} \bar c_{Np}(z_1,z_2,z_3)\mathcal{P}_{Np}\,,
\label{eq:PQ}
\end{eqnarray}
where $N$ is the total amount of derivatives
(which specifies the operator dimension) and $p$ is a multi-index that enumerates different operators.

Substituting this expansion in the RG equation (\ref{RGQP}) one finds that
the coefficients of non-quasipartonic local operators $b_{Np}(z_1,z_2,z_3)$ ($\bar b_{Np}(z_1,z_2,z_3)$)
can be found as solutions of the integral equation
that only involves $\mathbb{H}_{QQ}$ ($\mathbb{H}_{\bar Q\bar Q}$):
\begin{eqnarray}
 \mathbb{H}_{QQ}\, b_{Np}(z_1,z_2,z_3) &=& \gamma_{Np}\, b_{Np}(z_1,z_2,z_3)\,,
\nonumber\\
 \mathbb{H}_{\bar Q\bar Q}\, \bar b_{Np}(z_1,z_2,z_3) &=& \gamma_{Np}\, \bar b_{Np}(z_1,z_2,z_3)\,,
\label{eq:Hb}
\end{eqnarray}
where $\gamma_{Np}$ are the corresponding anomalous dimensions.
The two equations in Eq.~(\ref{eq:Hb}) are related by hermitian conjugation, cf.~(\ref{eq:Hb}),
so that $\bar b_{Np}(z_1,z_2,z_3) =  \pm b_{Np}(z_3,z_2,z_1)$, depending on the parity of
the operator $\mathcal{Q}_{Np}$ under the CP-conjugation.

In the context of this work we are interested in a particular set of contributions that
are related to the divergence of the leading twist conformal operator $\mathcal{O}_N$,
which we denote, schematically, $(\partial\mathcal{O})_N$, and its descendants obtained by
adding arbitrary amount of ``plus'' derivatives,
$\partial_+^k(\partial\mathcal{O})_N$. These operators can be identified as the
non-quasipartonic twist-four operators with
the anomalous dimensions coinciding with the anomalous dimensions of leading-twist operators.
Hence the corresponding coefficient functions $b_{Nk}(z_1,z_2,z_3)$ correspond to the subset
of solutions of the integral equation in Eq.~(\ref{eq:PQ}) with the eigenvalues $\gamma_N$
that are usual flavor-nonsinglet leading-twist anomalous dimensions:
\begin{equation}
   \gamma_N = C_F\left(1-\frac{2}{(N+1)(N+2)} + 4\sum_{m=2}^{N+1}\frac{1}{m}\right)
= 2 C_F\Big[\psi(N+3)+\psi(N+1)-\psi(3)-\psi(1)\Big]\,.
\label{eq:gammaN}
\end{equation}
Note that the quasipartonic operators completely decouple and their contributions can be ignored
at any stage of the calculation.

The operator $\mathbb{H}_{QQ}$ is a $3\times 3$ matrix with the entries being two-particle integral
operators:
\begin{eqnarray}
   \mathbb{H}_{QQ}\, \overrightarrow{\Psi} &=&
\begin{pmatrix}
H_{11}& H_{12} & H_{13}\\H_{21}& H_{22} & H_{23} \\ H_{31}& H_{32} & H_{33}
\end{pmatrix}
\begin{pmatrix}
       \Psi_1^{(1,1,1)}\\  \Psi_2^{(1,\frac32,\frac12)} \\ \Psi_3^{(\frac32,\frac32,1)}
\end{pmatrix}\,,
\end{eqnarray}
where the superscripts on the three components of the ``wave function'' indicate the corresponding
$SL(2)$ representations. Explicit expressions~\cite{Braun:2009vc} are given in Appendix~\ref{App:renorm}.

Unfortunately, Eq.~(\ref{eq:Hb}) appears to be too complicated to be solved directly.
The way out is that the operator $\mathbb{H}_{QQ}$ turns out to be self-adjoint (hermitian)
with respect to the following scalar product:
\begin{align}\label{eq:<<>>}
\langle\!\langle\overrightarrow{\Phi},\overrightarrow{\Psi}\rangle\!\rangle=
2\langle{\Phi}_1,{\Psi}_1\rangle_{111}
+\langle{\Phi}_2,{\Psi}_2\rangle_{1\frac32\frac12}
+\frac12\langle{\Phi}_3,{\Psi}_3\rangle_{\frac32\frac321}\,,
\end{align}
which can be verified by the explicit calculation, cf. Appendix~\ref{App:renorm}.

Thanks to hermiticity it proves to be sufficient to solve, instead of Eq.~(\ref{eq:Hb}), the
(much simpler) inverse problem: find the expansion of the divergence of
the leading-twist conformal operator $(\partial\mathcal{O})_N$ in terms of quark-gluon non-quasipartonic operators.
Examples of this expansion for $N=1,2$ are given in the Introduction,
cf.~Eqs.~(\ref{eq:puzzle}), (\ref{eq:Omualphabeta}).
 We have been able to obtain this expansion for arbitrary $N$ and bring the answer to the form
\begin{equation}
(\mu\lambda) (\bar\lambda\bar\mu) (\partial\mathcal{O})_N = \frac{ig\rho_N}{(N+1)^2}\Big[
 \langle\!\langle \overrightarrow{\Psi}_N, \overrightarrow{Q}\rangle\!\rangle
 - \langle\!\langle \overrightarrow{\bar\Psi}_N, \overrightarrow{\bar Q}\rangle\!\rangle\Big]
 + \ldots
\label{eq:dO1}
\end{equation}
where the functions
\begin{equation}\label{PsiBPsi}
 \overrightarrow{\bar\Psi}_N(z_1,z_2,z_3) = (-1)^N \overrightarrow{\Psi}_N(z_3,z_2,z_1)
\end{equation}
are known explicitly and the ellipses stand for (irrelevant) contributions of quasipartonic operators.
The corresponding calculation  is presented in Sect.~3. The symmetry relation~(\ref{PsiBPsi})
follows from properties of the operators under CP-transformations, see Eqs.~(\ref{QpmPC}) and (\ref{QNPC}).

Note that the two contributions in the square bracket in Eq.~(\ref{eq:dO1}) are renormalized
multiplicatively with the same anomalous dimension $\gamma_N$ (\ref{eq:gammaN}). Thus we can construct
another multiplicatively renormalizable twist-four operator
\begin{equation}
(\mu\lambda) (\bar\lambda\bar\mu) T_N = \frac{ig\rho_N}{(N+1)^2}\Big[
 \langle\!\langle \overrightarrow{\Psi}_N, \overrightarrow{Q}\rangle\!\rangle
 + \langle\!\langle \overrightarrow{\bar\Psi}_N, \overrightarrow{\bar Q}\rangle\!\rangle\Big]
 + \ldots
\label{eq:TN}
\end{equation}
which has the same leading twist anomalous dimension and is \emph{not}
reduced to total derivatives of the conformal operators. Hence it contributes, e.g., to
the total cross section of deep-inelastic scattering. An example will be given below.

The same expansion holds for the full conformal tower
obtained by adding ``plus" derivatives to $(\partial\mathcal{O})_N$, cf. Eq.~(\ref{eq:Cotower}):
\begin{equation}
 (\mu\lambda) (\bar\lambda\bar\mu) \partial_+^k(\partial\mathcal{O})_N =
\frac{ig\rho_N}{(N+1)^2}\Big[
\langle\!\langle (S_+)^k\overrightarrow{\Psi}_N, \overrightarrow{Q}\rangle\!\rangle
-  \langle\!\langle (S_+)^k\overrightarrow{\bar\Psi}_N, \overrightarrow{\bar Q}\rangle\!\rangle
\Big]  + \ldots
\label{eq:dO2}
\end{equation}
where the operators
$S_+^{(j_1,j_2,j_3)} = z_1^2\partial_{z_1} + z_2^2\partial_{z_2} + z_3^2\partial_{z_3} + 2j_1 z_1 + 2j_2z_2 + 2j_3z_3$
acting on the three components of $\overrightarrow{\Psi}_N$ have to be taken in the corresponding
representations, i.e.
\begin{equation}
 S^+ \overrightarrow{\Psi} =
\begin{pmatrix}
       S_+^{(1,1,1)} \Psi_1^{(1,1,1)}\\  S_+^{(1,\frac32,\frac12)}\Psi_2^{(1,\frac32,\frac12)}
\\ S_+^{(\frac32,\frac32,1)}\Psi_3^{(\frac32,\frac32,1)}
\end{pmatrix}\,.
\label{SPsi}
\end{equation}
The functions $\overrightarrow{\Psi}_{Nk} = (S^+)^k\overrightarrow{\Psi}_N$ are mutually orthogonal, and
also orthogonal to the coefficient functions of the other existing multiplicatively renormalizable operators.
It is easy to show that
\begin{equation}
 ||\overrightarrow{\Psi}_{Nk}||^2 = p^{-1}_{Nk} ||\overrightarrow{\Psi}_{N}||^2
\label{eq:pNk1}
\end{equation}
with the same coefficient $p_{Nk}$ as for the two-particle leading-twist operators, Eq.~(\ref{eq:pNk}).
It follows that the light-ray non-quasipartonic operators $\overrightarrow{Q}$ can be written as
\begin{eqnarray}
 ig\overrightarrow{Q}(z_1,z_2,z_3) &=&(n\tilde n) \sum_{N=1}^\infty\sum_{k=0}^\infty
\frac{p_{Nk}(N+1)^2}{\rho_N ||\Psi_N||^{2}}\,
\overrightarrow{\Psi}_{Nk} (z_1,z_2,z_3)\,\partial_+^k(\partial\mathcal{O})_N  + \ldots
\\
 ig\overrightarrow{\bar Q}(z_1,z_2,z_3) &=& (n\tilde n)\sum_{N=1}^\infty\sum_{k=0}^\infty
 \frac{p_{Nk}(N+1)^2}{\rho_N ||\Psi_N||^{2}}
(-1)^{N+1}\overrightarrow{ \Psi}_{Nk} (z_3,z_2,z_1)\,\partial_+^k(\partial\mathcal{O})_N  + \ldots\nonumber
\label{eq:victory}
\end{eqnarray}
which is the desired result. The ellipses stand for the contributions of all other existing twist-four
operators that cannot be reduced to the ``plus'' derivatives of $(\partial\mathcal{O})_N$.
As mentioned above, one of such ``genuine'' twist-four contributions has the same anomalous
dimension as in the leading twist.

\subsection{Reduction of the operator basis}
\label{subsec:reduction}

The description of the contribution of ``kinematic`` operators $\partial_+^k(\partial\mathcal{O})_N$
in terms of the three-component ``wave function'' $\overrightarrow{\Psi} = \{\Psi_1,\Psi_2,\Psi_3\}$
as in Eq.~(\ref{eq:victory}) is in fact redundant.

As discussed in~\cite{Braun:2008ia}, the expansion of light-ray operators in terms
of local multiplicatively renormalizable operators~(\ref{eq:PQ}) runs over all operators of
{\it collinear} twist-4. Thus together with the operators of {\it geometric}
twist-4, this expansion contains the operators  which are descendants of the
{\it geometric} twist-3 operators.%
\footnote{We remind that geometric twist of an operator is defined as ``dimension minus spin''
whereas collinear twist is given by the difference in dimension and spin projection on the light cone.
The classical example of the contribution of geometric twist-2 operators to the collinear twist-3
observable is the Wandzura-Wilczek contribution~\cite{Wandzura:1977qf} to the structure function
$g_2(x,Q^2)$ in the polarized deep-inelastic scattering.}
The latter ones are not interesting in the present context since they have autonomous scale dependence
and decouple in the similar way as the quasipartonic geometric twist-4 operators.

 It can be shown (see below) that the functions
$\overrightarrow{\Psi}_{Np}$ which correspond to the operators of geometric twist-4
satisfy the following conditions:
\begin{subequations}\label{2eq}
\begin{align}
&
\label{2eqa}
\frac{1}{z_{21}}\partial_{z_2} z_{21}^2\Psi_1(z_1,z_2,z_3)+\partial_{z_3} z_{31}\Psi_2(z_1,z_2,z_3)=0\,,
\\
&
\label{2eqb}
\frac1{z_{23}}\partial_{z_2} z_{23}^2 \Psi_1(z_1,z_2,z_3)+
\Psi_2(z_1,z_2,z_3)+z_{13}\Psi_3(z_1,z_2,z_3)=0\,.
\end{align}
\end{subequations}
These relations are $SL(2)$-invariant, i.e. if Eqs.~(\ref{2eq}) hold for
the components of $\overrightarrow{\Psi}_{Np}$, they hold for the functions
$S_{\pm,0}\overrightarrow{\Psi}_{Np}$ as well, where it is assumed that the generators
are taken in the corresponding representation, cf.~Eq.~(\ref{SPsi}).

The relations in Eq.~(\ref{2eq}) hold for the coefficient functions of an \emph{arbitrary}
operator of geometric twist-4, and in particular they hold for the coefficient functions
of the divergence of the leading twist operators, Eq.~(\ref{eq:victory}).
In principle, one can resolve these equations and express two of the
functions in terms of the third one, e.g. $\Psi_1$ and $\Psi_3$ in terms of $\Psi_2$:
\begin{equation}
z_{21}^2 \Psi_1(z_1,z_2,z_3) = \partial_{z_3} z_{13} \int_{z_2}^{z_1}dw\, (z_1-w)\Psi_2(z_1,w,z_3)\,,
\label{eq:killQ1}
\end{equation}
however, using the ``three-component'' formulation may be more convenient
because the scalar product (\ref{eq:<<>>}) has a simple form.

In order to derive Eq.~(\ref{2eq}), consider the quasipartonic twist-3
light-ray operator $\mathcal{Q}_3(z_1,z_2,z_3)=\bar\psi_+(z_1)f_{++}(z_3)\psi_+(z_3)$.
Similar to the above, $\mathcal{Q}_3$ can be expanded in contributions of
multiplicatively renormalizable local operators:
\begin{align}\label{TW3}
\mathcal{Q}_3(z_1,z_2,z_3)=\sum_{Nq} {\Psi}^{tw-3}_{Nq}(z_1,z_2,z_3)\mathcal{Q}^{tw-3}_{Nq}\,.
\end{align}
The trick is to calculate the derivatives
$\mu^\alpha(\partial/\partial\lambda^\alpha)\mathcal{Q}_3(z)$ and $[i\mathbf{P}_{\mu\bar\lambda},\mathcal{Q}_3(z)]$
in two different ways:
On the one hand, making use of the expansion in Eq.~(\ref{TW3}) one obtains
\begin{eqnarray}\label{Pmu}
\mu\partial_\lambda\mathcal{Q}_3(z_1,z_2,z_3)&=&\sum_{Nq} {\Psi}^{tw-3}_{Nq}(z_1,z_2,z_3)
\mu\partial_\lambda\mathcal{Q}^{tw-3}_{Nq}\,,
\notag\\
{}[i\mathbf{P}_{\mu\bar\lambda}\mathcal{Q}_3(z_1,z_2,z_3)]&=&
\sum_{Nq} {\Psi}^{tw-3}_{Nq}(z_1,z_2,z_3) [i\mathbf{P}_{\mu\bar\lambda},\mathcal{Q}^{tw-3}_{Nq}]\,.
\end{eqnarray}
The sums on the r.h.s. of~(\ref{Pmu}) do not contain genuine geometric twist-4 operators,
but rather the descendants (derivatives) of the twist-3 operators only.
On the other hand, one can express
$\mu\partial_\lambda\mathcal{Q}_3(z)$ and $[i\mathbf{P}_{\mu\bar\lambda},\mathcal{Q}_3(z)]$
directly in terms of the $\overrightarrow{Q}$-operators:
\begin{eqnarray}
(\bar\lambda\bar\mu) \mu\partial_\lambda\mathcal{Q}_3(z_1,z_2,z_3)&=&z_2^{-1}\partial_{z_2} z_2^2Q_1(z_1,z_2,z_3)
+\partial_{z_3}z_3Q_2(z_1,z_2,z_3)+ z_1 Q_3(z_1,z_2,z_3)\,,
\notag\\
(\bar\lambda\bar\mu)[i\mathbf{P}_{\mu\bar\lambda}\mathcal{Q}_3(z_1,z_2,z_3)]&=&\partial_{z_2}Q_1(z_1,z_2,z_3)+
\partial_{z_3}Q_2(z_1,z_2,z_3)+ Q_3(z_1,z_2,z_3)\,.
\label{eq:redun}
\end{eqnarray}
Since the expansion of $Q_k$ over the local operators~(\ref{eq:PQ})
contains {\it all} operators of collinear twist-4,
it means that the geometric twist-4 operators must drop out in the particular
combinations appearing in (\ref{eq:redun}) and only the descendants of
geometric twist-3 operators
survive. If follows that the coefficient functions of the
geometrical twist-4 operators in this expansion
must satisfy the equations
\begin{align}\label{eq:Psi123}
(z_2\partial_{z_2}+2)\Psi_1(z_1,z_2,z_3)+(z_3\partial_{z_3}+1)\Psi_2(z_1,z_2,z_3)
  +z_1 \Psi_3(z_1,z_2,z_3)=0\,,
\notag\\
\partial_{z_2}\Psi_1(z_1,z_2,z_3)+\partial_{z_3}\Psi_2(z_1,z_2,z_3)+\Psi_3(z_1,z_2,z_3)=0\,,
\end{align}
which are equivalent to Eqs.~(\ref{2eq}).

\section{Divergence of a conformal operator}

Let $\mathcal{O}_N^{\mu\mu_1 \ldots\mu_{N}}$ be a multiplicatively renormalizable
leading-twist operator with $N$ covariant derivatives, symmetric and traceless over all open indices.
The conformal operator $\mathcal{O}_N$ defined in Eq.~(\ref{ON}) is obtained by the
projection
\begin{equation}
 \mathcal{O}_N=n^\mu n^{\mu_1}\ldots n^{\mu_{N}} (\mathcal{O}_N)_{\mu\mu_1\ldots\mu_{N}}\,,
\end{equation}
where $n$ is an auxiliary light-like vector. We define a divergence of the conformal
operator as
\begin{eqnarray}
 (\partial\mathcal{O})_N &=& n^{\mu_1}\ldots n^{\mu_{N}}\partial^\mu (\mathcal{O}_N)_{\mu\mu_1\ldots\mu_{N}}
 = n^{\mu_1}\ldots n^{\mu_{N}}\,\Big[i\mathbf{P}^{\mu},
(\mathcal{O}_N)_{\mu \mu_1\mu_2\ldots\mu_{N}}\Big]\,,
\label{def:partialO}
\end{eqnarray}
where $\mathbf{P}_{\mu}$ is the usual four-momentum operator
\begin{equation}
   \mathbf{P}_{\mu} |p\rangle = p_\mu |p\rangle\,, \qquad i[\mathbf{P}_\mu,\Phi(x)]
  =\frac{\partial}{\partial x^\mu}\Phi(x)\,.
\label{eq:P}
\end{equation}
Taking into account that $n^{\mu}=\frac12(\lambda\sigma^\mu\bar\lambda)$ the same definition
can be rewritten as
\begin{equation}\label{div}
(\partial\mathcal{O})_N=
\frac1{N+1}\Big[i\mathbf{P}^{\mu},\frac{\partial}{\partial n^\mu}
\mathcal{O}_N(n)\Big]
=\frac1{(N+1)^2}\Big[i\bar{\mathbf{P}}^{\dot\alpha\alpha},\frac{\partial}{\partial \lambda^\alpha}
\frac{\partial}{\partial \bar\lambda^{\dot\alpha}}
\mathcal{O}_N(\lambda,\bar\lambda)\Big]\,,
\end{equation}
where the argument of $\mathcal{O}_N$ indicates that it has to be considered as
a function of $n^\mu$ or $\lambda^\alpha,\bar\lambda^{\dot\alpha}$, respectively.

\subsection{Expression in terms of non-quasipartonic operators}

Our first task is to rewrite this expression in terms of quark-gluon operators using QCD
equations of motion (EOM). To this end it is convenient to use
the integral representation~(\ref{eq:Copera}) for the conformal operator $\mathcal{O}_N$ in terms
of the light-ray operator
\begin{equation}\label{o++}
O_{++}(z_1,z_2)=\bar \psi_+(z_1)[z_1,z_2]\psi_+(z_2) = \bar \psi(z_1n) \bar n [z_1n,z_2n]\psi(z_2n)\,.
\end{equation}
A subtlety is that this operator is defined on the light-cone $n^2=0$ and in order to
take derivatives with respect to $n^\mu$ one has to extend this definition to $n^2\neq 0$.
This can be done by applying the trace subtraction operator (see the next section), but
a more elegant way out is to use the last (spinor) representation in Eq.~(\ref{div}).
Indeed, taking derivatives with respect to the auxiliary spinors
$\lambda,\bar\lambda$ one stays on the surface $n^2=0$.

Let
\begin{equation}
 (\partial O)_{++}(z_1,z_2)=\Big[i\bar{\mathbf{P}}^{\dot\alpha\alpha},\frac{\partial}{\partial \lambda^\alpha}
\frac{\partial}{\partial \bar\lambda^{\dot\alpha}}
\mathcal{O}_{++}(z_1,z_2)\Big]
\label{def:dO++}
\end{equation}
so that
\begin{equation}\label{whO++}
(\partial\mathcal{O})_N=
\frac{\convent_N }{(N+1)^2}\big\langle z_{12}^N, (\partial O)_{++}(z_1,z_2)\big\rangle
\equiv\frac{\convent_N }{(N+1)^2} \iint_{|z_i|<1} \mathcal{D}z_1 \mathcal{D}{z_2}\,
\bar z_{12}^N\, (\partial O)_{++}(z_1,z_2)\,.
\end{equation}
Taking into account that
\begin{equation}
\frac{\partial}{\partial
\lambda^\alpha}=\frac12(\sigma^\mu\bar\lambda)_{\alpha}\dfrac{\partial}{\partial n^\mu}\,,
\qquad
\frac{\partial}{\partial \bar\lambda^{\dot\alpha}}
=\frac12(\lambda\sigma^\mu)_{\dot\alpha}\frac{\partial}{\partial n^\mu}
\end{equation}
one obtains
\begin{equation}
\label{lambda-n}
(\partial O)_{++}(z_1,z_2)=
 \partial^\mu\Big[i\mathbf{P}_\mu, ((n\partial)+1)O_{++}(z_1,z_2)\Big]
-\frac12 \partial^2 \Big[(i\mathbf{P} n), O_{++}(z_1,z_2)\Big]\,,
\end{equation}
where $\partial_\mu=\frac{\partial}{\partial n^\mu}$.
Making use of the identities
\begin{eqnarray}
((n\partial)+1)O_{++}(z_1,z_2)&=&(z_1\partial_{z_1}+z_2\partial_{z_2}+2)O_{++}(z_1,z_2)\,,
\notag\\
\Big[(i\mathbf{P} n),O_{++}(z_1,z_2)\Big]&=&(\partial_{z_1}+\partial_{z_2})O_{++}(z_1,z_2)
\end{eqnarray}
this becomes
\begin{equation}\label{NN}
(\partial O)_{++}(z_1,z_2)=(z_1\partial_{z_1}+z_2\partial_{z_2}+2)
\Big[i\mathbf{P}_\mu, \partial^\mu O_{++}(z_1,z_2)\Big]-\frac12(\partial_{z_1}+\partial_{z_2})
\partial^2O_{++}(z_1,z_2)\,.
\end{equation}
Note that the definition in Eq.~(\ref{def:dO++}) involves
the light-ray operator $O_{++}(z_1,z_2)$ at strictly light-like separations $n^2=0$
and the calculation leading to Eq.~(\ref{NN}) is pure algebra. Hence the expression
in~(\ref{NN}) effectively does not contain derivatives in the directions "orthogonal"
to the light-ray, so that one can treat $n^\mu $ as a generic four-vector and put $n^2=0$
at the end of calculation.

The derivatives of the light-ray operator in Eq.~(\ref{NN}) include
contributions from the Wilson lines~\cite{Balitsky:1987bk,Braun:1989iv}:
\begin{subequations}
\begin{align}\label{Plink}
\Big[i\mathbf{P}_\mu,[z_1n,z_2n]\Big]\,=\,&\,igA_{\mu}(z_1)[z_1n,z_2n]-
ig [z_1n,z_2n]A_{\mu}(z_2)
\\
&+ig z_{12}\int_0^1du\,  [z_1n,z_{21}^u n]n^\nu F_{\mu\nu}(z_{21}^u)[z_{21}^un,z_2n]\,,
\notag\\
\label{Dlink}
\frac{\partial}{\partial n_{\mu}}[z_1n,z_2n]\,=\,&\,ig z_1  A_{\mu}(z_1)[z_1n,z_2n]-
ig z_2[z_1n,z_2n]A_{\mu}(z_2)
\notag\\
&+ig z_{12}\int_0^1du\,z_{21}^u  [z_1n,z_{21}^u n]n^\nu F_{\mu\nu}(z_{21}^u)[z_{21}^un,z_2n]\,.
\end{align}
\end{subequations}
In particular, taking into account Eq.~(\ref{Dlink}) one obtains
\begin{eqnarray}\label{DO++}
\partial^\mu O_{++}(z_1,z_2)&=&\bar\psi(z_1n)\bar\sigma^\mu\psi(z_2n)+
z_1[D^\mu\bar\psi](z_1n)\bar n\psi(z_2n)+z_2\bar\psi(z_1n)\bar n[D^\mu\psi](z_2n)
\notag\\
&&{}+ig z_{12}\int_0^1du\,z_{21}^u\,  \bar\psi(z_1n)\bar n n_\nu F^{\mu\nu}(z_{21}^u)\psi(z_2n)\,,
\end{eqnarray}
where for brevity we do not show the Wilson lines in the operators on the r.h.s.
The covariant derivatives are defined as $D_\mu \psi=(\partial_\mu-ig A^a_\mu t^a) \psi$,
$D_\mu\bar\psi=(\partial_\mu+ig A^a_\mu (t^a)^T)\bar\psi$.

Calculating the commutator of~(\ref{DO++}) with $i\mathbf{P}_\mu$ (or taking the second
derivative $\partial_\mu$) one gets contributions of the following type:
\begin{align}\label{set0}
D_\mu \bar\psi \bar n D^\mu\psi, \quad D^2\bar\psi \bar n \psi,\quad \bar\psi \bar n
D^2\psi, \quad ig\bar\psi F \psi\,,\quad\bar\psi FF\psi,\quad\bar \psi D^\mu F_{\mu\nu}\psi\,.
\end{align}
The last term $\sim \bar \psi D^\mu F_{\mu\nu}\psi$
can be reduced to the four-quark quasipartonic operator using EOM.
Also the terms $\sim \bar\psi FF\psi$ are quasipartonic. As explained above,
such contributions are irrelevant for our discussion and will be omitted.

The contributions of the first type,  $\sim D_\mu \bar\psi \bar n D^\mu\psi$, are,
on the other hand, the only ones that do not vanish in the free theory. One should expect that
such terms do not contribute to a divergence of the conformal operator by virtue
of the FGPG theorem~\cite{Ferrara:1972xq}). It is instructive to see how this
result is recovered in our formalism.

The contribution to $(\partial O)_{++}(z_1,z_2)$~(\ref{NN})
due to the terms $\sim\! D_\mu \bar\psi \bar n D^\mu\psi$ takes the form
\begin{eqnarray}
(\partial O)^{\rm free}_{++}(z_1,z_2)&=&\Big[(z_1\partial_{z_1}+z_2\partial_{z_2}+2)(z_1\!+\!z_2)
-\frac12(\partial_{z_1}\!+\!\partial_{z_2}) 2 z_1
z_2\Big][D_\mu \bar\psi](z_1n) \bar n [D^\mu\psi](z_2 n)
\nonumber\\
&=&
(z_1^2\partial_{z_1}+z_2^2\partial_{z_2}+2z_1+2z_2)[D_\mu \bar\psi](z_1n) \bar n [D^\mu\psi](z_2 n)
\nonumber\\&=& S_{12}^+[D_\mu \bar\psi](z_1n) \bar n [D^\mu\psi](z_2 n)\,,
\end{eqnarray}
so it is not zero, but proportional to the two-particle ``step-up'' operator
 (\ref{generators}). It follows that
\begin{equation}
 (\partial\mathcal{O})^{\rm free }_N \sim
 \big\langle z_{12}^N, S_{12}^+[D_\mu \bar\psi](z_1) \bar n [D^\mu\psi](z_2)\big\rangle
  = - \big\langle S_{12}^-z_{12}^N, [D_\mu \bar\psi](z_1) \bar n [D^\mu\psi](z_2)\big\rangle
  =0\,,
\end{equation}
as expected.

The remaining terms $\sim  D^2\bar\psi \bar n \psi$, $\sim \bar\psi \bar n D^2\psi$ and
$\sim ig\bar\psi F \psi $ give rise to the contributions of the non-quasipartonic operators
of interest (\ref{eq:Q}), (\ref{eq:barQ}).
The corresponding calculation is detailed in Appendix~\ref{App:B}.
The answer has the form
\begin{equation}
  (\mu\lambda)(\bar\lambda\bar\mu) (\partial\mathcal{O})_{++}(z_1,z_2) =
   ig\Big[ A(z_1,z_2) -  \bar A(z_1,z_2)\Big]+\ldots\,,
\end{equation}
where
\begin{eqnarray}
A(z_1,z_2)&=&\partial_{z_2} z_{12}^2 \biggl\{Q_1(z_1,z_1,z_2)+\int_0^1du \, u\Big[
\,Q_2(z_1,z_{21}^u,z_2)+z_{12}Q_3(z_1,z_{21}^u,z_2)\Big]\biggr\}
\notag\\
&&{}+
\partial_{z_1}\partial_{z_2} z_{12}^3
\int_0^1 du\,\Big[-Q_1(z_1,z_{21}^u,z_2)
+ \bar u\,Q_2(z_1,z_{21}^u,z_2)\Big],
\label{Aw0}\\
\bar A(z_1,z_2)&=&\partial_{z_2} z_{21}^2 \biggl\{\bar Q_1(z_1,z_2,z_2)+\int_0^1du \, u\,
\Big[\bar Q_2(z_1,z_{12}^u,z_2)+z_{21}\bar Q_3(z_1,z_{12}^u,z_2)\Big]\biggr\}
\notag\\
&&{}+\partial_{z_1}\partial_{z_2} z_{21}^3
\int_0^1 du\,\Big[-\bar Q_1(z_1,z_{12}^u,z_2)+ \bar u\,\bar Q_2(z_1,z_{12}^u,z_2)\Big]\,.
\label{Abar}
\end{eqnarray}
The ellipses stand for EOM, contributions of quasipartonic operators and terms
proportional to $S_{12}^+$ which do not contribute to the projection (\ref{whO++})
that defines the conformal operator:
\begin{eqnarray}\label{AO}
(\mu\lambda)(\bar\lambda\bar\mu)(\partial\mathcal{O})_N&=&\frac{ig\rho_N}{(N+1)^2}\Biggl[
\langle z_{12}^N,  A(z_1,z_2)\rangle_{11}-\langle z_{12}^N,\bar A(z_1,z_2)\rangle_{11}\Biggr]
\notag\\
&=&\frac{ig\rho_N}{(N+1)^2}
\iint_{|z_k|<1} \mathcal{D}z_1 \mathcal{D}z_2\, \bar z_{12}^N
\left[ A(z_1,z_2)- \bar A(z_1,z_2)\right].
\end{eqnarray}
%

\subsection{Coefficient functions}
Our answer for the divergence of the conformal operator~(\ref{AO}) involves the sum
of terms of the type, schematically
\begin{equation}
 (\partial\mathcal{O})_N \sim \langle z_{12}^N, [\mathbb{B}_iQ_i](z_1,z_2)\rangle_{11}\,,
\end{equation}
where $\mathbb{B}_i$ are some integral operators, and the subscript $\langle,\rangle_{11}$ indicates that
the scalar product is calculated for conformal spins $j_1=j_2=1$.
The idea is to rewrite this answer as a sum of terms
\begin{equation}
(\partial\mathcal{O})_N \sim  \langle \Psi_i^N(z_1,z_2,z_3),Q_i(z_1,z_2,z_3)\rangle_{j_1j_2j_3}\,,
\end{equation}
where $j_1,j_2,j_3$ are the conformal spins of the $Q_i$,
so that the functions $ \Psi_i^N(z_1,z_2,z_3)$ can be identified with the
coefficient functions of the $Q$-operators, cf.~Eq.~(\ref{eq:dO1}).

For illustration let us consider contribution of the first two terms in (\ref{Aw0})
  which we write as
\begin{equation}
A= \partial_{z_2} z_{12}^2 \big[\phi_1(z_1,z_2) + \phi_2(z_1,z_2)+\ldots \big] + \ldots\,,
\label{eq:example2}
\end{equation}
where
\begin{eqnarray}
\phi_1(z_1,z_2) &=& Q_1(z_1,z_1,z_2)\,,
\nonumber\\
\phi_2(z_1,z_2) &=& \int_0^1du \,u\,Q_2(z_1,z_{21}^u,z_2)\,.
\end{eqnarray}

As the first step, notice that any operator $Q_i(z_1,z_2,z_3)$ can be written as
\begin{align}
Q_i(z_1,z_2,z_3)=\big\langle\prod_{k=1}^3\mathcal{K}_{j_k}(z_k,\bar w_k),Q_i(w_1,w_2,w_3)\big\rangle_{j_1j_2j_3}\,,
\end{align}
where $\mathcal{K}_{j_k}$ are the reproducing kernels defined in Eq.~(\ref{repr}) and the
scalar product corresponds to the spins $j_1,j_2,j_3$. Thus
%
\begin{eqnarray}\label{VF1}
\varphi_1(z_1,z_2)&=&
            \big\langle\mathcal{K}_{1}(z_1,\bar w_1) \mathcal{K}_{1}(z_1,\bar w_2)\mathcal{K}_{1}(z_2,\bar w_3)
             Q_1(w_1,w_2,w_3)\big\rangle_{111}\,,
\\
\label{VF2}
\varphi_2(z_1,z_2)&=&
            \int_0^1du u\,\big\langle\mathcal{K}_{1}(z_1,\bar w_1)
            \mathcal{K}_{3/2}(z_{21}^u,\bar w_2)\mathcal{K}_{1/2}(z_2,\bar w_3)
             Q_1(w_1,w_2,w_3)\big\rangle_{1\frac32\frac12}
\notag\\
            &=&\frac12\big\langle\mathcal{K}_{1}(z_1,\bar w_1)
               \mathcal{K}_{1}(z_1,\bar w_2)
                \mathcal{K}_{1/2}(z_2,\bar w_2)
                 \mathcal{K}_{1/2}(z_2,\bar w_3)
                   Q_1(w_1,w_2,w_3)\big\rangle_{1\frac32\frac12}\,,
\end{eqnarray}
%
where we used that
\begin{eqnarray}
\int_0^1 \!du u\,\mathcal{K}_{3/2}(z_{21}^u,\bar w_2)&=&\int_0^1 du u\frac1{(1-(z_2\bar u+z_1 u)\bar w_2)^3}
=\int_0^1\! du \frac{u}{(u(1-z_1\bar w_2)+\bar u(1+z_2\bar w_2))^3}
\notag\\
&=&
\frac12\frac1{(1-z_1\bar w_2)^2}\frac1{(1-z_2\bar w_2)}
=\frac12 \mathcal{K}_{1}(z_{1},\bar w_2)
\mathcal{K}_{1/2}(z_{2},\bar w_2)\,.
\end{eqnarray}
Making the conformal transformation $z\to z'$,  $w\to w'$ and
taking into account that the reproducing kernels become
\begin{align}\label{kernelstrafo}
\mathcal{K}_j(z',\bar w')=(\bar b z+\bar a)^{2j}\mathcal{K}_j(z,\bar w)( b \bar w+ a)^{2j},
\end{align}
it is straightforward to check that both $\varphi_1(z_1,z_2)$ and $\varphi_2(z_1,z_2)$
transform according to the representation $T^{j_1=2}\otimes T^{j_2=1}$.
Using this result, it is easy to show that the both contributions in Eq.~(\ref{eq:example2}),
$\partial_2 z_{12}^2\varphi_1(z_1,z_2)$ and $\partial_2 z_{12}^2\varphi_2(z_1,z_2)$, transform
according to the representation $T^{j_1=1}\otimes T^{j_2=1}$, as they should.

The transformation properties follow immediately from the
following statements which can be checked by a direct calculation:
\begin{itemize}
\item
if a function $\varphi(z_1,z_2)$ transforms according to the representation $T^{j_1}\otimes T^{j_2}$,
then the function $\psi(z_1,z_2)=z_{12}\varphi(z_1,z_2)$ transforms according to the
representation
$T^{j_1-1/2}\otimes T^{j_2-1/2}$, i.e. multiplication by $z_{12}=z_1-z_2$ intertwines the representations
\begin{align}
z_{12}\,T^{j_1}\otimes T^{j_2}=T^{j_1-1/2}\otimes T^{j_2-1/2}\,z_{12}
\end{align}
\item if a function $\varphi(z)$ transforms according to the representation $T^{j=0}$ then its derivative
$\partial_z\varphi(z)$ transforms according to the representation $T^{j=1}$, i.e.
\begin{align}
\partial_z\,T^{j=0}=T^{j=1}\,\partial_z.
\end{align}
\end{itemize}
We have verified that all contributions to the functions $A(z_1,z_2)$ (\ref{Aw0}) and
$\bar A(z_1,z_2)$ (\ref{Abar}) transform separately according to the same representation,
$T^{j_1=1}\otimes T^{j_2=1}$, which provides a strong check of the calculation.

The contribution of $\varphi_{1,2}(z_1,z_2)$ to $(\partial\mathcal{O})_N$ has the form
\begin{equation}
 (\partial\mathcal{O})_N \sim \langle z_{12}^N,\mathbb{B}\,\varphi(z_1,z_2)\rangle_{11}\,,
\label{sc11}
\end{equation}
where the operator $\mathbb{B}=\partial_2z_{12}^2$ intertwines the representations
$\mathbb{B}\,T^{j_1=2}\otimes T^{j_2=1}=T^{j_1=1}\otimes T^{j_2=1}\,\mathbb{B}$, i.e.
\begin{align}
\mathbb{B}\, S_{0,\pm}^{(j_1=2,j_2=1)} =
                    {S}_{0,\pm}^{(j_1=1,j_2=1)}\mathbb{B}\,.
\end{align}
As the second step, we want to rewrite~(\ref{sc11}) in the form
\begin{align}
\langle z_{12}^N,\mathbb{B}\,\varphi_1(z_1,z_2)\rangle_{j_1=j_2=1}=
\langle \mathbb{B}^\dagger z_{12}^N,\varphi_1(z_1,z_2)\rangle_{j_1=2,j_2=1}\,.
\end{align}
It is easy to see that $\mathbb{B}^\dagger z_{12}^N= b_Nz_{12}^{N-1}$ where $b_N$ is a numerical
coefficient which can be fixed as follows:
\begin{align}
\langle z_{12}^N,\mathbb{B}z_{12}^{N-1}\rangle_{11}=-(N+1)||z_{12}^N||^2_{11}=
\langle \mathbb{B}^\dagger z_{12}^N,z_{12}^{N-1}\rangle_{21}=
b_N||z_{12}^{N-1}||^2_{21}\,.
\end{align}
Taking into account~(\ref{eq:z12n}) one gets
\begin{align}\label{}
b_N=-\frac1{6}N(N+2)(N+3)\,,
\end{align}
so that
\begin{equation}
 \langle z_{12}^N,\partial_2z_{12}^2\,\varphi_1(z_1,z_2)\rangle_{11}
= b_N \langle z_{12}^{N-1},\varphi_1(z_1,z_2)\rangle_{21}
= \langle\Psi_{1a}(w_1,w_2,w_3),Q_1(w_1,w_2,w_3)\rangle_{111}\,,
\end{equation}
where~\footnote{The subscript $1a$ serves to remind that this is only one of the two
contributions to the coefficient function of the $Q_1$ operator; the second contribution
comes from the term in $Q_1$ in the second line in (\ref{A}).}
\begin{equation}
\Psi_{1a}(w_1,w_2,w_3)=b_N\iint_{|z_k|<1} \mathcal{D}_{2}z_1\mathcal{D}_{1}z_2 \,z_{12}^{N-1}\,
\mathcal{K}_{1}(w_1,\bar z_1) \mathcal{K}_{1}(w_2,\bar z_1)\mathcal{K}_{1}( w_3,\bar z_2)\,.
\end{equation}
Combining the two reproducing kernels with the help of Feynman parametrization
\begin{equation}
\mathcal{K}_{1}(w_1,\bar z_1) \mathcal{K}_{1}(w_2,\bar z_1)=6\int_0^1d\alpha\alpha\bar\alpha\,
\mathcal{K}_{2}(w_{12}^\alpha,\bar z_1)
\end{equation}
and using that
\begin{equation}
 \iint_{|z_k|<1} \mathcal{D}_{2}z_1\mathcal{D}_{1}z_2 \,z_{12}^{N-1}
 \mathcal{K}_{2}(w_{12}^\alpha,\bar z_1) \mathcal{K}_{1}( w_3,\bar z_2)
 = (w_{12}^\alpha - w_3)^{N-1}
\end{equation}
one obtains
\begin{equation}
\Psi_{1a}(w_1,w_2,w_3)=-N(N+2)(N+3)\int_0^1d\alpha\,\alpha\bar\alpha\,(w_{12}^\alpha-w_3)^{N-1}\,.
\end{equation}
Similarly
\begin{equation}
 \langle z_{12}^N,\partial_2z_{12}^2\,\varphi_2(z_1,z_2)\rangle_{11}
= b_N \langle z_{12}^{N-1},\varphi_1(z_1,z_2)\rangle_{21}
= \langle\Psi_{2a}(w_1,w_2,w_3),Q_2(w_1,w_2,w_3)\rangle_{111}\,,
\end{equation}
where
\begin{equation}
\Psi_{2a}(w_1,w_2,w_3)=-\frac12 N(N+2)(N+3)\int_0^1d\alpha\,\alpha\bar\alpha\,
\int_0^1 d\beta (w_{12}^\alpha-w^\beta_{32})^{N-1}\,.
\end{equation}
All other contributions to (\ref{Aw0}), (\ref{Abar}) can be managed in the same manner.
Collecting all terms we obtain the final result:
\begin{align}
\langle z_{12}^N,\, A(z_1,z_2)\rangle_{11}=
\langle\langle \overrightarrow{\Psi}_N(w_1,w_2,w_3),\overrightarrow{Q}(w_1,w_2,w_3)\rangle\rangle\,,
\end{align}
where the invariant scalar product $\langle\langle\ldots\rangle\rangle$ is defined in Eq.~(\ref{eq:<<>>})
and the coefficient functions $\overrightarrow{\Psi}_N  = \Big\{\Psi^{(1)}_N,\Psi^{(2)}_N,\Psi^{(3)}_N\Big\}$
are given by the following expressions:
\begin{eqnarray}\label{Psi1}
\Psi_N^{(1)}(w)&=&a_N\iint \mathcal{D}_{3/2}z_1\,\mathcal{D}_{3/2}z_2\,\,z_{12}^{N-1}\,
\mathcal{K}_1(w_1,\bar z_1)\mathcal{K}_{1/2}(w_2,\bar z_1)\mathcal{K}_{1/2}(w_2,\bar z_2)\mathcal{K}_1(w_3,\bar z_2)
\notag\\
&&{}
+\frac12 b_N\iint \mathcal{D}_{2}z_1\, \mathcal{D}_{1}z_2\,\,z_{12}^{N-1}\,
\mathcal{K}_1(w_1,\bar z_1)\mathcal{K}_{1}(w_2,\bar z_1)\mathcal{K}_1(w_3,\bar z_2)
\\
&=&
4a_N\left[\int_0^1d\alpha\,\bar\alpha\int_0^1d\beta\,\bar\beta\,
(w_{12}^\alpha-w_{32}^\beta)^{N-1}
-\frac1{N+1}
\int_0^1d\alpha\,\alpha\bar\alpha\,(w_{12}^\alpha-w_3)^{N-1}
\right]\,,
\nonumber
\end{eqnarray}
\begin{eqnarray}\label{Psi2}
\Psi_N^{(2)}(w)&=&-a_N
\iint \mathcal{D}_{3/2}z_1\,\mathcal{D}_{3/2}z_2\,\,z_{12}^{N-1}\,
\mathcal{K}_1(w_1,\bar z_1)\mathcal{K}_{1/2}(w_2,\bar z_1)\mathcal{K}_{1}(w_2,\bar z_2)\mathcal{K}_{1/2}(w_3,\bar z_2)
\notag\\
&&{}
+\frac12 b_N
\iint \mathcal{D}_{2}z_1\, \mathcal{D}_{1}z_2\,\,z_{12}^{N-1}\,
\mathcal{K}_1(w_1,\bar z_1)\mathcal{K}_{1}(w_2,\bar z_1)\mathcal{K}_{1/2}(w_2,\bar z_2)\mathcal{K}_{1/2}(w_3,\bar z_2)
\notag\\
&=&-4a_N\int_0^1d\alpha\,\bar\alpha\int_0^1d\beta\,\left(\beta+\frac1{N+1}\alpha\right)
(w_{12}^\alpha-w_{32}^\beta)^{N-1}\,,
\end{eqnarray}
\begin{eqnarray}\label{Psi3}
\Psi_N^{(3)}(w)&=&-c_N\iint \mathcal{D}_{5/2}z_1\, \mathcal{D}_{3/2}z_2\,\,z_{12}^{N-2}
\mathcal{K}_{3/2}(w_1,\bar z_1)\mathcal{K}_{1}(w_2,\bar z_1)\mathcal{K}_{1/2}(w_2,\bar z_2)\mathcal{K}_{1}(w_3,\bar z_2)
\notag\\
&=&-24c_N\int_0^1d\alpha\,\bar\alpha^2\alpha\int_0^1d\beta\,\bar\beta\,
(w_{12}^\alpha-w_{32}^\beta)^{N-2}\,.
\end{eqnarray}
Here
\begin{align}
a_N=&\frac18(N+3)(N+2)(N+1)N, \notag\\
 b_N=&-\frac1{6}(N+3)(N+2)N\,,
\notag\\
c_N=&\frac1{48}\frac{(N+4)!}{(N+1)(N-2)!}\,.
\end{align}
We have checked that these expressions satisfy the relations~(\ref{2eq}).

\subsection{Calculation of the norm $||{\Psi}_N||^2$}\label{sect:norm}
To pick up the contribution of the operator
$(\partial\mathcal{O})_N$ to the light-ray operators $Q_k$ (\ref{eq:victory})
we need to know the norm of the ``wave function'' $\overrightarrow{\Psi}_N$:
\begin{align}\label{PsiNnorm}
||\overrightarrow{\Psi}_N||^2=2||{\Psi}^{(1)}_N||^2_{111}+||{\Psi}^{(2)}_N||^2_{1\frac32\frac12}+
\frac12||{\Psi}^{(3)}_N||^2_{\frac32\frac321}\,.
\end{align}
Consider the last  term in~(\ref{PsiNnorm}) as an example.
To evaluate
\begin{align}
\langle{\Psi}^{(3)}_N,\Psi^{(3)}_N\rangle_{\frac32\frac321}
\end{align}
we use the parametric representation for ``ket'' function (second line in (\ref{Psi3})), and the reproducing kernel
representation for the ``bra'' function (first line in (\ref{Psi3})), respectively.
Rewriting the product of  reproducing kernels as
\begin{align}\label{feynman1}
\mathcal{K}_{1}(w_2,\bar z_1)\mathcal{K}_{1/2}(w_2,\bar z_2)=2\int_0^1 du \,u \,\mathcal{K}_{3/2}(w_2,\bar z_{21}^u)
\end{align}
and using Eq.~(\ref{repr}) one can easily take the integrals over $w_1,w_2,w_3$ to get
\begin{eqnarray}
||{\Psi}^{(3)}_N||^2_{\frac32\frac321}
&=&48 c_N^2 ||z_{12}^{N-2}||^2_{\frac52\frac32}
\int_0^1d\alpha\,\bar\alpha^2\alpha\int_0^1d\beta\,\bar\beta\,\int_0^1 du\, u\,
(\bar\alpha+u(\alpha-\beta))^{N-2}
\notag\\
&=&8a_N\,||z_{12}^{N}||^2_{11}\left[
\frac13-\frac12\frac1{(N+1)(N+2)}-\frac{2[\psi(N+2)-\psi(2)]}{N(N+3)}
\right]\,.
\end{eqnarray}
The other two contributions are calculated similarly:
\begin{eqnarray}
||\Psi_N^{(1)}||^2&=&4a_N||z_{12}^{N}||^2_{11}
\Big[
\frac{(N+1)(N+2)}{N(N+3)}[\psi(N+2)-\psi(2)]-\frac5{12}\Big]\,,
\notag\\
||\Psi_N^{(2)}||^2&=&4a_N||z_{12}^{N}||^2_{11}
\Big[
\frac{2[\psi(N+2)-\psi(2)]}{(N+3)N}+1+\frac1{2(N+1)(N+2)}\Big]\,.
\end{eqnarray}
Collecting all terms we obtain
\begin{align}
||\overrightarrow{\Psi}_N||^2=\frac12||z_{12}^{N}||^2_{11}\,(N+2)^2(N+1)^2\,
\Big[\psi(N+3)+\psi(N+1)-\psi(3)-\psi(1)\Big]\,.
\label{eq:norm}
\end{align}
Note that the expression in the square bracket coincides with the leading twist anomalous
dimension $\gamma_N$ (\ref{eq:gammaN}) up to the color factor $2C_F$.
Proportionality $||\overrightarrow{\Psi}_N||^2 \sim \gamma_N$ is certainly not accidental.
Exploring this interesting connection goes, however, beyond the tasks of this paper.

\subsection{Example: $N=1$ }

In the simplest case $N=1$ one obtains
\begin{equation}
\Psi^{(1)}_{N=1} =  2\,, \qquad \Psi^{(2)}_{N=1} = -4\,, \qquad \Psi^{(3)}_{N=1}=0\,.
\end{equation}
Inserting these expressions in Eqs.~(\ref{eq:dO1}) and (\ref{eq:TN}) yields
\begin{eqnarray}
 (\partial\mathcal{O})_{N=1}&=&-\frac{3ig}{(n\tilde n)}\Big[(Q_2-Q_1) +
(\bar Q_2-\bar Q_1)\Big]_{z_k=0}=-6i\bar q_L gF_{+\mu} \gamma^\mu q_L\,,
\nonumber\\
 T_{N=1}&=&-\frac{3ig}{(n\tilde n)}\Big[(Q_2-Q_1) -
(\bar Q_2-\bar Q_1)\Big]_{z_k=0}= - 6\bar q_L g\widetilde F_{+\mu} \gamma^\mu \gamma_5q_L\,,
\label{eq:N=1}
\end{eqnarray}
where $q_L=\frac12(1-\gamma_5)q$.

The first equation in Eq.~(\ref{eq:N=1} is exactly the contribution of left-handed
quarks to the identity (\ref{eq:puzzle})~\cite{Kolesnichenko:1984dj,Braun:2004vf}
which is easy to derive by direct computation.
The coefficient is correct because in our normalization
\begin{align}
 [\mathcal{O}_{N=1}]_{\mu\nu} = -\frac32\Bigl[\bar q \gamma_\mu\!\stackrel{\leftrightarrow}{D}_\nu \!q
  + (\mu\leftrightarrow\nu)-\text{traces}\Bigr].
\end{align}
In its turn, the operator $T_{N=1}$ appearing in the second line in Eq.~(\ref{eq:N=1}) is familiar
from studies of power corrections $\sim 1/Q^2$ to the deep-inelastic scattering~\cite{Shuryak:1981kj}.
Its anomalous dimension equals $\gamma_{T_1} = 4/3(N_c-1/N_c)$~\cite{Shuryak:1981kj} and indeed coincides
with the leading-twist anomalous dimension $\gamma_{N=1}$ (\ref{eq:gammaN}),
in agreement with our general result.%
\footnote{\small It appears that the coincidence $\gamma_{T_1} = \gamma_{N=1}$ remained unnoticed
over the 30 years.}

\section{Intermediate summary: kinematic projection operators }

{}For the reader's convenience we reiterate here the main result.

As explained in the introduction, calculation of kinematic corrections $\sim t/Q^2, m^2/Q^2$
to hard reactions in off-forward kinematics requires taking into account contributions
of higher-twist operators that can be reduced to total derivatives of leading-twist
operators~$\mathcal{O}_N$.
In particular picking up the contribution of the divergence of leading-twist operators
$(\partial\mathcal{O})_N$ 
is complicated because its matrix elements
on free quarks vanish.
Using QCD equations of motion this operator can be expressed in terms of quark-gluon
operators as follows:
\begin{equation}
 2 (n\tilde n) (\partial\mathcal{O})_N = \frac{ig\rho_N}{(N+1)^2}\Big[
 \langle\!\langle \overrightarrow{\Psi}_N, \overrightarrow{Q}\rangle\!\rangle
 - \langle\!\langle \overrightarrow{\bar\Psi}_N, \overrightarrow{\bar Q}\rangle\!\rangle\Big]
 + \ldots\,,
\label{eq:dO3}
\end{equation}
where $\overrightarrow{Q} = \Big\{Q_1,Q_2,Q_3\Big\}$ 
and
$\overrightarrow{\bar Q} = \{\bar Q_1,\bar Q_2,\bar Q_3\}$ 
are the light-ray operators with ``good'' conformal transformation properties,
Eqs.~(\ref{eq:Q}) and (\ref{eq:barQ}),
the scalar product $\langle\!\langle\ldots\rangle\!\rangle$ is defined in Eq.~(\ref{eq:<<>>})
and the coefficient $\rho_N$ in Eq.~(\ref{eq:rhoN}). Explicit expressions
for the ``wave functions'' $\overrightarrow{\Psi}_N$ are given in Eqs.~(\ref{Psi1}), (\ref{Psi2}), (\ref{Psi3}).
The ellipses stand for contributions of quasipartonic operators which are irrelevant in the present context.

Using this representation and orthogonality of the ``wave functions'' of different operators with respect
to our scalar product we derive the following ``kinematic projections'' for the complete set
of flavor-nonsinglet non-quasipartonic light-ray operators:
\begin{eqnarray}\label{eq:victory+coef}
 ig\overrightarrow{Q}(z_1,z_2,z_3) &=&(n\tilde n) \sum_{N=1}^\infty\sum_{k=0}^\infty
\frac{p_{Nk}(N+1)^2}{\rho_N ||\Psi_N||^{2}}\,
\overrightarrow{\Psi}_{Nk} (z_1,z_2,z_3)\,\partial_+^k(\partial\mathcal{O})_N  + \ldots\,,
\\
 ig\overrightarrow{\bar Q}(z_1,z_2,z_3) &=& (n\tilde n)\sum_{N=1}^\infty\sum_{k=0}^\infty
 \frac{p_{Nk}(N+1)^2}{\rho_N ||\Psi_N||^{2}}
(-1)^{N+1}\overrightarrow{ \Psi}_{Nk} (z_3,z_2,z_1)\,\partial_+^k(\partial\mathcal{O})_N  + \ldots\,,\nonumber
\end{eqnarray}
where $||\Psi_N||^2$ is given by Eq.~(\ref{eq:norm}), the coefficients $p_{Nk}$ are defined
in (\ref{eq:pNk1}), (\ref{eq:pNk}) and the functions $\overrightarrow{\Psi}_{Nk} (z_1,z_2,z_3)$
are given by
\begin{equation}
 \overrightarrow{\Psi}_{Nk} (z_1,z_2,z_3) = (S_+)^k \overrightarrow{\Psi}_{N} (z_1,z_2,z_3)\,,
\label{def:psiNk}
\end{equation}
where the differential operator $S_+$
depends implicitly on the conformal spins $j_1,j_2,j_3$ of the  components of $\overrightarrow{\Psi}_N$,
cf. Eq.~(\ref{SPsi}).
The ellipses in Eq.~(\ref{eq:victory+coef}) stand for the contributions of other twist-four operators
which are not reduced to total derivatives of leading twist ones, hence their matrix elements are not reduced
to leading twist parton distributions.

As a byproduct of this study, we have found a series of
``genuine'' twist-four flavor-nonsinglet quark-antiquark-gluon operators which have the same anomalous
dimensions as the leading twist quark-antiquark operators:
\begin{equation}
 2 (n\tilde n) T_N = \frac{ig\rho_N}{(N+1)^2}\Big[
 \langle\!\langle \overrightarrow{\Psi}_N, \overrightarrow{Q}\rangle\!\rangle
 +  \langle\!\langle \overrightarrow{\bar\Psi}_N, \overrightarrow{\bar Q}\rangle\!\rangle\Big]
 + \ldots
\label{eq:T2}
\end{equation}
These operators are not reduced to total derivatives, so that they contribute e.g. to the total cross section
of deep inelastic scattering.

\section{T-product of electromagnetic currents (I): Twist expansion}\label{sec:T1}

In the remaining Sections we apply the general framework developed above to the calculation of the
time-ordered product of two electromagnetic currents
\begin{equation}
 j_\mu(x) = \bar q(x) \gamma_\mu Q q(x)\,,
\end{equation}
where $Q$ is a matrix of the quark electromagnetic charges $Q=\text{diag}\{e_u,e_d,\ldots\}$.
Our goal will be to obtain the complete expression for
\begin{equation}
   T_{\mu\nu}(z_1,z_2) = i T\{j_\mu(z_1x)j_\nu(z_2x)\}\,,
\qquad
T_{\alpha\beta\dot\alpha\dot\beta}= \sigma^{\mu}_{\alpha\dot\alpha}\sigma^{\nu}_{\beta\dot\beta}T_{\mu\nu}
\end{equation}
to the twist-four accuracy, including the
higher-twist operators related to total derivatives of the leading-twist ones.
Here $x^\mu$ is a four-vector and $z_1$, $z_2$ are real numbers.
Schematically
\begin{equation}
 T_{\mu\nu}(z_1,z_2) = T^{t=2}_{\mu\nu}(z_1,z_2) + T^{t=3}_{\mu\nu}(z_1,z_2) + T^{t=4}_{\mu\nu}(z_1,z_2) + \ldots
\end{equation}
Conservation of the electromagnetic current implies the Ward identities
\begin{eqnarray}
    \partial^\mu  T_{\mu\nu}(z_1,z_2) &=& z_2\big[ i\mathbf{P}^\mu, T_{\mu\nu}(z_1,z_2)\big]\,,
\nonumber\\
    \partial^\nu  T_{\mu\nu}(z_1,z_2) &=& z_1\big[ i\mathbf{P}^\nu, T_{\mu\nu}(z_1,z_2)\big]\,.
\label{Ward}
\end{eqnarray}
In addition, translation invariance along the line connecting the currents implies that
\begin{eqnarray}
      e^{iz\mathbf{P}\cdot x}\, T_{\mu\nu}(z_1,z_2)\, e^{-iz\mathbf{P}\cdot x}  &=& T_{\mu\nu}(z_1+z,z_2+z)\,.
\label{translation}
\end{eqnarray}
Both relations, Eqs.~(\ref{Ward}) and (\ref{translation}), are only valid in the sum of all twists
but not for each twist separately. For the Ward identity, this property
was noticed in Refs.~\cite{Anikin:2000em,Kivel:2000rb,Radyushkin:2000ap,Belitsky:2000vx}.
Implementation of the electromagnetic gauge invariance beyond the leading twist accuracy has been at the center of
many discussions. By contrast, violation of the translation invariance condition (\ref{translation}) for the
contributions of a given twist has never been emphasized, to the best of our knowledge. As a manifestation of this
problem, the structure of the twist-four contribution depends in a nontrivial way on the positions of the currents.
For example we find that the kinematic twist-four part of the $T$-product with symmetric positions of the currents,
$iT\{j_\mu(x)j_\nu(-x)\}^{t=4}$, is considerably more complicated compared to the case when one of the
currents is taken at the origin, $iT\{j_\mu(2x)j_\nu(0)\}^{t=4}$. In the calculation of physical observables the
difference must be compensated by different structure of the Nachtmann-type kinematic corrections that originate
from the subtraction of traces in leading-twist operators. An example will be given in Sec.~\ref{sec:app}.
The lesson is that the distinction between finite-$t$ and
target mass corrections $\propto t/Q^2$, $\propto m^2/Q^2$ that originate from the operators of different twist
has no physical meaning. In particular the existing estimates of kinematic corrections to DVCS from the contributions
of twist-two operators alone can be misleading.

\begin{figure}[t]
\begin{center}
 \includegraphics[width=14.0cm]{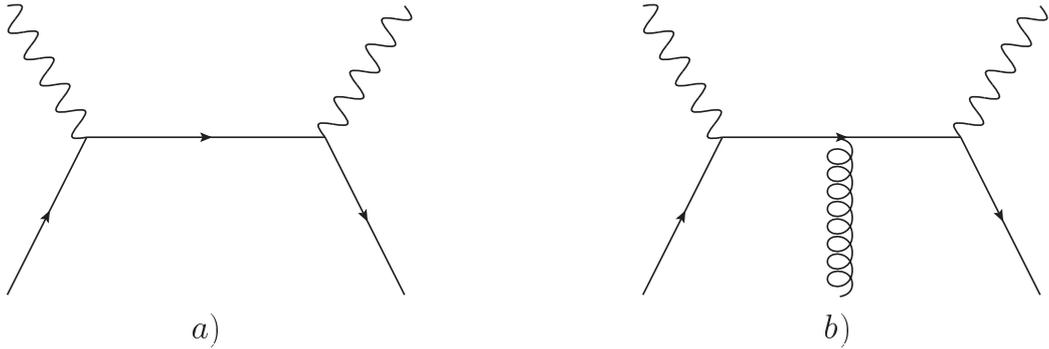}
\caption{Leading-order contributions to the time-ordered product of two electromagnetic currents}
\label{fig:1}
\end{center}
\end{figure}

The leading-order contributions to the $T$-product are given by the two Feynman diagrams
shown in Fig.~\ref{fig:1}a,b (and crossing-symmetric diagrams which are not shown for
brevity):
\begin{equation}
 iT\big\{j_\mu(z_1x)j_\nu(z_2x)\big\} = T^{(a)}_{\mu\nu}(z_1,z_2) + T^{(b)}_{\mu\nu}(z_1,z_2)+\ldots
\end{equation}
The ellipses stands for quasipartonic operators of twist-four and operators of higher twist
$t>4$.

A simple calculation yields (see e.g. Ref.~\cite{Balitsky:1987bk})
\begin{eqnarray}
 T^{(a)}_{\mu\nu}(z_1,z_2)&=&-\frac{1}{2\pi^2 x^4 z_{12}^3}
\bar q(z_1x)\gamma_\mu\slashed{x}\gamma_{\nu} Q^2 q(z_2x)+
\Big(
\mu\leftrightarrow\nu,\,\,
z_1\leftrightarrow z_2
\Big)\,,
\nonumber\\
 T^{(b)}_{\mu\nu}(z_1,z_2)&=&
 - \frac{g}{8\pi^2 x^2z_{12}}\int_{0}^1du\, x^\alpha\,\bar q(z_1) Q^2\gamma_\mu
 \Big[ \widetilde F_{\alpha\beta}(z_{21}^u)\gamma_5 +
 i(2u-1) F_{\alpha\beta}(z_{21}^u)\Big]\gamma^\beta\gamma_\nu\, q(z_2)
\nonumber\\
&&{}+\Big(
\mu\leftrightarrow\nu,\,\,
z_1\leftrightarrow z_2
\Big)\,,
\end{eqnarray}
The Wilson lines between the quarks are always implied.
Going over to spinor notation we obtain
\begin{eqnarray}\label{Tmn1}
T^{(a)}_{\alpha\beta\dot\alpha\dot\beta}(z_1,z_2)&=&-\frac2{\pi^2 x^4 z_{12}^3}
\Big\{
x_{\alpha\dot\beta}
\Big[\bar\psi_{\dot\alpha}(z_1x) \psi_\beta(z_2x)
-\chi_\beta(z_2x)\bar\chi_{\dot\alpha}(z_1x)
\Big]
\notag\\
&&
\phantom{-\frac2{\pi^2(x^2)^2 z_{12}^3}}
+x_{\beta\dot\alpha}
\Big\{\chi_\alpha(z_1x)\bar\chi_{\dot\beta}(z_2x)-
\bar\psi_{\dot\beta}(z_2x)\psi_\alpha(z_1x)\Big]\Big\}
\end{eqnarray}
and
\begin{eqnarray}\label{Tmn2}
T^{(b)}_{\alpha\beta\dot\alpha\dot\beta}(z_1,z_2)&=&
\frac{ig}{\pi^2x^2z_{12}}\int_0^1du\,\Big\{
 \bar\psi_{\dot\alpha}(z_1)\Big[ u (\bar x f)_{\dot\beta\alpha}(z_{21}^u)
 +
\bar u(x\bar f)_{\alpha\dot\beta}(z_{21}^u)\Big]\psi_\beta(z_2)
\notag\\
&&
\phantom{\frac{ig}{\pi^2x^2z_{12}}\int_0^1du}
-
\chi_{\alpha}(z_1)\Big[ \bar u (\bar x f)_{\dot\alpha\beta}(z_{21}^u)+
 u(x\bar f)_{\beta\dot\alpha}(z_{21}^u)\Big]\bar\chi_{\dot\beta}(z_2)
\notag\\
&&
\phantom{\frac{ig}{\pi^2x^2z_{12}}\int_0^1du}
-\bar\psi_{\dot\beta}(z_2)\Big[ u (\bar x f)_{\dot\alpha\beta}(z_{12}^u)+
\bar u(x\bar f)_{\beta\dot\alpha}(z_{12}^u)\Big]\psi_\alpha(z_1)
\notag\\
&&
\phantom{\frac{ig}{\pi^2x^2z_{12}}\int_0^1du}
+
\chi_{\beta}(z_2)\Big[ \bar u (\bar x f)_{\dot\beta\alpha}(z_{12}^u)+
 u(x\bar f)_{\alpha\dot\beta}(z_{12}^u)\Big]\bar\chi_{\dot\alpha}(z_1)\Big\}\,,
\end{eqnarray}
where for brevity we omitted the matrix of quark charges $Q^2$.

The operators in Eqs.~(\ref{Tmn1}), (\ref{Tmn2}) do not have a definite twist yet.
The twist separation is our first goal, where, in difference to the existing results
(e.g. \cite{Balitsky:1987bk}) we have to take into account operators containing
total derivatives.

\subsection{Handbag diagram, Fig.~1a}
For definiteness, let us consider the operator built of the left-handed quark and antiquark:
\begin{equation}\label{defO}
O_{\alpha\dot\alpha}(z_1x,z_2x) = \bar\psi_{\dot\alpha}(z_1x)[z_1x,z_2x]\psi_\alpha(z_2x)\,.
\end{equation}
Our task in this section is to expand $O_{\alpha\dot\alpha}$ in contributions of different twist
\begin{equation}
  O_{\alpha\dot\alpha}(z_1x,z_2x) = O^{t=2}_{\alpha\dot\alpha}(z_1x,z_2x)
                            +O^{t=3}_{\alpha\dot\alpha}(z_1x,z_2x)
                            +O^{t=4}_{\alpha\dot\alpha}(z_1x,z_2x)+\ldots
\end{equation}
To this end we start from the Taylor expansion
\begin{equation}\label{taylorO}
O_{\alpha\dot\alpha}(z_1x,z_2x)=\sum_{k=0}^\infty \frac1{2^k k!}
\bar x^{\dot\alpha_1\alpha_1}\ldots \bar x^{\dot\alpha_k\alpha_k}
O^{(k)}_{\alpha\alpha_1\ldots\alpha_k,\dot\alpha\dot\alpha_1\ldots\dot\alpha_k}\,,
\end{equation}
where
\begin{equation}\label{Oalpha}
O^{(k)}_{\alpha\alpha_1\ldots\alpha_k,\dot\alpha\dot\alpha_1\ldots\dot\alpha_k}(z_1,z_2)=
\partial_{\alpha_1\dot\alpha_1}\ldots\partial_{\alpha_k\dot\alpha_k}O_{\alpha\dot\alpha}(z_1x,z_2x)\Big|_{x=0}\,.
\end{equation}
Here and below
\begin{equation}
  \partial_{\alpha\dot\alpha} = \sigma^\mu_{\alpha\dot\alpha} \frac{\partial}{\partial x^\mu}
                   \equiv \sigma^\mu_{\alpha\dot\alpha} \partial_\mu\,.
\label{partialaba}
\end{equation}
Note that the operator
$O^{(k)}_{\alpha\alpha_1\ldots\alpha_k,\dot\alpha\dot\alpha_1\ldots\dot\alpha_k}$ is symmetric
under  interchange of the pairs of indices
$(\alpha_i,\dot\alpha_i)\leftrightarrow(\alpha_j,\dot\alpha_j)$, $i,j=1,\ldots,k$.

In order to simplify the notations we will use multi-indices
$A_k ,\dot A_k$:
\begin{align}
A_k=\{\alpha,\alpha_1\ldots\alpha_k\}, &&
\dot A_k=\{\dot\alpha,\dot\alpha_1\ldots\dot\alpha_k \}\,,
\end{align}
so that
\begin{equation}
O^{(k)}_{A_k\dot A_k}\equiv O^{(k)}_{\alpha\alpha_1\ldots\alpha_k,\dot\alpha\dot\alpha_1\ldots\dot\alpha_k}(z_1,z_2)\,.
\end{equation}
We also write
\begin{align}
\lambda^{A_k}\equiv\lambda^{\alpha}\lambda^{\alpha_1}\ldots\lambda^{\alpha_k}\,, &&
\frac{\partial}{\partial\lambda^{A_k}}\equiv\frac{\partial}{\partial\lambda^{\alpha}}
\frac{\partial}{\partial\lambda^{\alpha_1}}
\ldots
\frac{\partial}{\partial\lambda^{\alpha_k}}\,.
\end{align}
and similar in the ``dotted'' sector.

\subsubsection{Leading-twist projection operator}
An important advantage of the spinor formalism for the twist separation is that symmetrization
and trace subtraction in Lorentz indices are replaced by \emph{one} requirement: picking up
the leading-twist contribution corresponds to symmetrization separately in dotted and undotted
indices. The symmetric (alias leading-twist-two) part of~(\ref{Oalpha}) can be written as
\begin{equation}\label{symO}
\text{Sym}\,O^{(k)}_{A_k\dot A_k}(z_1,z_2)=
\frac1{[(k+1)!]^2}\frac{\partial}{\partial\lambda^{A_k}}
\frac{\partial}{\partial\bar\lambda^{\dot A_k}}
O_{k}(\lambda,\bar\lambda)\Big|_{\lambda,\bar\lambda=0}\,,
\end{equation}
where
\begin{equation}
O_{k}(\lambda,\bar\lambda)=\lambda^{A_k}\bar\lambda^{\dot A_k}
O^{(k)}_{A_k\dot A_k}(z_1,z_2)
=
(\lambda\partial\bar\lambda)^{k}O_{++}(z_1x,z_2x)\Big|_{x=0}\,
\end{equation}
and
\begin{equation}
O_{++}(z_1x,z_2x)=\lambda^{\alpha}\bar\lambda^{\dot\alpha}O_{\alpha\dot\alpha}(z_1x,z_2x)
   =\bar\psi_+(z_1x)[z_1x,z_2x]\psi_+(z_2x)\,.
\end{equation}
Inserting (\ref{symO}) into (\ref{taylorO}) we obtain
\begin{align}\label{ot2f}
O^{t=2}_{\alpha\dot\alpha}(z_1,z_2)=\partial_\alpha\bar\partial_{\dot\alpha}\sum_{k=0}^\infty \frac1{2^k k!}\frac1{[(k+1)!]^2}(\bar\partial \bar x\partial)^k
\Big[\Big[(\lambda\partial\bar\lambda)^{k}O_{++}(z_1x,z_2x)\Big]_{x=0}\Big]_{\lambda,\bar\lambda=0}\,.
\end{align}
Here
\begin{align}\label{}
(\bar\partial \bar x\partial)=\frac{\partial}{\partial\bar\lambda^{\dot\alpha}}\bar x^{\dot\alpha\alpha}
\frac{\partial}{\partial\lambda^{\alpha}}\equiv\bar\partial_{\dot\alpha}\bar x^{\dot\alpha\alpha}\partial_\alpha
\,,
&&
(\lambda\partial\bar\lambda)=\lambda^\alpha\partial_{\alpha\dot\alpha}\bar\lambda^{\dot\alpha}\,.
\end{align}
Note that we are using  shorthand notations
\begin{align}
\partial_\alpha=\frac{\partial}{\partial\lambda^{\alpha}}\,, &&
\bar\partial_{\dot\alpha}=\frac{\partial}{\partial\bar\lambda^{\bar\alpha}}\,,
\end{align}
for the derivatives with respect to the auxiliary spinors,
whereas $\partial_{\alpha\dot\alpha}=\sigma^{\mu}_{\alpha\dot\alpha}\partial_\mu$
stands for the derivative over the four-vector $x^\mu$.
We hope that this similarity will not lead to a confusion.

The sum in~(\ref{ot2f}) can be rewritten as follows
\begin{eqnarray}
O^{t=2}_{\alpha\dot\alpha}(z_1,z_2)&=&
\partial_\alpha\bar\partial_{\dot\alpha}\sum_{k=0}^\infty\frac1{[(k+1)!]^2}
(\bar\partial \bar x\partial)^k \left(\sum_{m=0}^\infty\frac1{2^m m!}
(\lambda\partial\bar\lambda)^{m}O_{++}(z_1x,z_2x)\Big]_{x=0}
\right)\Big|_{\lambda,\bar\lambda=0}
\notag\\
&=&\partial_\alpha\bar\partial_{\dot\alpha}
\sum_{k=0}^\infty\frac1{[(k+1)!]^2}(\bar\partial \bar x\partial)^kO_{++}(z_1n,z_2n)\Big|_{\lambda,\bar\lambda=0}\,,
\end{eqnarray}
where $n_{\alpha\dot\alpha}=\lambda_{\alpha}\bar\lambda_{\dot\alpha}$ and we took into
account that only the terms with $k=m$ survive.
Finally, using that
$\partial_{\alpha\dot\alpha}(\bar\partial \bar x\partial)^{k+1}=2(k+1)\partial_\alpha\bar\partial_{\dot\alpha}
(\bar\partial \bar x\partial)^{k}$ one obtains
\begin{eqnarray}\label{Ot2}
O^{t=2}_{\alpha\dot\alpha}(z_1x,z_2x)&=&\frac12\partial_{\alpha\dot\alpha}
\int_0^1du \sum_{k=1}^\infty u^{k}
\frac{(\bar\partial \bar x\partial)^{k+1}}{[(k+1)!]^2}O_{++}(z_1n,z_2n)\Big|_{\lambda,\bar\lambda=0}
\notag\\
&=&\frac12\partial_{\alpha\dot\alpha}\Pi(x,\lambda)\int_0^1 du\,
O_{++}(uz_1n,uz_2n)
\,,
\end{eqnarray}
where the projector $\Pi(x,\lambda)$ is given by the following expression
\begin{align}\label{PI}
[\Pi f](x)\equiv\Pi(x,\lambda)f(\lambda,\bar\lambda)=\sum_{k=0}^\infty \frac{(\bar\partial \bar x\partial)^{k}}{[k!]^2}
 f(\lambda,\bar\lambda)\Big|_{\lambda,\bar\lambda=0}.
\end{align}
Note that the operator $O_{++}(uz_1n,uz_2n)$ on the r.h.s. of Eq.~(\ref{Ot2}) ``lives'' on
the light ray, $n=\lambda\otimes\bar\lambda$, thus $n^2=0$. Hence the leading-twist part
of the nonlocal quark-antiquark operator at $x^2\slashed{=}0$ is expressed in terms of the
light-ray operator.

$\Pi(x,\lambda)$  is indeed a projector since
\begin{align}
\Pi^2=\Pi \Longleftrightarrow\Pi(x,\lambda)\Pi(n=\lambda\otimes\bar\lambda,\eta)=\Pi(x,\eta)\,.
\end{align}
It satisfies a number of relations that are useful in practical calculations:
\begin{eqnarray}\label{P-properties}
\partial_\mu\partial^\mu\,\Pi(x,\lambda)&=&0\,,
\notag\\
\Pi(x,\lambda)\,\lambda_\gamma\partial_\alpha&=&-\frac12 (x\bar\partial)_{\gamma\alpha}\,\Pi(x,\lambda)\,,
\notag\\
\Pi(x,\lambda)\,\bar\lambda_{\dot\gamma}\bar\partial_{\dot\alpha}&=&\frac12(\bar x\partial)_{\dot\gamma\dot\alpha}
\,\Pi(x,\lambda)\,,
\notag\\
\Pi(x,\lambda)\lambda^\alpha\bar\lambda^{\dot\alpha}&=&
\bar x^{\dot\alpha\alpha}\,\Pi(x,\lambda)-\frac12x^2\,\bar\partial^{\dot\alpha\alpha}\,\int_0^1 du\, \Pi(ux,\lambda)\,.
\end{eqnarray}
These relations can easily be verified making use of Eq.~(\ref{PI}).

It is instructive to compare the expression for $\Pi(x,\lambda)$ with the leading-twist projection operator
in the conventional vector notation. It can be shown that (see Appendix~\ref{app:projector})
\begin{align}\label{PIVO}
\Pi(x,\lambda) f(n=\lambda\otimes\bar\lambda)=
\sum_{k=0}^\infty P_k(x,\partial_y) f(y)\big|_{y=0}\,,
\end{align}
where
\begin{align}
P_k(x,y)=\frac1{2^k k!}(x^2 y^2)^{k/2}C_k^{(1)}\left(\frac{(xy)}{\sqrt{x^2y^2}}\right)\,
\end{align}
and $C^{(1)}_k(x)$ is the Gegenbauer polynomial. This form is not unique. Equivalent representations
can be found in Refs.~\cite{Belitsky:2001hz,Geyer:2004bx,Balitsky:1987bk}, see also the book
\cite{Vilenkin}.
The above expression for the projector $\Pi$ in the spinor formalism is much simpler and,
therefore, easier to handle.

\subsubsection{Twist three}

The next task is to separate the twist-three contribution, $O^{t=3}_{\alpha\dot\alpha}(z_1x,z_2x)$.
To this end one has to contract one pair of indices in the operator~(\ref{Oalpha}) and symmetrize
over all the others. Since
$$\epsilon^{\alpha_i\alpha_j}\partial_{\alpha_i\dot\alpha_i}\partial_{\alpha_j\dot\alpha_j}=
-\epsilon_{\dot\alpha_i\dot\alpha_j}\partial^2$$
a contraction involving two derivatives gives rise to a twist-four operator. The only possibility
to get twist three is therefore to contract the quark field index with the one of a derivative.
Hence one can construct two operators
\begin{align}
\epsilon^{\alpha_k\alpha}
[\partial_{\alpha_1\dot\alpha_1}\ldots\partial_{\alpha_k\dot\alpha_k}\mathcal{O}_{\alpha\dot\alpha}(z_1x,z_2x)]|_{x=0}\,,
&&
\epsilon^{\dot\alpha\dot\alpha_k}
[\partial_{\alpha_1\dot\alpha_1}\ldots\partial_{\alpha_k\dot\alpha_k}\mathcal{O}_{\alpha\dot\alpha}(z_1x,z_2x)]|_{x=0}\,,
\end{align}
and symmetrize over all remaining open indices multiplying with the auxiliary spinors:
\begin{eqnarray}
\mathcal{O}^{(t=3)}_k(\lambda,\bar\lambda)&=&
(\lambda\partial\bar\lambda)^{k-1}\partial_{\alpha+}{{{O}_k}^{\alpha}}_ +(z_1x,z_2x)|_{x=0}\,,
\notag\\
\mathcal{O}^{(t=3)}_k(\lambda,\bar\lambda)&=&
(\lambda\partial\bar\lambda)^{k-1}\partial_{+\dot\alpha}
{{{O}_k}_+}^{\dot\alpha}(z_1x,z_2x)|_{x=0}\,,
\label{Ott3}
\end{eqnarray}
where for brevity we do not show the dependence of $\mathcal{O}^{(t=3)}_k$,
$\bar{\mathcal{O}}^{(t=3)}_k$ on the coordinates $z_i$.

Since the operators in (\ref{Ott3}) are the only existing independent twist-three operators,
the twist-three part of the local operator with open indices can be written as
\begin{align}\label{OT3}
O_{A_k\dot A_k}^{(t=3)}=\Pi_{A_k\dot A_k}\mathcal{O}_k^{(t=3)}+{\bar\Pi}_{A_k\dot A_k}\bar{\mathcal{O}}_k^{(t=3)}\,.
\end{align}
Explicit expressions for the coefficients $\Pi_{A_k\dot A_k}$, $\Pi_{A_k\dot A_k}$ can be found from the
requirement that $O_{A_k\dot A_k}^{(t=3)}$ is symmetric
under the permutation of the pairs of indices
$(\alpha_i,\dot\alpha_i)\leftrightarrow(\alpha_j,\dot\alpha_j)$, $i,j=1,\ldots,k$
and reduces to either $\mathcal{O}^{(t=3)}_k$ or $\bar{\mathcal{O}}^{(t=3)}_k$ by a
contraction of one pair of indices and multiplication with auxiliary spinors for the
remaining ones. One obtains
\begin{eqnarray}\label{PPform}
\Pi_{A_k\dot A_k}\mathcal{O}_k^{(t=3)}&=&
\frac{k}{[(k+1)!]^2}\bar\partial_{\dot\alpha}\left(\sum_{j=1}^k\epsilon_{\alpha_j\alpha}\bar\partial_{\dot\alpha_j}
\prod_{\substack{i=1,\\ i\neq j}}^k\partial_{\alpha_i}\bar \partial_{\dot \alpha_i}\right)
\mathcal{O}_k^{(t-3)}(\lambda,\bar\lambda)\Big|_{\lambda=0}\,,
\notag\\
{\bar\Pi}_{A_k\dot A_k}\bar{\mathcal{O}}_k^{(t=3)}&=&
\frac{k}{[(k+1)!]^2}
\partial_{\alpha}\left(\sum_{j=1}^k\epsilon_{\dot \alpha\dot \alpha_j}\partial_{\alpha_j}
\prod_{\substack{i=1,\\ i\neq j}}^k\partial_{\alpha_i}\bar \partial_{\dot \alpha_i}\right)
\bar{\mathcal{O}}_k^{(t-3)}(\lambda,\bar\lambda)\Big|_{\lambda=0}\,.
\end{eqnarray}
It remains to rewrite this result in terms of nonlocal operators. Let
\begin{eqnarray}\label{OBO}
\mathcal{O}^{(t=3)}(z_1n, z_2n)&=&
{}[\partial_{\alpha+}{{{O}}^{\alpha}}_ +](z_1n,z_2n)\,,
\notag\\
\bar{\mathcal{O}}^{(t=3)}(z_1n,z_2n)&=&
{}[\partial_{+\dot\alpha}{{O}_+}^{\dot\alpha}](z_1n,z_2n)\,.
\end{eqnarray}
Inserting~(\ref{OT3}) in~(\ref{taylorO}) and using explicit expressions from~(\ref{PPform})
we end up with
\begin{eqnarray}\label{tw3sumform}
{O}^{t=3}_{\alpha\dot\alpha}(z_1,z_2)&=&
\frac12
\sum_{k=1}^\infty \frac{k(\bar\partial \bar x\partial)^{k-1}}{(k+1)!^2}
\Bigl[(\bar\partial\bar x)_\alpha\bar\partial_{\dot\alpha}\mathcal{O}^{(t=3)}(z_1n,z_2n)+
(\bar x\partial)_{\dot\alpha} \partial_{\alpha}\bar{\mathcal{O}}^{(t=3)}(z_1n,z_2n)
\Bigr]\Big|_{\lambda,\bar\lambda=0}\,,
\nonumber\\&=&
\frac12
\sum_{k=1}^\infty \frac{(\bar \partial \bar x \partial)^{k}}{(k+1)!^2}
\Bigl[\lambda_\alpha\bar\partial_{\dot\alpha}\mathcal{O}^{(t=3)}(z_1n,z_2n)+
\bar\lambda_{\dot\alpha} \partial_{\alpha}\bar{\mathcal{O}}^{(t=3)}(z_1n,z_2n)
\Bigr]\Big|_{\lambda,\bar\lambda=0}\,.
\end{eqnarray}
{}Finally, using
\begin{align}
  \frac{\partial}{\partial \bar x^{\dot\alpha\gamma}} (\bar\partial\bar x \partial)^k
 (\bar\partial\bar x)^\gamma =
2(k+2) (\bar\partial\bar x \partial)^k \bar\partial_{\dot\alpha}\,,
&&
 \quad \frac{\partial}{\partial \bar x^{\dot\gamma\alpha}} (\bar\partial\bar x \partial)^k
 (\bar x \partial)^{\dot\gamma} =
2(k+2) (\bar\partial\bar x \partial)^k \partial_{\alpha}
\end{align}
this expression can  be rewritten as
\begin{eqnarray}\label{tw3intform}
{O}^{t=3}_{\alpha\dot\alpha}(z_1x,z_2x)&=&
-\frac14\int_0^1 du \, u\bar u\Big[{\bar \partial_{\dot\alpha}{} }^{\gamma}\Pi(x,\lambda)
\lambda_\gamma\lambda_\alpha\mathcal{O}^{(t=3)}(z_1un,z_2un)
\notag\\
&&{}\hspace*{2cm}+{\partial_{\alpha}}^{\dot\gamma}\Pi(x,\lambda)
\bar\lambda_{\dot\gamma}\bar\lambda_{\dot\alpha}\bar{\mathcal{O}}^{(t=3)}(z_1un,z_2un)\Big]\,,
\end{eqnarray}
where $\Pi(x,\lambda)$ is the projection operator defined in Eq.~(\ref{PI}).
We remind that $n^{\alpha\bar\alpha} = \lambda^\alpha\bar\lambda^{\dot\alpha}$ and application of
the projection operator involves sending $\lambda,\bar\lambda\to 0$ in the final
expression.

As the next step, we have to unravel the twist-three light-ray operators on the r.h.s.
of Eq.~(\ref{tw3intform}) to separate the contributions of interest.
Omitting EOM terms that contain $D_{\alpha+}\psi^\alpha$ or
$D_{+\dot\alpha}\bar\psi^{\dot\alpha}$ one obtains
\begin{eqnarray}\label{OBO-ex}
\mathcal{O}^{(t=3)}(z_1n,z_2n)&=&z_1[D_{\alpha +}\bar\psi_+](z_1n)\psi^\alpha(z_2n)
-igz_{12}\int_{0}^1du\, z_{21}^u\,\bar\psi_+(z_1n)\,\bar f_{++}(z_{21}^un)\,
\psi_+(z_2n)\,,
\notag\\
\bar{ \mathcal{O}}^{(t=3)}(z_1n,z_2n)&=&
z_2\bar\psi^{\dot\alpha}(z_1n)[D_{+\dot\alpha }\psi_+](z_2n)-igz_{12}
\int_{0}^1du\, z_{21}^u\,\bar\psi_+(z_1n)\, f_{++}(z_{21}^un)\,
\psi_+(z_2n)\,,
\nonumber\\
\end{eqnarray}
where the terms in $f_{++}, \bar f_{++}$ originate from differentiation of the Wilson line,
cf. Eq.~(\ref{Dlink}). Such terms give rise to quasipartonic twist-three operators
which are not relevant in the present context, so that we will discard them altogether.
The remaining contributions (i.e. without the integral) can be rewritten
in terms of the momentum operator (\ref{eq:P}), e.g.
\begin{eqnarray}\label{OPP}
(\mu\lambda)\mathcal{O}^{(t=3)}(z_1n,z_2n)&=&
z_1(\mu\lambda)\big[i\mathbf{P}_{\alpha +},\bar\psi_+(z_1n)\psi^\alpha(z_2n)\big]
\notag\\
&=&
 z_1\big[i\mathbf{P}_{++},\bar\psi_+(z_1n)\psi_-(z_2n)\big]
-z_1\big[i\mathbf{P}_{-+},\bar\psi_+(z_1n)\psi_+(z_2n)\big]\,.
\hspace{0.5cm}{}
\end{eqnarray}
The second term on the r.h.s. of~(\ref{OPP}) has the desired form:
it is a total transverse  derivative of the leading-twist operator.
The first term can be related to descendants of twist-two operators using
the following identities:
\begin{eqnarray}\label{muP}
\mu^{\alpha}\frac{\partial}{\partial\lambda^\alpha}O_{++}(z_1n,z_2n)&=&
O_{+-}(z_1n,z_2n)+ \frac12 z_1[D_{-+}\psi_+](z_1n)\psi_+(z_2n)
\notag\\
&&+
z_2\partial_{z_2} O_{+-}(z_1,z_2)+\ldots\,,
\notag\\
{}[i\mathbf{P}_{-+},O_{++}(z_1n,z_2n)]&=&[D_{-+}\psi_+](z_1n)\psi_+(z_2n)+2\partial_{z_2}O_{+-}(z_1n,z_2n)
+\ldots\,,
\end{eqnarray}
where
\begin{equation}
O_{+-}(z_1n,z_2n)=\bar\psi_+(z_1n)\psi_-(z_2n)\,.
\end{equation}
and the ellipses stand for the contributions of quasipartonic operators.
Excluding the term~$[D_{-+}\psi_+](z_1n)\psi_+(z_2n)$ from Eqs.~(\ref{muP})
one obtains an equation for the operator $O_{+-}(z_1n,z_2n)$
\begin{align}\label{}
\partial_{z_2} z_{21}O_{+-}(z_1n,z_2n)=(\mu\partial_\lambda)O_+(z_1n,z_2n)-\frac{z_1}2[i\mathbf{P}_{-+},O_+(z_1n,z_2n)]
\end{align}
which is easily solved to give
\begin{align}\label{Opmsol}
O_{+-}(z_1n,z_2n)=&(\mu\partial_\lambda)\int_0^1du\,O_+(z_1n,z_{21}^un)-
\frac{z_1}2\int_0^1du\,[i\mathbf{P}_{-+},O_+(z_1n,z_{21}^un)]\,.
\end{align}
Using this expression in~(\ref{OPP}) we obtain
\begin{eqnarray}
\lambda_\alpha\mathcal{O}^{(t=3)}(z_1n,z_2n)&=&
-z_1[i\mathbf{P}_{\alpha+},O_{++}(z_1,z_2)]
+
\frac{z_1}{z_{12}}\int_{z_2}^{z_1}dw\Big[i\mathbf{P}_{++},
\partial_\alpha O_{++}(z_1,w)\Big]
\notag\\
&&{}
-\frac{z^2_1}{2 z_{12}}\int_{z_2}^{z_1}dw
\Big[i\mathbf{P}_{++},\Big[i\mathbf{P}_{\alpha+},O_{++}(z_1,w)\Big]
\Big]\,
\end{eqnarray}
and similarly
\begin{eqnarray}
\bar\lambda_{\dot\alpha}\bar{\mathcal{O}}^{(t=3)}(z_1n,z_2n)&=&
-z_2[i\mathbf{P}_{+\dot\alpha},O_{++}(z_1,z_2)]
+
\frac{z_2}{z_{12}}\int_{z_2}^{z_1}dw\Big[i\mathbf{P}_{++},
\bar\partial_{\dot\alpha} O_{++}(w,z_2)\Big]
\notag\\
&&{}
-
\frac{z^2_2}{2z_{12}}\int_{z_2}^{z_1}dw\Big[i\mathbf{P}_{++},
\Big[i\mathbf{P}_{+\dot\alpha},O_{++}(w,z_2)\Big]
\Big]\,.
\end{eqnarray}
Finally, inserting these expressions in~(\ref{tw3intform}) one obtains after some algebra
\begin{eqnarray}
O_{\alpha\dot\alpha}^{t=3}(z_1x,z_2x)&=&\frac14\partial_{\alpha\dot\alpha} \Pi(x,\lambda)
\int_0^1du\, u\int_{z_2}^{z_1}\frac{dw}{z_{12}}\,
\Big[i\mathbf{P}_{++}, z_1O_{++}(z_1u, wu)+z_2O_{++}(wu, z_2u)\Big]
\notag\\
&&{}+\frac14{\bar\partial_{\dot\alpha}{}}^{\gamma} \Pi(x,\lambda)
\int_0^1du\, u^2\int_{z_2}^{z_1}\frac{dw}{z_{12}}\,
\Big[i(\mathbf{P}\bar n)_{\alpha\gamma}, z_1O_{++}(z_1u, wu)\Big]
\notag\\
&&{}+\frac14{\partial_{\alpha}{}}^{\dot\gamma} \Pi(x,\lambda)
\int_0^1du\, u^2\int_{z_2}^{z_1}\frac{dw}{z_{12}}\,
\Big[i(\bar n{\mathbf{P}})_{\dot\gamma\dot\alpha}, z_2O_{++}(wu, z_2u)\Big].
\end{eqnarray}
Note that the operators $O_{++}$ on the r.h.s. of this expression are on the light cone,
i.e. $O_{++}(wu, z_2u) \equiv O_{++}(wun, z_2un)$. Making use of Eqs.~(\ref{P-properties})
one can move all $\lambda$-dependent factors to the left
of the leading-twist projector $\Pi(x,\lambda)$ and rewrite the answer
in terms of the operator
\begin{equation}\label{def:O++}
 \mathcal{O}^{t=2}_{++}(z_1x,z_2x) = \Pi(x,\lambda){O}_{++}(z_1n,z_2n)=\big[\Pi O_{++}\big](z_1x,z_2x)\,,
\end{equation}
i.e. the leading-twist part of the nonlocal quark-antiquark operator $\bar\psi_+(z_1x)\psi_+(z_2x)$
off the light cone.
In this way one obtains
\begin{eqnarray}\label{O-TW3}
O_{\alpha\dot\alpha}^{t=3}(z_1x,z_2x)&=&
\frac14\int_0^1du\, u\int_{z_2}^{z_1}\frac{dw}{z_{12}}
\notag\\
&& {}\times \biggl[
\Big[i\bar{\mathbf{P}}_{\mu},(x\bar\sigma^\mu\partial)_{\alpha\dot\alpha}z_1\mathcal{O}^{t=2}_{++}(z_1u, wu)
+(\bar x\sigma^\mu\bar \partial)_{\dot\alpha\alpha}z_2\mathcal{O}^{t=2}_{++}(wu, z_2u)\Big]
\nonumber\\
&&+
\ln u\partial_{\alpha\dot\alpha}x^2\partial^\nu
\Big[i\mathbf{P}_\nu,
z_1\mathcal{O}^{t=2}_{++}(z_1u, wu)+z_2\mathcal{O}^{t=2}_{++}(wu, z_2u)\Big]
\biggr]+\ldots
\end{eqnarray}
which is our final result.
It is easy to check that
$$
x^{\alpha\dot\alpha}O_{\alpha\dot\alpha}^{t=3}(z_1x,z_2x)=
\partial^{\alpha\dot\alpha}O_{\alpha\dot\alpha}^{t=3}(z_1x,z_2x)=0.
$$
Note that the terms in the last line of Eq.~(\ref{O-TW3}) are by themselves twist-four.
They must be included in order to achieve the proper separation of twist-three and twist-four
contributions (i.e. their role is to subtract unwanted twist-four contributions that are present in the
other terms), but can be omitted if the calculation is done to the twist-three accuracy only.
The ellipses stand for the contributions of twist-three quasipartonic operators and EOM terms.

\subsubsection{Twist four}\label{subsec:twist4}
Taking into account that $O^{(k)}_{\alpha\alpha_1\ldots\alpha_k,\dot\alpha\dot\alpha_1\ldots\dot\alpha_k}$~(\ref{Oalpha})
is symmetric under permutations of the pairs of indices
$(\alpha_i,\dot\alpha_i)\leftrightarrow(\alpha_j,\dot\alpha_j)$, $i,j=1,\ldots,k$,
one has three possibilities to project a twist-four operator:
\begin{eqnarray}
 P^{A_k\dot A_k}_1 &=& \frac12\epsilon^{\alpha_1\alpha_2}\epsilon^{\dot\alpha_1\dot\alpha_2}
          (\lambda^{\alpha}\lambda^{\alpha_3}\ldots\lambda^{\alpha_k})
          (\bar\lambda^{\dot\alpha}\bar\lambda^{\dot\alpha_3}\ldots\bar\lambda^{\alpha_k})\,,
\nonumber\\
 P^{A_k\dot A_k}_2 &=& \epsilon^{\alpha_1\alpha}\epsilon^{\dot\alpha_1\dot\alpha} (\lambda^{\alpha_2}\ldots\lambda^{\alpha_k})
         (\bar\lambda^{\dot\alpha_2}\ldots\bar\lambda^{\alpha_k})\,,
\nonumber\\
 P^{A_k\dot A_k}_3 &=& \epsilon^{\alpha_1\alpha}\epsilon^{\dot\alpha_2\dot\alpha} (\lambda^{\alpha_2}\ldots\lambda^{\alpha_k})
           (\bar\lambda^{\dot\alpha_1}\bar\lambda^{\dot\alpha_3}\ldots\bar\lambda^{\alpha_k})\,,
\label{P123}
\end{eqnarray}
so that
\begin{eqnarray}
P_1\cdot O^{(k)} &=& -(\lambda\partial\bar\lambda)^{k-2}\partial^2 {O}_{++}(z_1x,z_2x)]\big|_{x=0}\,,
\notag\\
P_2\cdot O^{(k)} &=& -(\lambda\partial\bar\lambda)^{k-1} \partial_{\alpha\dot\alpha}
                          {O}^{\alpha\dot\alpha}(z_1x,z_2x)]\big|_{x=0}\,,
\notag\\
P_3 \cdot O^{(k)}&=& -(\lambda\partial\bar\lambda)^{k-2}\partial_{\alpha+}\partial_{+\dot\alpha}{O}^{\alpha\dot\alpha}
(z_1x,z_2x)]\big|_{x=0}\,.
\label{eq:PcdotO}
\end{eqnarray}
However, because of the Fierz identity
\begin{equation}
 \partial^2 \mathcal{O}_{++} = (\lambda\partial\bar\lambda)
 \partial_{\alpha\dot\alpha} \mathcal{O}^{\alpha\dot\alpha}
 -
\bar\partial_{+\alpha}\partial_{+\dot\alpha} \mathcal{O}^{\alpha\dot\alpha}
\end{equation}
only two projections are independent, e.g. $P_3 \cdot O^{(k)}$ can be
eliminated in favor of the two others.
Introducing the operators
\begin{eqnarray}
\Pi^{(1)}_{A_k\dot A_k}&=&
\sum_{j<m}\epsilon_{\alpha_j\alpha_m}\epsilon_{\dot\alpha_j\dot\alpha_m} \partial_\alpha\bar\partial_{\dot\alpha}
\prod_{i\neq \{j,m\}} \partial_{\alpha_i}\bar\partial_{\dot\alpha_i}\,,
\notag\\
\Pi^{(2)}_{A_k\dot A_k}&=&\sum_{j=1}^k\epsilon_{\alpha_j\alpha}\epsilon_{\dot\alpha_j\dot\alpha}
\prod_{i\neq j} \partial_{\alpha_i}\bar\partial_{\dot\alpha_i}
\end{eqnarray}
we can write the twist-four part of the quark-antiquark operator with free indices as
\begin{equation}\label{twist4O}
O_{A_k\dot A_k}^{(t=4)}=\sum_{a,b=1}^2\Pi^{(a)}_{A_k\dot A_k} M_{ab} \, P_b\cdot O^{(k)}\,.
\end{equation}
The matrix of coefficients $M_{ab}$ can be found applying the projection operators
$P^{A_k\dot A_k}_1$, $P^{A_k\dot A_k}_2$ to this equation and requiring that the result reproduces
Eq.~(\ref{eq:PcdotO}). A short calculation gives
\begin{equation}
M = \frac{k^2}{[(k+1)!]^2}
\begin{pmatrix}k+3&-1\\
1-k& k\end{pmatrix}\,.
\end{equation}
Inserting this representation, Eq.~(\ref{twist4O}),  in Eq.~(\ref{taylorO})
and using that
\begin{align}
\prod_{q=1}^k \bar x^{\dot\alpha_q\alpha_q} \Pi^{(1)}=
-\frac12 k\, x^2\frac{\partial}{\partial \bar x^{\dot\alpha\alpha}}(\bar \partial \bar x \partial)^{k-1}\,,
&&
\prod_{q=1}^k \bar x^{\dot\alpha_q\alpha_q} \Pi^{(2)}=
 -k\, x_{\alpha\bar\alpha} (\bar \partial \bar x \partial)^{k-1}\,
\end{align}
one obtains after a little algebra
\begin{eqnarray}\label{4sumO}
\mathcal{O}^{t=4}_{\alpha\dot\alpha}(z_1,z_2)&=&
\frac12\sum_{k=1}^\infty\frac{k^2}{[(k+1)!]^2}
\Big\{k\,x_{\alpha\dot\alpha}-\frac12x^2\partial_{\alpha\dot\alpha}\Big\}
(\partial x\bar\partial)^{k-1} O^{(A)}(n;z_1,z_2)\Big|_{\lambda,\bar\lambda=0}
\notag\\
&-&\frac1{4}\sum_{k=2}^\infty\frac{k^2}{[(k+1)!]^2}\Big\{x_{\alpha\dot\alpha}
-\frac{k+3}{k-1}\,\frac12x^2\partial_{\alpha\dot\alpha}
\Big\}(\partial x\bar\partial)^{k-1}O^{(B)}(n;z_1,z_2)\Big|_{\lambda,\bar\lambda=0}\,,
\nonumber\\
\end{eqnarray}
where
\begin{eqnarray}
O^{(A)}(x;z_1,z_2)&=&[\partial_{\beta\dot\beta} O^{\beta\dot\beta}](z_1x,z_2x)\,,
\notag\\[2mm]
O^{(B)}(x;z_1,z_2)&=&\lambda^{\alpha}\bar\lambda^{\dot\alpha}[\partial^2_x O_{\alpha\dot\alpha}](z_1x,z_2x)\,.
\end{eqnarray}
The result in Eq.~(\ref{4sumO}) can be rewritten in an integral form
which is more convenient in applications:
\begin{equation}
O_{\alpha\dot\alpha}^{(t=4)}(z_1,z_2)=\mathcal{O}_{\alpha\dot\alpha}^{(A)}(z_1,z_2)+
                                      \mathcal{O}_{\alpha\dot\alpha}^{(B)}(z_1,z_2)\,,
\end{equation}
where
\begin{eqnarray}
\mathcal{O}^{(A)}_{\alpha\dot\alpha}(z_1,z_2)&=&\frac12 \int_0^1\!du\,
\Big[(1+\ln u)\,x_{\alpha\dot\alpha}+\frac12 \ln u\, x^2\partial_{\alpha\dot\alpha}\Big]
\Big[\Pi O^{(A)}\Big](x;uz_1,uz_2)\,,
\notag\\
\mathcal{O}^{(B)}_{\alpha\dot\alpha}(z_1,z_2)&=&\frac14\int_0^1\!du
\Big[\ln u\,x_{\alpha\dot\alpha}
+\frac12\Big(\frac{1}{u^2}\!-\!1\! +\! \ln u\Big)
x^2\partial_{\alpha\dot\alpha}\Big]
\Big[\Pi O^{(B)}\Big](x;uz_1,uz_2)\,,\hspace*{0.3cm}
\end{eqnarray}
Using Eq.~(\ref{Plink}) the operators $O^{(A)}, O^{(B)}$ can be expressed in terms of
quark-gluon operators. Neglecting the quasipartonic contributions one obtains
\begin{eqnarray}
O^{(A)}(n;z_1,z_2)&=&-\int_{z_2}^{z_1} dw\, w\,\Big[\mathcal{F}+\bar{\mathcal{F}}\Big](z_1,w,z_2)\,,
\notag
\\
O^{(B)}(n;z_1,z_2)&=& z_1\,z_2\big[i\mathbf{P}_\mu,\big[i\mathbf{P}^\mu,O_{++}(z_1,z_2)\big]\big]-
2z_{12}\Big[z_2\mathcal{F}(z_1,z_2,z_2)+z_1\bar{\mathcal{F}}(z_1,z_1,z_2)\Big]
\notag\\
&&{}+\int_{z_2}^{z_1}dw\Big\{ z_1(w-z_2) \Bigl[\mathcal{D}(z_1,w,z_2)-2\partial_{z_1}
{}[\mathcal{F}+\bar{\mathcal{F}}](z_1,w,z_2)\Bigr]
\notag\\
&&
\phantom{\int_{z_2}^{z_1}dw\Biggl\{}
+z_2(w-z_1) \Bigl[\bar{\mathcal{D}}(z_1,w,z_2)-2\partial_{z_2}[\mathcal{F}+\bar{\mathcal{F}}](z_1,w,z_2)\Bigr]
\Big\}\,,
\end{eqnarray}
where $\mathcal{F},\bar{\mathcal{F}},{\mathcal{D}},\bar{\mathcal{D}}$ are defined in
Eq.~(\ref{DQdef}). Using Eq.~(\ref{app:calnoncal}) the result can further be rewritten
in terms of operators of the conformal basis (\ref{eq:Q}), (\ref{eq:barQ}).

\subsection{Gluon emission diagram, Fig.~1b}\label{subsec:ge}
The  gluon emission diagram shown in Fig~\ref{fig:1}b is written in terms of quark-antiquark-gluon
operators, Eq.~(\ref{Tmn2}), and does not contain twist-two terms.
The leading twist-three contributions
to this diagram are due to quasipartonic operators of the type $\bar\psi_+f_{++}\psi_+$
which are irrelevant for the present study, as we repeatedly stressed above.
The goal is, therefore, to isolate the
subleading twist-four contributions. The expression in Eq.~(\ref{Tmn2}) contains eight
terms which are all similar so that it is enough to work out twist separation on one example.

Expanding the nonlocal operator in the Taylor series
\begin{equation}\label{FTexp}
\bar\psi_{\dot\alpha}(z_1x)
 f_{\gamma\alpha}(wx)\psi_\beta(z_2x)=\sum_{n=0}^\infty \frac1{2^n n!}
 \left[\prod_{i=1}^n\bar x^{\dot\alpha_i\alpha_i}\partial_{\alpha_i\dot\alpha_i}\right]\bar\psi_{\dot\alpha}(z_1y)
 f_{\gamma\alpha}(wy)\psi_\beta(z_2y)\Big|_{y=0}
\end{equation}
the task is reduced to picking up the twist-four part of the local operators
\begin{equation}\label{operatorT}
T_{\alpha\beta\gamma\alpha_1\ldots\alpha_n,\dot\alpha_1\ldots\alpha_n}=
\left[\prod_{i=1}^n\bar \partial_{\alpha_i\dot\alpha_i}\right]\bar\psi_{\dot\alpha}(z_1y)
 f_{\gamma\alpha}(wy)\psi_\beta(z_2y)\Big|_{y=0}\,.
\end{equation}
By construction, $T_{\alpha\beta\gamma\alpha_1\ldots\alpha_n,\dot\alpha_1\ldots\alpha_n}$ is symmetric
in $\alpha\leftrightarrow\gamma$ and under the interchange of the pairs of indices
$(\alpha_i\dot\alpha_i)\leftrightarrow(\alpha_j\dot\alpha_j)$.
Hence there exist three apriory inequivalent possibilities to contract one pair of indices
and symmetrize over the other ones:%
\footnote{The $\epsilon^{\alpha\alpha_1}$ contraction can be rewritten in terms of $P_1$, $P_2$
because $\epsilon^{\alpha\alpha_1}\lambda^{\beta} =
\epsilon^{\alpha\beta}\lambda^{\alpha_1} +\epsilon^{\beta\alpha_1}\lambda^\alpha$}
\begin{eqnarray}
P_1\cdot T &=&\epsilon^{\beta\alpha}\lambda^{\gamma}\bar\lambda^{\dot\alpha}
\prod_{k=1}^n\lambda^{\color{red}\alpha_k}\bar\lambda^{\color{red}\dot\alpha_k}
T_{\alpha\beta\gamma\alpha_1\ldots\alpha_n,\dot\alpha_1\ldots\alpha_n}\,,
\notag\\
P_2\cdot T &=&
\epsilon^{\beta\alpha_1}\lambda^{\gamma}\lambda^{\alpha}\bar\lambda^{\dot\alpha}\bar\lambda^{\dot\alpha_1}
\prod_{k=2}^n\lambda^{\color{red}\alpha_k}\bar\lambda^{\color{red}\dot\alpha_k}
T_{\alpha\beta\gamma\alpha_1\ldots\alpha_n,\dot\alpha_1\ldots\alpha_n}\,,
\notag\\
P_3\cdot T &=& \epsilon^{\dot\alpha\dot\alpha_1}
  \lambda^{\gamma}\lambda^{\alpha}{\color{red}\lambda^{\beta}\lambda^{\alpha_1}}
\prod_{k=2}^n\lambda^{\color{red}\alpha_k}\bar\lambda^{\color{red}\dot\alpha_k}
T_{\alpha\beta\gamma\alpha_1\ldots\alpha_n,\dot\alpha_1\ldots\alpha_n}\,.
\end{eqnarray}
Obviously, the contraction of two ``undotted'' indices as in $P_1$, $P_2$ and the
contraction of two ``dotted'' indices, $P_3$, give rise to tensors with
different Lorentz spin. It is easy to check that $P_3$ results in operators that
do not contribute to $(\partial\mathcal{O})$ and can be omitted.
Hence only the first two projections, $P_1$ and $P_2$ are relevant.

The rest of the calculation is very similar to the separation of higher-twist contributions
to the quark-antiquark operator.
Let
\begin{eqnarray}
\Pi_1&=&\left(\epsilon_{\beta\alpha}\partial_\gamma+\epsilon_{\beta\gamma}\partial_\alpha\right)\bar\partial_{\dot\alpha}
\prod_{j=1}^n\partial_{\alpha_j}\bar\partial_{\dot\alpha_j}\,,
\notag\\
\Pi_2&=&\sum_{j=1}^n\epsilon_{\beta\alpha_j}\partial_\alpha\partial_\gamma\bar\partial_{\dot\alpha}
\bar\partial_{\dot\alpha_j}\prod_{i\neq j}^n\partial_{\alpha_i}\bar\partial_{\dot\alpha_i}\,.
\end{eqnarray}
Note that the operators $\Pi_{1,2}$ have the same symmetry properties under permutation of indices as $T$.
Suppressing the open indices,
the twist-four part of the tensor $T_{\alpha\beta\gamma\alpha_1\ldots\alpha_n,\dot\alpha_1\ldots\alpha_n}$ can be written as
(cf. Eq.~(\ref{twist4O}))
\begin{equation}
   T^{(t=4)} =
\frac{1}{(n+3){[(n+1)!]^2}}\Big(\Pi_1,\Pi_2\Big)
\begin{pmatrix} n+1& -n \\ -2 & 3 \end{pmatrix}
\begin{pmatrix}P_1\cdot T\\ P_2\cdot T\end{pmatrix}.
\label{FT4}
\end{equation}
It is easy to check that $P_i\cdot[T-T^{(t=4)}]=0$ i.e. $T^{(t-4)}$ is indeed the twist-four part of the
tensor $T$.

Inserting~(\ref{FT4}) in~(\ref{FTexp}) and taking into account the identities
\begin{eqnarray}
{\bar x_{\dot\beta}\,\!}^\gamma \prod_{i=1}^n \bar x^{\dot\alpha_i\alpha_i}\Pi_1&=&
\frac{1}{2(n+1)}
(\epsilon_{\beta\alpha}(\bar x\partial)_{\dot\beta\dot\alpha}-\bar x_{\dot\beta\beta}\partial_{\alpha\dot\alpha})
(\bar\partial\bar x\partial)^{n+1}\,,
\notag\\
{\bar x_{\dot\beta}\,\!}^\gamma \prod_{i=1}^n \bar x^{\dot\alpha_i\alpha_i}\Pi_2&=&
-\frac1{2(n+1)}\left(x_{\beta\dot\beta}(x\partial)\partial_{\alpha\dot\alpha}
-\frac12x^2\partial_{\beta\dot\beta}\partial_{\alpha\dot\alpha}\right)(\bar\partial\bar x\partial)^{n+1}
\end{eqnarray}
one finds
\begin{eqnarray}
{\bar x_{\dot\beta}\,\! }^{\gamma}[\bar\psi_{\dot\alpha}(z_1x)
 f_{\gamma\alpha}(wx)\psi_\beta(z_2x)]^{t=4}
 &=&
-\frac12x_{\alpha\dot\beta}\partial_{\beta\dot\alpha}
\Pi(x,\lambda)\int_0^1 du\,
 \mathcal{F}(uz_1,uw,u z_2)
\notag\\
&+&
\frac18\Big(x_{\alpha\dot\beta}\partial_{\beta\dot\alpha}+x_{\beta\dot\beta}\partial_{\alpha\dot\alpha}\Big)\Pi(x,\lambda)
\int_0^1 du\,(1-u^2)\,
\notag \\
&\times&\Big\{z_1 u \mathcal{D}(z_1u,wu,z_2u)+
 2(w\partial_w+2)\mathcal{F}(uz_1,uw,u z_2)\Big\}\,.
\nonumber\\
\end{eqnarray}
Similarly, for the operator that involves the anti-selfdual gluon field one obtains
\begin{eqnarray}
{ x_{\alpha}}^{\dot\gamma}[\bar\psi_{\dot\alpha}(z_1x)
 \bar f_{\dot\gamma\dot\beta}(wx)\psi_\beta(z_2x)]^{t=4}
 &=&
-\frac12x_{\alpha\dot\beta}\partial_{\beta\dot\alpha}\Pi(x,\lambda)\int_0^1 du\,
\bar{\mathcal{F}}(uz_1,uw,u z_2)
\notag\\
&+&
\frac18\Big(x_{\alpha\dot\beta}\partial_{\beta\dot\alpha}+x_{\alpha\dot\alpha}\partial_{\beta\dot\beta}\Big)
\Pi(x,\lambda)
\int_0^1 du\,(1-u^2)\,
\notag \\
&\times&\Big\{z_2 u \bar{\mathcal{D}}(z_1u,wu,z_2u)+
 2(w\partial_w+2)\bar{\mathcal{F}}(uz_1,uw,u z_2)\Big\}\,.
\nonumber\\
\end{eqnarray}
The operators $\mathcal{F},\bar{\mathcal{F}},{\mathcal{D}},\bar{\mathcal{D}}$ are defined in
Eq.~(\ref{DQdef}).

\section{T-product of electromagnetic currents (II): Kinematic twist-four}


Adding together the contributions of the two diagrams in Fig.~\ref{fig:1} we can write the twist-four
contribution to the $T$-product of two electromagnetic currents in the following form:
\begin{eqnarray}
T_{\alpha\beta\dot\alpha\dot\beta}^{(t=4)}(z_1,z_2)&=&-\frac{2}{\pi^2 x^4 z_{12}^3}
\Big\{x_{\alpha\dot\beta} x_{\beta\dot\alpha}\, \Big[\mathbb{A}(x;z_1,z_2)-\mathbb{A}(x;z_2,z_1)\Big]
\notag\\
&&{}+
x^2\Big[x_{\alpha\dot\beta}\partial_{\beta\dot\alpha}\mathbb{B}(x;z_1,z_2)-
x_{\beta\dot\alpha}\partial_{\alpha\dot\beta}\mathbb{B}(x;z_2,z_1)\Big]
\notag\\
&&{}
+x^2\Big[x_{\beta\dot\beta}\partial_{\alpha\dot\alpha}\mathbb{C}(x;z_1,z_2)-
x_{\alpha\dot\alpha}\partial_{\beta\dot\beta}\mathbb{C}(x;z_2,z_1)\Big]
\Big\} + \ldots\,,
\label{T-twist4}
\end{eqnarray}
where the ellipses stand for the contributions of the right-handed quarks and
quasipartonic (four-particle) operators. Note that this expression is explicitly symmetric
under the interchange of the pairs of spinor indices
$(\alpha,\dot\alpha)\leftrightarrow(\beta\dot\beta)$ and, simultaneously, the replacement
$z_1\leftrightarrow z_2$.

The invariant functions $\mathbb{A}$, $\mathbb{B}$ and  $\mathbb{C}$ are given by the leading-twist
projection of nonlocal quark-antiquark-gluon operators and can be written as
\begin{equation}
 \mathbb{A}(x;z_1,z_2)=\Pi(x,\lambda)A(n;z_1,z_2)
\label{A-offcone}
\end{equation}
(and similar for the other two), where $A(n;z_1,z_2)$, $B(n;z_1,z_2)$, $C(n;z_1,z_2)$ are light-ray operators,
i.e. they only involve the fields on the light-cone.
The $A$-term originates from the handbag diagram in Fig.~\ref{fig:1}a only,
the $C$-term is due to the contribution of the gluon emission from the hard propagator,
Fig.~\ref{fig:1}b, and the $B$-term receives both contributions.

In order to get explicit expressions, we have to collect the results of
Sec.~\ref{subsec:twist4} and Sec.~\ref{subsec:ge} and rewrite them
using operators from the conformal basis, Eq.~(\ref{QF-relation}).
This calculation is described in Appendix~\ref{app:ABC1}.
At this stage it proves to be advantageous to use the relations
in Sec.~\ref{subsec:reduction} to rewrite all contributions of
$Q_1$ and $Q_3$ operators in terms of $Q_2$.
Remarkably, all results can be written in terms of the light-ray
operators $R, \bar R$ that involve one and the same integral over the  light-cone
coordinate of the gluon field:
\begin{align}\label{PhibarPhi}
R(z_1,z_2)=\frac{ig}{(n\tilde n)}\int_{z_2}^{z_1}dw\, (w-z_2)\, Q_2(z_1,w,z_2)\,,
\notag\\
\bar R(z_1,z_2)=\frac{ig}{(n\tilde n)}\int_{z_2}^{z_1}dw\, (z_1-w)\, \bar Q_2(z_1,w,z_2)\,.
\end{align}
This property is only true for the sum of both contributions in Fig.~\ref{fig:1} but not
for each diagram separately.

We obtain:
\begin{eqnarray}\label{ABC1}
A(n;z_1,z_2)&=&\frac14\int_0^1du \,\biggl\{ u^2\ln u\,
z_1z_2\,\left[i\mathbf{P}^\mu \left[i\mathbf{P}_\mu, O_{++}\right]\right](z_1u,z_2u)
\notag\\
&&{}+\left(z_2\partial_{z_2}-\frac{z_1}{z_{12}}-\ln u\, z_2\partial_{z_2}^2 z_{12}\right) R(uz_1,uz_2)
\notag\\&&{}
-\left(z_1\partial_{z_1}-\frac{z_2}{z_{21}}-\ln u\, z_1\partial_{z_1}^2 z_{21}\right) \bar R(uz_1,uz_2)
\biggr\}\,,
\nonumber\\
B(n;z_1,z_2)&=&\frac18\int_0^1\frac{du}{u^2} \,\biggl\{u^2(1\!-\!u^2\!+\!u^2\ln u)\,
z_1z_2\,\left[i\mathbf{P}^\mu \left[i\mathbf{P}_\mu, O_{++}\right]\right](z_1u,z_2u)
\notag\\
&&{}-\left[
(1-u^2)\left(z_2\partial_{z_2}-\frac{z_1}{z_{12}}\right)+(1\!-\!u^2\!+\!u^2\ln u)\,
z_2\partial_{z_2}^2 z_{12}\right] R(uz_1,uz_2)
\notag\\
&&{}
+\left[
(1-u^2)\left(z_1\partial_{z_1}-\frac{z_2}{z_{21}}\right)+(1\!-\!u^2\!+\!u^2\ln u)\,
z_1\partial_{z_1}^2 z_{21}\right] \bar R(uz_1,uz_2)
\biggl\},
\nonumber\\
%
C(n;z_1,z_2)&=&-\frac{1}{8}\int_0^1\frac{du}{u^2}\left[R(uz_1,uz_2)+\bar R(uz_2,uz_1)\right]\,.
\end{eqnarray}
The last step is to obtain a ``kinematic projection'' for the operators $R(z_1,z_2)$,
$\bar R(z_1,z_2)$.
Since the contributions of $Q_1$ and $Q_3$ operators
are no more present, we only need to use
the expansion~(\ref{eq:victory+coef}) for $Q_2$ which involves the functions
\begin{equation}
 {\Psi}^{(2)}_{Nk} (z_1,z_2,z_3) = \big(S_+^{(1,\frac32,\frac12)}\big)^k\Psi^{(2)}_{N} (z_1,z_2,z_3)\,.
\end{equation}
We remind that $S_+^{(1,\frac32,\frac12)}$ is the three-particle step-up operator for conformal
spins $j_1=1, j_2=3/2, j_3 = 1/2$. Explicit expression for
 $\Psi^{(2)}_{N} (z_1,z_2,z_3)$ is given in Eq.~(\ref{Psi2}).
The possibility to represent the result in a simple form is based on the following
identity (see Appendix~\ref{app:Id}):
\begin{align}\label{SS20}
\int_{z_2}^{z_1}\!dw\, (w-z_2) \big(S_+^{(1,\frac32,\frac12)}\big)^k\Psi_{N}^{(2)}(z_1,w,z_2)=
\big(S_+^{(1,0)}\big)^k\int_{z_2}^{z_1}\!dw\, (w-z_2) \Psi_{N}^{(2)}(z_1,w,z_2)\,,
\end{align}
where $S_+^{(1,\frac32,\frac12)}
=z_1^2\partial_{z_1}+2z_1+w^2\partial_w+3w+z_2^2\partial_{z_2}+z_2$  and
$S_{+}^{(1,0)}=z_1^2\partial_{z_1}+2z_1+z_2^2\partial_{z_2}$.
In other words, a simplification arises because
the quark-antiquark-gluon operator gets integrated over the light-cone
coordinate of the gluon field in the expression for the T-product of two
electromagnetic currents with a particular weight, Eq.~(\ref{PhibarPhi}).

Making use of the explicit expression for $\Psi_N^{(2)}$, Eq.~(\ref{Psi2}), we get
\begin{equation}
\int_{z_2}^{z_1}dw\, (w-z_2) \Psi_{N}^{(2)}(z_1,w,z_2)=r_N \,(z_1-z_2)^{N+1}\,,
\label{eq:rN}
\end{equation}
where
\begin{align}
r_N=&-\frac12(N+2)\big[\psi(N+3)+\psi(N+1)-\psi(3)-\psi(1)\big]\,.
\end{align}
Substituting this expression in Eq.~(\ref{eq:victory+coef}) one obtains
(cf. Eq.~(\ref{ONk}))
\begin{eqnarray}\label{phi-ex}
R(z_1,z_2)&=&\sum_{N=1}^\infty\sum_{k=0}^\infty\frac{\varkappa_N p_{Nk}(N+1)^2}{\rho_N||\Psi_N||^2}
\big(S_+^{(1,0)}\big)^k z_{12}^{N+1}\, \partial_+^k (\partial \mathcal{O})_N+\ldots
\notag\\
&=&
-\sum_{N=1}^\infty\sum_{k=0}^\infty\frac{\omega_{Nk}}{N+2}
\big(S_+^{(1,0)}\big)^k z_{12}^{N+1}\, \partial_+^k (\partial \mathcal{O})_N+\ldots\,,
\end{eqnarray}
and similarly
\begin{align}\label{barphi-ex}
\bar R(z_1,z_2)=-\sum_{N=1}^\infty\sum_{k=0}^\infty\frac{\omega_{Nk}}{N+2}
\big(S_+^{(0,1)}\big)^k z_{12}^{N+1}\, \partial_+^k (\partial \mathcal{O})_N+\ldots\,,
\end{align}
where the ellipses stand for the ``dynamical'' contributions of ``genuine''
quark-antiquark-gluon operators that we will omit hereafter.
The coefficients  $\omega_{Nk}$ are defined in Eq.~(\ref{omegaNk}).

It remains to substitute the expressions~(\ref{phi-ex}), (\ref{barphi-ex}) into
Eqs.~(\ref{ABC1}) and perform the integration over $u$, which becomes trivial.

Let
\begin{equation}
    \Psi^{(j_1,j_2)}_{Nk}(z_1,z_2) = \big(S_+^{(j_1,j_2)}\big)^k z_{12}^N\,.
\label{psiz1z2}
\end{equation}
Using the identities
\begin{align}
{S}_{0,\pm}^{(j_1,j_2)} z_{12}=z_{12}{S}_{0,\pm}^{(j_1+\frac12,j_2+\frac12)}\,,&&
\partial^2_z {S}_{0,\pm}^{(j=-\frac12)}={S}_{0,\pm}^{(j=\frac32)}\partial_z^2\,
\end{align}
the final results can be written in terms of two functions,
$\Psi^{(1,1)}_{Nk}(z_1,z_2)$ and $\Psi^{(\frac32,\frac12)}_{Nk}(z_1,z_2)$,
where the first one is already familiar from the expansion of twist-two light-ray operators,
$\Psi^{(1,1)}_{Nk}(z_1,z_2)\equiv \Psi^{(t=2)}_{Nk}(z_1,z_2)$,  cf.~Eq.~(\ref{ONk}). We obtain
\begin{eqnarray}\label{ABC2}
\hspace*{-1.5mm}\Delta A(n;z_1,z_2)&\equiv&A(n,z_1,z_2)-A(n,z_2,z_1)
\notag\\
&=&\frac14 z_1 z_2 \int_0^1du \, u^2\ln u\,
\left[i\mathbf{P}^\mu \left[i\mathbf{P}_\mu, O_{++}(z_1u,z_2u)-O_{++}(z_2u,z_1u)\right]\right]
\notag\\
&&{}+\frac12\sum_{N,\text{odd}}\sum_{k=0}^\infty \frac{\omega_{Nk}}{(N+2)(N+k+2)}\biggl\{
z_1\Psi^{(\frac32,\frac12)}_{Nk}(z_1,z_2)-z_2\Psi^{(\frac32,\frac12)}_{Nk}(z_2,z_1)
\notag\\
&&{}+\frac{N+1}{N+k+2} 2z_1z_2(\partial_{z_1}+\partial_{z_2})\Psi^{(1,1)}_{Nk}(z_1,z_2)
\biggr\}\,\partial_+^k\left(\partial \mathcal{O}\right)_N\,,
\nonumber\\
B(n;z_1,z_2)&=&\frac18 z_1 z_2\int_0^1du \,(1-u^2+u^2\ln u)\,
 \left[i\mathbf{P}^\mu \left[i\mathbf{P}_\mu, O_{++}\right]\right](z_1u,z_2u)
\notag\\
&&{}\hspace*{-3mm}-\frac14\sum_{N,\text{odd}}\sum_{k=0}^\infty \frac{\omega_{Nk}}{(N\!+\!2)(N\!+\!k)(N\!+\!k\!+\!2)}
\biggl\{
z_1\Psi^{(\frac32,\frac12)}_{Nk}(z_1,z_2)-z_2\Psi^{(\frac32,\frac12)}_{Nk}(z_2,z_1)
\notag\\
&&{}\hspace*{-3mm}-\frac{2(N\!+\!1)}{N\!+\!k\!+\!2}
\Big[z_1\!+\!z_2\!+\!\frac14(N\!+\!k\!+\!4)z_{12}(z_1\partial_{z_1}\!-\!z_2\partial_{z_2})\Big]\Psi^{(1,1)}_{Nk}(z_1,z_2)
\!\biggr\}\partial_+^k\!\left(\partial \mathcal{O}\right)_N,
\nonumber\\
C(n,z_1,z_2)&=&\frac14 z_{12}\sum_{N,\text{odd}}\sum_{k=0}^\infty \frac{\omega_{Nk}}{(N+2)(N+k)}\Psi^{(\frac32,\frac12)}_{Nk}(z_1,z_2)
\,\partial_+^k\left(\partial \mathcal{O}\right)_N\,,
\end{eqnarray}
which is the final result. The functions $\Psi^{(1,1)}_{Nk}(z_1,z_2)$ and $\Psi^{(\frac32,\frac12)}_{Nk}(z_1,z_2)$
can be written more explicitly with the help of the integral representation in Eq.~(\ref{eq:itrafo}).
In this way the results can be brought to the form similar to Eq.~(\ref{Oconf-ex}), cf.~\cite{Braun:2011zr}.

\subsection{Light-ray operator representation for $R,\bar R$}

Kinematic projection for the operators $R(z_1,z_2)$ (\ref{phi-ex}) and $\bar R(z_1,z_2)$
(\ref{barphi-ex}) and as a consequence the results in Eq.~(\ref{ABC2})
are written as Wilson OPE in contributions of local operators of increasing spin and dimension.
In applications to hard exclusive reactions a nonlocal representation in terms of light-ray
operators is often preferable, as it allows one to calculate physical amplitudes directly
in terms of parton distributions rather than their moments. For this purpose we need to find a
light-ray representation for $R,\bar R$ operators which we can rewrite as
\begin{eqnarray}
     R(z_1,z_2)&=&-z_{12}\sum_{N=1}^\infty\sum_{k=0}^\infty\frac{\omega_{Nk}}{N+2}
\Psi^{(\frac32,\frac12)}_{Nk}(z_1,z_2)\, \partial_+^k (\partial \mathcal{O})_N,
\nonumber\\
\bar R(z_1,z_2)&=&-z_{12}\sum_{N=1}^\infty\sum_{k=0}^\infty\frac{\omega_{Nk}}{N+2}
\Psi^{(\frac12,\frac32)}_{Nk}(z_1,z_2) \, \partial_+^k (\partial \mathcal{O})_N\,.
\label{RbarR1}
\end{eqnarray}
%
 As the first step, using
$[i\bar{\mathbf{P}}^{\dot\alpha\alpha},\partial_\alpha\bar\partial_{\dot\alpha}\,\mathcal{O}_{N}]
= (N+1)^2 (\partial\mathcal{O})_{N} $ and applying $[i\mathbf P_+,\,\cdot\,]^k$ to both
sides one obtains after a short calculation
\begin{equation}
[i\bar{\mathbf{P}}^{\dot\alpha\alpha},\partial_\alpha\bar\partial_{\dot\alpha}\,
\partial_+^k\, \mathcal{O}_{N}] = (N+1)^2 \partial_+^k(\partial\mathcal{O})_{N}
+ {\color{red}\frac14} k(2N+k+3) [i\bar{\mathbf{P}}^{\dot\alpha\alpha},
[i{\mathbf{P}}_{\alpha\dot\alpha},\partial_+^{k-1} \mathcal{O}_{N}]]\,.
\label{DNO1}
\end{equation}
Multiplying both sides by $\omega_{Nk}\Psi^{(1,1)}_{Nk}(z_1,z_2)$ and summing over $N,k$ this becomes
\begin{eqnarray}\label{DNO}
\lefteqn{\hspace*{-1cm}
\sum_{Nk}\omega_{Nk}(N+1)^2\Psi_{Nk}(z_1,z_2)\partial_+^k(\partial\mathcal{O})_{N}=}
\nonumber\\&=&
[i\bar{\mathbf{P}}^{\dot\alpha\alpha},\partial_\alpha\bar\partial_{\dot\alpha} O_{++}(z_1,z_2)]\,
- {\color{red}\frac14} S_+^{(1,1)} [i\bar{\mathbf{P}}^{\dot\alpha\alpha},[i{\mathbf{P}}_{\alpha\dot\alpha},, O_{++}(z_1,z_2)]]\,,
\end{eqnarray}
where we took into account that $k(2N+k+3)\omega_{Nk} = \omega_{Nk-1}$
and used Eq.~(\ref{ONk}) to rewrite the sum over $\partial_+^k\, \mathcal{O}_{N}$
in terms of the light-ray operator $O_{++}(z_1,z_2)$.

The expression on the l.h.s. of Eq.~(\ref{DNO}) can be transformed into the sum for the
$R(z_1,z_2)$ operator, Eq.~(\ref{RbarR1}), using the following identity:
\begin{equation}
\int_0^1d\!\alpha\int_0^{\bar \alpha}\!d\beta\, \frac{\beta}{\bar \beta}\Psi^{(1,1)}_{Nk}(z_{12}^\alpha,z_{21}^\beta)=
\frac{1}{(N+1)^2(N+2)} \Psi^{(\frac32,\frac12)}_{Nk}(z_1,z_2)\,.
\label{intertwine2}
\end{equation}
The simplest way to verify this relation is to observe that
the integral operator
\begin{align}
[K\varphi](z_1,z_2)=\int_0^1d\alpha\int_0^{\bar \alpha}d\beta\, \frac{\beta}{\bar \beta}\,\varphi(z_{12}^\alpha,z_{21}^\beta)\,.
\end{align}
intertwines the representations $T^{j_1=1}\otimes T^{j_2=1}$ and $T^{j_1=3/2}\otimes T^{j_2=1/2}$, that is%
\footnote{A general approach for the construction of two-particle intertwining operators is described
in Appendix~B in Ref.~\cite{Braun:2009vc}.}
\begin{align}
K \,S_{+}^{(11)} = S_{+}^{(\frac32\frac12)}\, K\,.
\end{align}
As a consequence, since $\Psi^{(11)}_{Nk}(z_1,z_2)=(S^{(11)}_+)^k z_{12}^N$ one derives
\begin{equation}
 \int_0^1\!\!d\alpha\!\int_0^{\bar \alpha}\!\!\!\!d\beta\,
\frac{\beta}{\bar \beta} \Psi^{(11)}_{Nk}(z_{12}^\alpha,z_{21}^\beta)
=
 \big(S_{+}^{(\frac32,\frac12)}\big)^k\!\!\int_0^1\!d\alpha\!\!\int_0^{\bar \alpha}\!\!\!\!d\beta\,
  \frac{\beta}{\bar\beta} (z_{12}^\alpha-z_{21}^\beta)^N
= \frac{\big(S_{+}^{(\frac32,\frac12)}\big)^k  z_{12}^N}{(N+1)^2(N+2)},
\end{equation}
which is nothing but Eq.~(\ref{intertwine2}).

Applying this operator to both sides of Eq.~(\ref{DNO}) we obtain
\begin{eqnarray}
R(z_1,z_2)&=&
-z_{12}\int_0^1d\alpha\int_0^{\bar \alpha}d\beta\, \frac{\beta}{\bar \beta}
\biggr\{
[i\bar{\mathbf{P}}^{\dot\gamma\gamma},\partial_\gamma\bar\partial_{\dot\gamma}
O_{++}(z_{12}^\alpha,z_{21}^\beta)]\,
\notag\\
&&{}\hspace*{3.5cm}
- {\color{red}\frac14} S_+^{(\frac32\frac12)} [i\bar{\mathbf{P}}^{\dot\gamma\gamma},
[i{\mathbf{P}}_{\gamma\dot\gamma}, O_{++}(z_{12}^\alpha,z_{21}^\beta)]]\biggl\}\,
\label{eq:Rlightray}
\end{eqnarray}
and, similarly
\begin{eqnarray}
\bar R(z_1,z_2)&=&
-z_{12}\int_0^1d\beta\int_0^{\bar \beta}d\alpha\, \frac{\alpha}{\bar \alpha}
\biggr\{
[i\bar{\mathbf{P}}^{\dot\gamma\gamma},\partial_\gamma\bar\partial_{\dot\gamma}
O_{++}(z_{12}^\alpha,z_{21}^\beta))]\,
\notag\\
&&{}\hspace*{3.5cm}
- {\color{red}\frac14} S_+^{(\frac12\frac32)} [i\bar{\mathbf{P}}^{\dot\gamma\gamma},
[i{\mathbf{P}}_{\gamma\dot\gamma}, O_{++}(z_{12}^\alpha,z_{21}^\beta))]]\biggl\}\,.
\label{eq:barRlightray}
\end{eqnarray}

The relations in Eqs.~(\ref{eq:Rlightray}), (\ref{eq:Rlightray})
are between the light-ray operators, i.e. they involve fields ``living''
on the light-ray $n^2=0$. The OPE of the product of two electromagnetic
currents (\ref{T-twist4}) involves the leading-twist projector
$\Pi$ acting on the light-ray operators (\ref{A-offcone}).
For the calculation of matrix elements it is convenient to rewrite
\begin{equation}
 \mathcal{R}(x;z_1,z_2)=\Pi(x,\lambda)R(z_1,z_2)\,,\qquad \bar{\mathcal{R}}(x;z_1,z_2)=\Pi(x,\lambda)\bar R(z_1,z_2)
\end{equation}
directly in terms of $\mathcal{O}^{t=2}_{++}(z_1x,z_2x) = \Pi(x,\lambda){O}_{++}(z_1n,z_2n)$
(\ref{def:O++})
i.e. in terms of the leading-twist part of the nonlocal quark-antiquark operator $\bar\psi_+(z_1x)\psi_+(z_2x)$
off the light-cone.
Using
\begin{equation}
  \Pi(x,\lambda)\partial_\gamma\bar\partial_{\dot\gamma} O_{++}(z_1,z_2) =
\frac12\partial_{\gamma\dot\gamma}\big(z_1\partial_{z_1}+z_2\partial_{z_2}+1\big) \mathcal{O}^{t=2}_{++}(z_1x,z_2x)
\end{equation}
we obtain
\begin{eqnarray}
\mathcal{R}(x;z_1,z_2)&=&
z_{12}
\int_0^1d\alpha\int_0^{\bar \alpha}d\beta \frac{\beta}{\bar \beta}
\biggl\{{\color{red}\frac12}S_+^{(\frac32\frac12)}
[i{\mathbf{P}}^{\mu},[i{\mathbf{P}}_{\mu}, \mathcal{O}^{t=2}_{++}(z_{12}^\alpha x,z_{21}^\beta x)]]
\notag\\&&{}\hspace*{3.2cm}
-(S_0^{(\frac32\frac12)}-1) \Big[i{\mathbf{P}}^{\mu},\frac{\partial}{\partial x^\mu}
\mathcal{O}^{t=2}_{++}(z_{12}^\alpha x,z_{21}^\beta x)\Big]
\biggr\},
\notag\\
\bar{\mathcal{R}}(x;z_1,z_2)&=&
z_{12}
\int_0^1d\beta\int_0^{\bar \beta}d\alpha \frac{\alpha}{\bar \alpha}
\biggl\{{\color{red}\frac12}S_+^{(\frac12\frac32)}
[i{\mathbf{P}}^{\mu},[i{\mathbf{P}}_{\mu}, \mathcal{O}^{t=2}_{++}(z_{12}^\alpha x,z_{21}^\beta x)]]
\notag\\&&{}\hspace*{3.2cm}
-(S_0^{(\frac12\frac32)}-1) \Big[i{\mathbf{P}}^{\mu},\frac{\partial}{\partial x^\mu}
\mathcal{O}^{t=2}_{++}(z_{12}^\alpha x,z_{21}^\beta x)\Big]\biggr\}.
\end{eqnarray}
Substituting these expressions in Eqs.~(\ref{ABC1}), (\ref{A-offcone})
gives the kinematic part of the twist-four contribution to the T-product of two
electromagnetic currents (\ref{T-twist4}) in terms of the leading-twist operator
$\mathcal{O}^{t=2}_{++}(z_1x,z_2x)$.
To save space, let us introduce the notations
\begin{eqnarray}
\mathcal{O}_1(z_1,z_2) &=&{\color{red}\frac12}
 [i{\mathbf{P}}^{\mu},[i{\mathbf{P}}_{\mu}, \mathcal{O}^{t=2}_{++}(z_1 x,z_2 x)]]\,,
\nonumber\\
\mathcal{O}_2(z_1,z_2) &=&
\Big[i{\mathbf{P}}^{\mu},\frac{\partial}{\partial x^\mu}
\mathcal{O}^{t=2}_{++}(z_1 x,z_2 x)\Big]\,,
\label{calO12}
\end{eqnarray}
and
\begin{equation}
  \Delta \mathcal{O}_1(z_1,z_2) = \mathcal{O}_1(z_1,z_2) - \mathcal{O}_1(z_2,z_1)\,,
\hspace*{1cm}
  \Delta \mathcal{O}_2(z_1,z_2) = \mathcal{O}_2(z_1,z_2) - \mathcal{O}_2(z_2,z_1)\,.
\label{DeltacalO12}
\end{equation}
A short calculation gives:
\begin{eqnarray}\label{Af}
\Delta\mathbb{A}&=&
\frac14\!\int_0^1\!du\, \bigg\{{\color{red}2}u^2 \ln u\, z_1 z_2\, \Delta\mathcal{O}_1 (uz_{1},uz_{2})
- u^2
\int_0^1\!\!\!d\alpha\!\!\int_0^{\bar \alpha}\!\!\!d\beta \,
z^2_{12}\Delta\mathcal{O}_1 (uz_{12}^\alpha ,uz_{21}^\beta )
\nonumber\\ &&{}\hspace*{13mm}
-u^2\!
\int_0^1\!\!\!d\alpha\!\!\int_0^{\bar \alpha}\!\frac{d\beta}{\bar\beta}
\Big[
 z_{21}^\beta S_+^{(\frac32\frac12)} \Delta\mathcal{O}_1 (uz_{12}^\alpha ,uz_{21}^\beta )
+
 z_{12}^\beta S_+^{(\frac12\frac32)} \Delta\mathcal{O}_1 (uz_{12}^\beta ,uz_{21}^\alpha )
\Big]
\nonumber\\ &&{}\hspace*{13mm}
- u^2 \ln u\!
\int_0^1\!\!d\alpha\, \Big[
z_2 S_+^{(\frac12\frac32)}\Delta\mathcal{O}_1 (uz_{12}^\alpha ,uz_{2} )
+
z_1 S_+^{(\frac32\frac12)}\Delta\mathcal{O}_1 (uz_{1} , uz_{21}^\alpha )
\Big]
\nonumber\\&&{}\hspace*{13mm}
- u[1-\delta(\bar u)]
\int_0^1\!\!d\alpha\!\int_0^{\bar \alpha}\!\frac{d\beta}{\bar\beta}
\Big[
z_{21}^\beta\Delta\mathcal{O}_2 (uz_{12}^\alpha ,uz_{21}^\beta )
+
z_{12}^\beta\Delta\mathcal{O}_2 (uz_{12}^\beta ,uz_{21}^\alpha )
\Big]
\nonumber\\&&{}\hspace*{13mm}
- u(1+\ln u)\,
\int_0^1\!\!d\alpha\,
\Big[
z_2 \Delta\mathcal{O}_2 (uz_{12}^\alpha ,uz_{2} )
+
z_1 \Delta\mathcal{O}_2 (uz_{1} ,uz_{21}^\alpha )
\Big]\biggr\},
\end{eqnarray}
\begin{eqnarray}\label{Bf}
\mathbb{B}&=&\frac18\!\int_0^1\!\!\!du \biggl\{{\color{red}2}(1\!-\!u^2\!+\!u^2\ln u)
z_1z_2\mathcal{O}_1(uz_1,uz_2)
+ (1\!-\!u^2)\!
\int_0^1\!\!\!d\alpha\!\int_0^{\bar \alpha}\!\!\!\!d\beta\,
z_{12}^2 \mathcal{O}_1(uz_{12}^\alpha ,uz_{21}^\beta )
\nonumber\\ &&{}\hspace*{13mm}
+ (1\!-\!u^2)
\int_0^1\!\!d\alpha\int_0^{\bar \alpha}\!\!\frac{d\beta}{\bar\beta}
\Big[
  z_{21}^\beta S_+^{(\frac32\frac12)} \mathcal{O}_1(uz_{12}^\alpha , uz_{21}^\beta )
+
  z_{12}^\beta S_+^{(\frac12\frac32)} \mathcal{O}_1(uz_{12}^\beta , uz_{21}^\alpha )
\Big]
\nonumber\\ &&{}\hspace*{13mm}
-(1\!-\!u^2\!+\!u^2\ln u)\!
\int_0^1\!\!d\alpha\,
\Big[
 z_2S_+^{(\frac12\frac32)}\mathcal{O}_1(uz_{12}^\alpha, uz_{2} )
+
 z_1S_+^{(\frac32\frac12)}\mathcal{O}_1(uz_{1}, uz_{21}^\alpha )
\Big]
\nonumber\\ &&{}\hspace*{13mm}
- \frac{1}{u} (1\!+\!u^2)
\int_0^1\!\!d\alpha\int_0^{\bar \alpha}\!\!\frac{d\beta}{\bar\beta}\,
\Big[
z_{21}^\beta \mathcal{O}_2(uz_{12}^\alpha ,uz_{21}^\beta )
+
z_{12}^\beta \mathcal{O}_2(uz_{12}^\beta ,uz_{21}^\alpha )
\Big]
\nonumber\\ &&{}\hspace*{13mm}
+\frac{1}{u}(1\!-\!u^2\ln u)
\int_0^1\!\!d\alpha\,
\Big[
z_2 \mathcal{O}_2(uz_{12}^\alpha ,uz_{2} )
+
z_1 \mathcal{O}_2(uz_{1} ,uz_{21}^\alpha )
\Big]\biggr\},
\end{eqnarray}
\begin{eqnarray}\label{Cf}
\mathbb{C}
&=&
-\frac{1}{8}z_{12}\int_0^1\!{du}\!
\int_0^1\!d\alpha\!\int_0^{\bar \alpha}\!\!d\beta\, \frac{\beta}{\bar \beta}
\biggl\{S_+^{(\frac32\frac12)}
\Delta\mathcal{O}_1(uz_{12}^\alpha ,uz_{21}^\beta )
-\Big(\delta(\bar u)+\frac1u\Big) \Delta\mathcal{O}_2(uz_{12}^\alpha ,uz_{21}^\beta )
\biggr\},
\nonumber\\
\end{eqnarray}
where
\begin{equation}
  \Delta\mathbb{A}(x;z_1,z_2) = \mathbb{A}(x;z_1,z_2) - \mathbb{A}(x;z_2,z_1)\,.
\end{equation}
To arrive at these expressions we used the following identities:
\begin{eqnarray}
  \partial_{z_2}z_{12} \int_0^1\!\!d\alpha\!\int_0^{\bar \alpha}\!\!\!d\beta \frac{\beta}{\bar \beta}
 \,\mathcal{O}(z_{12}^\alpha x,z_{21}^\beta x)
 &=&  - \int_0^1\!\!d\alpha\!\int_0^{\bar \alpha}\!\!\!d\beta\,
 \mathcal{O}(z_{12}^\alpha x,z_{21}^\beta x)
\nonumber\\
  \partial^2_{z_2}z^2_{12} \int_0^1\!\!d\alpha\!\int_0^{\bar \alpha}\!\!\!d\beta \frac{\beta}{\bar \beta}
 \,\mathcal{O}(z_{12}^\alpha x,z_{21}^\beta x)
 &=&  \phantom{-}\int_0^1\!\!d\alpha\, \mathcal{O}(z_{12}^\alpha x,z_2 x)
\end{eqnarray}
which can be verified by examining their properties under conformal transformations.

Our final expression for the $T$-product of electromagnetic currents to the twist-four
accuracy is given in the next Section.

\section{T-product of electromagnetic currents (III): Final expressions}

In most of the discussion so far we considered the contribution of left-handed quarks~$\psi$ only.
Adding the contribution of the right-handed quarks, $\bar\chi$, is straightforward and reduces to a redefinition
of the leading-twist operator $O_{++}$. From now on it will be implied that
\begin{eqnarray}
\mathcal{O}^{t=2}_{++}(z_1,z_2)&=&\Pi(x,\lambda)O_{++}(z_1,z_2)\,,
\end{eqnarray}
with
\begin{eqnarray}\label{newO++}
O_{++}(z_1,z_2)&=&\bar\psi_+(z_1n)\psi_+(z_2n)-\chi_+(z_2n)\bar\chi_+(z_1n)
\notag\\
&=&
\frac12\big[\bar q(z_1n)\slashed{n}(1-\gamma_5)q(z_2n)-
\bar q(z_2n)\slashed{n}(1+\gamma_5)q(z_1n)\big]
\notag\\
&=&
\frac12\big[O_V(z_1,z_2)-O_V(z_2,z_1)-O_A(z_1,z_2)-O_A(z_2,z_1)\big]\,.
\end{eqnarray}
$O_V$ and $O_A$ are the usual vector and axial-vector light-ray operators
\begin{align}
O_V(z_1,z_2)=\bar q(z_1n)\slashed{n}q(z_2n)\,,&&O_A(z_1,z_2)=\bar q(z_1n)\slashed{n}\gamma_5 q(z_2n)\,.
\end{align}
Note that $O_V$ always enters antisymmetrized and $O_A$ symmetrized over the quark positions.

Now we are in a position to write down the final result.
The $T$-product of two electromagnetic currents including all contributions of leading-twist
quark-antiquark operators and their total derivatives to twist-four accuracy can be cast in the form
\begin{eqnarray}\label{Tt4}
T_{\alpha\beta\dot\alpha\dot\beta}(z_1,z_2)&=&-\frac{2}{\pi^2 x^4 z_{12}^3}
\biggl\{x_{\alpha\dot\beta} \mathfrak{B}_{\beta\dot\alpha}(z_1,z_2)-
x_{\beta\dot\alpha} \mathfrak{B}_{\alpha\dot\beta}(z_2,z_1)
+x_{\alpha\dot\beta} x_{\beta\dot\alpha}\, \Delta\mathbb{A}(z_1,z_2)
\notag\\
&&{}\hspace*{1.7cm}
+x^2\Big[x_{\beta\dot\beta}\partial_{\alpha\dot\alpha}\mathbb{C}(z_1,z_2)-
x_{\alpha\dot\alpha}\partial_{\beta\dot\beta}\mathbb{C}(z_2,z_1)\Big] + \ldots
\biggr\}\,,
\end{eqnarray}
or, equivalently, in vector notation~\cite{Braun:2011zr}%
\footnote{The function $\mathbb{W}^\beta$ in (\ref{T4vec})
   corresponds to $\mathbb{A}^\beta$ in the notation of Ref.~\cite{Braun:2011zr}.}
\begin{align}\label{T4vec}
T_{\mu\nu} &= -\frac{1}{\pi^2x^4z_{12}^3}\Big\{
x^{\alpha}\Big[S_{\mu\alpha\nu\beta} \mathbb{V}^\beta
-i\epsilon_{\mu\alpha\nu\beta} \mathbb{W}^\beta
\Big]
+x^2\Big[(x_\mu\partial_\nu +x_\nu\partial_\mu) \mathbb{X}
+(x_\mu\partial_\nu-x_\nu\partial_\mu) \mathbb{Y}
\Big]
\Big\}\,,
\end{align}
where
\begin{eqnarray}
\mathbb{V}_\mu(z_1,z_2)&=&\phantom{-}\mathfrak{B}_\mu(z_1,z_2)-\mathfrak{B}_\mu(z_2,z_1)+
x_\mu\Delta \mathbb{A}(z_1,z_2)\,,
\notag\\
\mathbb{W}_\mu(z_1,z_2)&=& - \mathfrak{B}_\mu(z_1,z_2)- \mathfrak{B}_\mu(z_2,z_1)\,,
\notag\\
\mathbb{X}(z_1,z_2)&=&\phantom{-}\mathbb{C}(z_1,z_2)-\mathbb{C}(z_2,z_1)\,,
\notag\\
\mathbb{Y}(z_1,z_2)&=&-\mathbb{C}(z_1,z_2)-\mathbb{C}(z_2,z_1)\,,
\end{eqnarray}
$\mathfrak{B}_{\beta\dot\alpha}(z_1,z_2) = \sigma^\mu_{\beta\dot\alpha}\mathfrak{B}_\mu(z_1,z_2)$,
$S_{\mu\alpha\nu\beta}=g_{\mu\alpha}g_{\nu\beta}+g_{\nu\alpha}g_{\mu\beta}-g_{\mu\nu}g_{\alpha\beta}$,
and a totally antisymmetric tensor is defined such that $\epsilon_{0123}=1$.

The invariant functions $\Delta\mathbb{A}(z_1,z_2)$ and $\mathbb{C}(z_1,z_2)$ are given by the
expressions~(\ref{Af}) and (\ref{Cf}), respectively, with
the redefined operator $O_{++}$ as shown in Eq.~(\ref{newO++}). Both these contributions are twist-four.
In addition, the function $\mathfrak{B}_{\alpha\dot\alpha}(z_1,z_2)$ contains all twists starting from the leading one:
\begin{align}
\mathfrak{B}_{\alpha\dot\alpha}(z_1,z_2)=\mathfrak{B}^{t=2}_{\alpha\dot\alpha}(z_1,z_2)+
\mathfrak{B}_{\alpha\dot\alpha}^{t=3}(z_1,z_2)+\mathfrak{B}^{t=4}_{\alpha\dot\alpha}(z_1,z_2)+\ldots
\end{align}
Collecting all expressions, we obtain:
\begin{eqnarray}
\mathfrak{B}^{t=2}_{\alpha\dot\alpha}(z_1,z_2)
&=&\frac12\partial_{\alpha\dot\alpha}\int_0^1 du\,
\mathcal{O}_{++}^{t=2}(uz_1x,uz_2x)\,,
\notag\\[2mm]
\mathfrak{B}^{t=3}_{\alpha\dot\alpha}(z_1,z_2)
&=&
\!\frac14\!\!
\int_0^1\!\!\!udu\!\int_{z_2}^{z_1}\!\!\frac{dv}{z_{12}}
\biggl\{
\Big[{i\mathbf{P}}_{\!\mu},(x\bar\sigma^\mu\partial)_{\alpha\dot\alpha}z_1\mathcal{O}^{t=2}_{++}(z_1u,\! vu)
\!+\!
(\bar x\sigma^\mu\bar \partial)_{\dot\alpha\alpha}z_2\mathcal{O}^{t=2}_{++}(vu,\! z_2u)\Big]
\notag\\
&&+\frac12\ln u\,\partial_{\alpha\dot\alpha}x^2\partial_{\beta\dot\beta}\,
\Big[i\bar{\mathbf{P}}^{\dot\beta\beta}, z_1\mathcal{O}^{t=2}_{++}(z_1u, vu)+z_2\mathcal{O}^{t=2}_{++}(vu, z_2u)\Big]
\biggr\}\,,
\nonumber\\[2mm]
\mathfrak{B}^{t=4}_{\alpha\dot\alpha}(z_1,z_2)&=&x^2\partial_{\alpha\dot\alpha}\mathbb{B}(z_1,z_2)\,,
\end{eqnarray}
with $\mathbb{B}(z_1,z_2)$ given by Eq.~(\ref{Bf}).
We have checked (using the representation in Eq.~(\ref{ABC2}) for the twist-four functions
$\Delta\mathbb{A},\mathbb{B},\mathbb{C}$)
that this expression satisfies the Ward identities (\ref{Ward}) up to
twist-five terms.

As mentioned in Sec.~\ref{sec:T1}, translation invariance relation (\ref{translation}) is only recovered
in the sum of all twists as well. As a consequence, dependence of the results on the positions of the
electromagnetic currents, $z_1$ and $z_2$, is nontrivial. We have found that
the expressions for higher-twist contributions can be simplified considerably for a special choice,
when one of the electromagnetic currents is at the origin, e.g. $z_1=1$ and $z_2=0$.
In this case we obtain
\begin{eqnarray}\label{V10}
\mathfrak{B}^{t=2}_{\mu}(1,0)
&=&\frac12\partial_{\mu}\int_0^1 du\,
\mathcal{O}_{++}^{t=2}(u,0)\,,
\notag\\[2mm]
\mathfrak{B}^{t=3}_{\mu}(1,0)
&=&
\frac14
\int_0^1du\int_{0}^{u}dv
\Big(\Big[S_{\mu\rho\nu\sigma}+i\epsilon_{\mu\rho\nu\sigma}\Big]x^{\rho}\partial^\sigma+
\ln u\,\partial_{\mu}x^2\partial_\nu\Big)
\Big[{i\mathbf{P}}^{\nu},\mathcal{O}^{t=2}_{++}(u, v)\Big]\,,
\notag\\[2mm]
\mathfrak{B}^{t=4}_{\mu}(1,0)
&=&\frac18x^2\partial_{\mu} \int_0^1du\int_0^u dv\biggl\{
-\left[1+\frac{3\bar u}{u}+\frac{2}{v}\ln\left(1-\frac{\bar u}{u}\frac{v}{\bar v}\right)\right]
  \mathcal{O}_1(u,v)
 \notag\\
 &&{}\hspace*{1.5cm}
+\left[1+\frac{\bar u}{u}\left(3+\frac{\bar u}{u}+\frac{\bar v}{v}\right)+
 \frac{1+v^2}{v^2}\ln\left(1-\frac{\bar u}{u}\frac{v}{\bar v}\right)\right]\mathcal{O}_2(u,v)
\biggr\}\,
\end{eqnarray}
and%
\footnote{The results for $\mathfrak{B}_{\mu}(1,0)$ and $\mathfrak{B}_{\mu}(0,1)$ are related by hermitian conjugation,
           $\mathfrak{B}_\mu(z_1,z_2)=(\mathfrak{B}_\mu(z_2,z_1))^\dagger$, taking into account that
           $\mathcal{O}_{++}^{t=2}(u,v)=\left(\mathcal{O}_{++}^{t=2}(v,u)\right)^\dagger$ and
$\mathcal{O}_{1,2}(u,v)=\left(\mathcal{O}_{1,2}(v,u)\right)^\dagger$.}
\begin{eqnarray}\label{V01}
\mathfrak{B}^{t=2}_{\mu}(0,1)
&=&\frac12\partial_{\mu}\int_0^1 du\,
\mathcal{O}_{++}^{t=2}(0,u)\,,
\notag\\[2mm]
\mathfrak{B}^{t=3}_{\mu}(0,1)
&=&
\frac14
\int_0^1du\int_{0}^{u}dv
\Big(\Big[S_{\mu\rho\nu\sigma}-i\epsilon_{\mu\rho\nu\sigma}\Big]x^{\rho}\partial^\sigma+
\ln u\,\partial_{\mu}x^2\partial_\nu\Big)
\Big[{i\mathbf{P}}^{\nu},\mathcal{O}^{t=2}_{++}(v, u)\Big]\,,
\notag\\[2mm]
\mathfrak{B}^{t=4}_{\mu}(0,1)
&=&\frac18x^2\partial_{\mu} \int_0^1du\int_0^u dv\biggl\{
-\left[1+\frac{3\bar u}{u}+\frac{2}{v}\ln\left(1-\frac{\bar u}{u}\frac{v}{\bar v}\right)\right]
  \mathcal{O}_1(v,u)
 \notag\\
 &&{}\hspace*{1.5cm}
+\left[1+\frac{\bar u}{u}\left(3+\frac{\bar u}{u}+\frac{\bar v}{v}\right)+
 \frac{1+v^2}{v^2}\ln\left(1-\frac{\bar u}{u}\frac{v}{\bar v}\right)\right]\mathcal{O}_2(v,u)
\biggr\}\,.
\end{eqnarray}
Also
\begin{equation}\label{A10}
\Delta\mathbb{A}(1,0) =
\frac{1}4  \int_0^1du\int_0^u dv\biggl\{-\frac{v}{\bar v} \Delta \mathcal{O}_1(u,v)
+\left(\frac{v}{\bar v}+\ln\left(1-\frac{\bar u}{u}\frac{v}{\bar v}\right)\right)\Delta \mathcal{O}_2(u,v)
\biggr\}
\end{equation}
and finally
\begin{eqnarray}\label{C10}
\mathbb{C}(1,0)&=&\frac{1}{8} \int_0^1du\int_0^u dv\biggl\{-\Big[\frac{v}{\bar v}-\frac{2\bar u}{u}
-\frac{2}{v}\ln\left(1-\frac{\bar u}{u}\frac{v}{\bar v}\right)\Big]
  \Delta\mathcal{O}_1(u,v)
 \notag\\
 &&{}\hspace*{1.5cm}
+\Big[\frac{v}{\bar v}-\frac{\bar u}{u}
\Big( 2+\frac{\bar v}{v}+\frac12\frac{\bar u}{u}\Big)
 -\frac1{v^2}\ln\left(1-\frac{\bar u}{u}\frac{v}{\bar v}\right)\Big]
 \Delta\mathcal{O}_2(u,v)
\biggr\},
\notag\\[2mm]
\mathbb{C}(0,1)&=&\frac{1}{8} \int_0^1du\int_0^u dv\Biggl\{\frac{\bar u}{u}\,
  \Delta\mathcal{O}_1(v,u)-\frac{\bar u}{u}\left[1+\frac12\frac{\bar u}{u}\right]\Delta\mathcal{O}_2(v,u)
\biggr\},
\end{eqnarray}
where $\Delta \mathcal{O}_{1,2}(u,v) = \mathcal{O}_{1,2}(u,v) - \mathcal{O}_{1,2}(v,u)$.

Operator product expansion in terms of nonlocal light-ray operators $\mathcal{O}_{1,2}(v,u)$ has an
advantage that the matrix elements are given directly by parton distributions rather than
their moments. The above expressions still appear to be rather complicated, however.
A~possible drawback of this representation is also that
the operator $\mathcal{O}_2 = \Big[i{\mathbf{P}}^{\mu},\partial_\mu \mathcal{O}^{t=2}_{++}\Big]$
contains both terms of the type $\sim \partial^2 \mathcal{O}_{\mu_1\ldots\mu_n}$
and $\sim \partial^\mu \mathcal{O}_{\mu\mu_1\ldots\mu_n}$, cf.~(\ref{DNO1}), (\ref{DNO}), and one
may like to have them separated.

This can be achieved by using another representation which is usually referred to as conformal
OPE (see e.g. the review \cite{Braun:2003rp}), in terms of integrals over light-ray positions
of local conformal operators, similar to Eq.~(\ref{Oconf-ex}).
To this end we define the (axial)vector conformal operators as
\begin{align}
\mathcal{O}_{N}^{V(A)}(y)=&
(\partial_{z_1}\!+\!\partial_{z_2})^NC_N^{3/2}\left(
\frac{\partial_{z_1}\!-\!\partial_{z_2}}{\partial_{z_1}\!+\!\partial_{z_2}}\right)
{O}_{V(A)}(z_1x +y,z_2x+y)\Big|_{z_i=0}\,
\label{eq:On}
\end{align}
and
\begin{eqnarray}
(\partial{\mathcal{O}})^{V(A)}_N(y)&=&\frac1{N+1}\frac{\partial}{\partial x^\mu}
\bigl[i\mathbf{P}^\mu,\mathcal{O}^{V(A)}_N(y)\bigr]
=
\bigl[i\mathbf{P}^\mu,\mathcal{O}^{V(A)}_{\mu\mu_1\ldots\mu_N}(y)\bigr] x^{\mu_1}\ldots x^{\mu_N}\,.
\end{eqnarray}
The twist-two contributions can be written as
\begin{eqnarray}
\mathbb{V}_\mu^{t=2}&=&\phantom{-}\mathfrak{B}^{t=2}_\mu(1,0)-\mathfrak{B}^{t=2}_\mu(0,1)=
\partial_\mu\!\sum_{N,\text{odd}}\frac{2(2N+3)}{(N+2)!}\int_0^1\!\!du\,u^N\bar u^{N+2}
\Big[\mathcal{O}_{N}^{V}(ux)\Big]_{lt}\,,
\hspace*{1cm}
\notag\\
\mathbb{W}_\mu^{t=2}&=&-\mathfrak{B}^{t=2}_\mu(1,0)-\mathfrak{B}^{t=2}_\mu(0,1)=
\partial_\mu\!\sum_{N,\text{even}}\frac{2(2N+3)}{(N+2)!}\int_0^1\!\!du\,u^N\bar u^{N+2}
\Big[\mathcal{O}_{N}^{V}(ux)\Big]_{lt},
\end{eqnarray}
where for arbitrary function the leading twist projection $[\ldots]_{lt}$ is defined as
\begin{equation}
       [f(x)]_{lt}=[\Pi f](x)=\Pi(x,\lambda)f(n)\,.
\end{equation}
The conformal expansion for the twist-three contributions reads
\begin{eqnarray}
\mathbb{V}_\mu^{t=3}&=&
S_{\mu\rho\nu\sigma} x^{\rho}\partial^\sigma\sum_{N,\text{odd}}\frac{2N+3}{(N+2)!(N+1)}
\int_0^1 du\, (u\bar u)^{N+1}\Big[i\mathbf{P}^\nu,\mathcal{O}_{N}^{V}(ux)\Big]_{lt}
\hspace*{2cm}
\notag\\
&&{}
-i\epsilon_{\mu\rho\nu\sigma} x^{\rho}\partial^\sigma\sum_{N,\text{even}}\frac{2N+3}{(N+2)!(N+1)}
\int_0^1 du\, (u\bar u)^{N+1}\Big[i\mathbf{P}^\nu,\mathcal{O}_{N}^{A}(ux)\Big]_{lt}
\notag\\
&&{}+\partial_\mu x^2\partial_\nu\sum_{N,\text{odd}}\frac{2N+3}{(N+2)!}
\int_0^1\!du\, u^{N+2}\bar u^N\int_0^1\!\!dv\, v^{N+1}\ln v
\Big[i\mathbf{P}^\nu,\mathcal{O}_{N}^{V}(uvx)\Big]_{lt},
\notag\\
\mathbb{W}_\mu^{t=3}&=&
S_{\mu\rho\nu\sigma} x^{\rho}\partial^\sigma\sum_{N,\text{even}}\frac{2N+3}{(N+2)!(N+1)}
\int_0^1 du\, (u\bar u)^{N+1}\Big[i\mathbf{P}^\nu,\mathcal{O}_{N}^{A}(ux)\Big]_{lt}
\notag\\
&&{}
-i\epsilon_{\mu\rho\nu\sigma} x^{\rho}\partial^\sigma\sum_{N,\text{odd}}\frac{2N+3}{(N+2)!(N+1)}
\int_0^1 du (u\bar u)^{N+1}\Big[i\mathbf{P}^\nu,\mathcal{O}_{N}^{V}(ux)\Big]_{lt}
\notag\\
&&{}+\partial_\mu x^2\partial_\nu\sum_{N,\text{even}}\frac{2N+3}{(N+2)!}
\int_0^1\!\!du\, u^{N+2}\bar u^N\!\!\int_0^1\!dv\, v^{N+1}\ln v
\Big[i\mathbf{P}^\nu,\mathcal{O}_{N}^{A}(uvx)\Big]_{lt},
\end{eqnarray}
and, finally,
for the twist-four terms one obtains in this representation a remarkably simple
result~\cite{Braun:2011zr}
\begin{eqnarray}
\mathbb{V}^{t=4}_\mu&=& \sum_{N,\text{odd}}\frac{2N+3}{(N+2)!(N+2)}\int_0^1 \!\!du \,(u\bar u)^{N+1}
\biggl\{
x_\mu\,[(\partial\mathcal{O})_N^V(u x)]_{lt}\,\hspace*{4cm}
\notag\\[-2mm]
&&{}\hspace*{4cm}
+\frac12{N(N+3)} \int_0^1\!\!dv\,v^{N-1}\,  x^2\partial_\mu\,
[(\partial\mathcal{O})_N^V(uv x)]_{l.t.}\biggr\},
\notag\\
\mathbb{W}^{t=4}_\mu&=&\frac12 x^2\partial_\mu \sum_{N,\text{even}}\frac{(2N+3)N(N+3)}{(N+2)!(N+2)}
\int_0^1 \!\!du \,(u\bar u)^{N+1}\int_0^1\!\!dv\,v^{N-1}\,
[(\partial\mathcal{O})_N^A(uv x)]_{lt}\,,
\nonumber\\
\mathbb{X}^{t=4}&=&\frac12\sum_{N,\textrm{odd}} \frac{(2N+3)(N+1)}{(N+2)!(N+2)}
\int_0^1\!\! du\, (u\bar u)^{N}(u-\bar u)\int_0^1\!\! dv\, v^{N-1}\,
[(\partial\mathcal{O})_N^V(uv x)]_{lt}\,,
\nonumber\\
\mathbb{Y}^{t=4}&=&-\frac12\sum_{N,\textrm{odd}} \frac{(2N+3)(N+1)}{(N+2)!(N+2)}
\int_0^1\!\! du\, (u\bar u)^{N}(u^2\!+\!\bar u^2)\!\int_0^1 \!\!dv\, v^{N-1}\,
[(\partial\mathcal{O})_N^V(uv x)]_{lt}\,.
\label{pobeda}
\end{eqnarray}
This representation reveals that
the operator $\partial^2 \mathcal{O}_{\mu_1\ldots\mu_n}$ which
corresponds to $[i\mathbf{P}_\mu[i\mathbf{P}^\mu, \mathcal{O}_N]$ in our
notation, does not contribute to the answer
for our special choice of the correlation function $T\{j_\mu(x)j_\nu(0)\}$.
Thus, for this choice, kinematic power corrections to twist-four accuracy are
related to the contribution of the divergence of the leading twist operators alone
(apart from Nachtmann-type corrections).
The T-product with symmetric positions of the currents, $T\{j_\mu(x)j_\nu(-x)\}$,
includes both operators. The corresponding expression turns out to be much more cumbersome.

\section{Sample applications}
\label{sec:app}

As we already mentioned, contributions of different twist in the OPE of the product of
currents in off-forward kinematics are intertwined by electromagnetic gauge and
Lorentz invariance.
Implementation of the electromagnetic gauge invariance in off-forward processes beyond the
leading twist accuracy has been at the center of
many discussions, see e.g.~\cite{Anikin:2000em}.
By contrast, importance of the translation invariance condition has never been emphasized,
to the best of our knowledge:
the distinction between the kinematic  corrections that originate from contributions of
the leading- and higher-twist operators
is not invariant under translations along the line connecting the currents and does not
have physical meaning. We want to illustrate this statement and also demonstrate
restoration of the translation invariance in the sum of all twists
on the simplest example: distribution amplitude (DA) of a
longitudinally polarized $\rho$-meson. Our discussion closely follows Ref.~\cite{Ball:1998ff}.

To this end we consider a nonlocal quark-antiquark operator sandwiched between vacuum and
the $\rho$-meson state~\cite{Ball:1998ff}
\begin{equation}\label{odin}
\vev{0|\bar u(z_1x)\slashed{x} d(z_2x)|\rho^-(p)}
=
f_\rho m_\rho (e\cdot x)\int_0^1du\, e^{-iz_{12}^u(px)}
\Big[\phi_{\parallel}(u)+\frac1{16} z_{12}^2x^2 \Phi(u)+O(x^4)\Big]\,,
\end{equation}
where $m_\rho$ and $f_\rho$ are the $\rho$-meson mass and decay constant, respectively,
$p_\mu$ is the meson momentum, $p^2=m_\rho^2$, and $e_\mu$ is the polarization vector, $e\cdot p=0$.
A Wilson line is implied between the quarks.
The dependence on quark coordinates $z_1$ and $z_2$ is fixed uniquely by the
translation invariance condition
\begin{equation}
 \vev{0|\bar u\big((z_1+\delta)x\big)\slashed{x} d\big((z_2+\delta)x\big)|\rho^-(p)}
  = e^{-i\delta(px)}  \vev{0|\bar u(z_1x)\slashed{x} d(z_2x)|\rho^-(p)}\,.
\end{equation}
It is easy to verify that this relation holds for the above parametrization of the matrix element.

The function $\phi_{\parallel}(u)$ has a physical meaning of the momentum fraction distribution
of the quark and antiquark in the longitudinally polarized $\rho$-meson
and is dubbed the leading-twist DA.
It is usually written as an expansion over Gegenbauer polynomials
\begin{equation}
  \phi_{\parallel}(u,\mu) = 6u(1-u) \sum_{N,\text{even}} a_N(\mu) C_N^{3/2}(2u-1)
\end{equation}
and is normalized by the condition
$
 \int_0^1 du\,\phi_{\parallel}(u) = 1\,
$,
so that $a_0=1$. The coefficients $a_N$ in the Gegenbauer expansion correspond to the reduced
matrix elements of local conformal operators $\mathcal{O}_N$ defined in Eq.~(\ref{eq:On}).

In turn, the function $\Phi(u)$ is called the twist-four two-particle DA. Our task is to find all
``kinematic'' contributions to $\Phi(u)$ which are proportional to $m^2_\rho$~~\cite{Ball:1998ff}.
{}For simplicity of the argument we inserted ``$\slashed{x}$'' between the quarks, in which
case twist-three contributions drop out
\begin{align}
\bar u(z_1x)\slashed{x} d(z_2x)=[\bar u(z_1x)\slashed{x} d(z_2x)]^{t=2}+[\bar u(z_1x)\slashed{x} d(z_2x)]^{t=4}+
O(x^4)\,,
\end{align}
and, more importantly, will assume that $\phi_{\parallel}(u)$ has its asymptotic form
at large scales, $\phi_{\parallel}(u) = 6u(1-u)$, i.e. put all coefficients $a_N$ due to the conformal operators
$\mathcal{O}_N$ with $N\ge 2$ to zero. In this case the matrix elements of twist-four operators
$(\partial \mathcal{O})_N$ obviously vanish as well and we are left with a simple expression
\begin{equation}
[\bar u(z_1x)\slashed{x} d(z_2x)]^{t=4}=\frac14 x^2\, z_1\,z_2
\int_0^1 \!dv\, v^2\,
\big[i\mathbf{P}_\mu,\big[i\mathbf{P}^\mu,[\bar u(vz_1x)\slashed{x} d(vz_2x)]\big]\big]+
\ldots
\end{equation}
or, for the matrix element
\begin{eqnarray}
\langle 0 |[\bar u(z_1x)\slashed{x} d(z_2x)]^{t=4}|\rho^-(p)\rangle
&=&- \frac{z_1z_2}{4} m^2_\rho x^2  \int_0^1 \!dv\, v^2\,
\langle 0 |\bar u(vz_1x)\slashed{x} d(vz_2x)|\rho^-(p)\rangle +
\ldots \hspace*{1cm}
\nonumber\\ &=&
 - \frac{z_1z_2}{4} f_\rho m^3_\rho (e\cdot x) x^2 \!\int_0^1 \!\!dv\,v^2\!\!
\int_0^1\!\! du\,e^{-ivz_{12}^u(px)} \phi_{\parallel}(u) +\ldots
\label{mel4}
\end{eqnarray}
where the ellipses stand for dynamic twist-four contributions which are of no interest to us at present.

One sees that the twist-four contribution alone does not have the form required by translational invariance.
Moreover, it vanishes if $z_1=0$ or $z_2=0$, i.e. if the quark of the antiquark field is at the origin.
The invariance must be recovered in the sum  with the term $\sim m^2_\rho x^2$ from the
leading-twist contribution
\begin{equation}
   [\bar u(z_1x)\slashed{x} d(z_2x)]^{t=2}\equiv
  [\bar u(z_1x)\slashed{x} d(z_2x)]_{lt} = \Pi(x,\lambda)[\bar u(z_1n)\slashed{n} d(z_2n)]\,,
\end{equation}
where $\Pi$ is the leading-twist projector, e.g. in the vector representation (\ref{PIVO})
\begin{eqnarray}
[\Pi f](x) &=& \sum_{k=0}^\infty \frac{1}{2^k k!}
 (x^2 \partial_n^2)^{l/2} C_l^{(1)}\left(\frac{x\cdot\partial_n}{\sqrt{x^2 \partial_n^2}}\right) f(n)\Big|_{n=0}
\nonumber\\
   &=& \sum_{k=0}^\infty \frac{1}{k!}
   \Big[(x\cdot\partial_n)^k -\frac{k-1}{4} (x\cdot\partial_n)^{k-2} x^2\partial_n^2 + O(x^4)
   \Big]f(n)\Big|_{n=0}\,.
\end{eqnarray}
Here $\partial_n = \partial/\partial n_\mu$.
One obtains after a simple algebra (cf.~\cite{HD-THEP-90-38,Ball:1998ff})
\begin{equation}
 [\Pi (e\cdot x) e^{-ipn}](x) \equiv [(e\cdot x)e^{-ipx}]_{lt} =
(e\cdot x)\Big[e^{-ipx} + \frac14 p^2 x^2 \int_0^1 \!\!dv\, v^2 e^{-ivpx} + O(x^4)\Big]
\end{equation}
and therefore
\begin{eqnarray}\label{mel2}
\vev{0|[\bar u(z_1x)\slashed{x} d(z_2x)]^{t=2}|\rho^-}
&=&
f_\rho m_\rho (e\cdot x)\int_0^1du \biggl[e^{-i z_{12}^u(px)}
\phi_{\parallel}(u)
\notag\\
&&{}
+\frac14{m_\rho^2 x^2}\int_0^1\!\!dv\, v^2 e^{-i v z_{12}^u (px)}
(z_{12}^u)^2\phi_{\parallel}(u)+O(x^4)
\biggr].
\end{eqnarray}
The second term ($O(x^2)$) in this equation is the Nachtmann-type mass correction that originates from
subtraction of traces in the definition of the leading-twist operator.
Adding (\ref{mel4}) and (\ref{mel2}) one gets
\begin{eqnarray}\label{melS}
\vev{0|\bar u(z_1x)\slashed{x} d(z_2x)|\rho^-}&=&
f_\rho m_\rho (e\cdot x)\int_0^1\! du \biggl[e^{-i(px) z_{12}^u}
\phi_{\parallel}(u) \hspace*{5cm}
\notag\\
&&{}
+\frac14{m_\rho^2 x^2}\int_0^1\!\!dv\, v^2 e^{-i(px) v z_{12}^u}
[(z_{12}^u)^2-z_1z_2]\phi_{\parallel}(u)+O(x^4)
\biggr]\,.
\end{eqnarray}
Finally, using explicit expression $\phi_{\parallel}(u) = 6u\bar u$
and rewriting $[(z_{12}^u)^2-z_1z_2]=z_{12}[z_{12} u\bar u+(1-2u)z_{12}^u]$,
the second line can be simplified using
\begin{equation}\label{dva}
\int_0^1\!du\int_0^1\!dv\, v^2 e^{-i(px) v z_{12}^u}
[(z_{12}^u)^2-z_1z_2] u\bar u
= \frac{z_{12}^2}2\int_0^1\!du\, (u\bar u)^2\,e^{-i(px)z_{12}^u}\,.
\end{equation}
Thus, translation invariance is indeed restored and we obtain,
comparing~(\ref{melS}) with the ansatz in Eq.~(\ref{odin})
\begin{align}
\Phi(u)=12 m^2_\rho u^2\bar u^2\,,
\end{align}
which agrees with~\cite{Ball:1998ff}.

We emphasize that the the translational invariance is only rescued in our calculation
for the asymptotic distribution amplitude $\phi_{\parallel}(u) = 6u\bar u$, which is
consistent with neglecting twist-four contribitions due to divergence of the leading
twist (conformal) operators. For a generic $\phi_{\parallel}(u)$ translation invariance
will be recovered by adding the terms in $(\partial \mathcal{O})_N$.
The lesson is that kinematic corrections that originate from the contributions of operators of
different twist in the OPE are all intertwined by Lorentz (and eventually also gauge) invariance.
Estimates based on the contributions of leading twist alone can be misleading.

In the rest of this section we consider matrix elements of the operators in question
sandwiched between proton states with different momenta, which are relevant for the
description of hard exclusive scattering reactions in Bjorken limit, such that
the DVCS. To leading-twist accuracy, the relevant nonperturbative input is encoded in
the generalized parton distributions (GPDs) which are defined by the matrix elements
of nonlocal quark-antiquark light-ray operators. For example, for the vector operator
(see, e.g. Ref.~\cite{Diehl:2003ny})
\begin{eqnarray}
\vev{p'|{O}_V(z_1n,z_2n)|p}&=&
\int_{-1}^1\!\!
 du\, e^{-iP_+((z_2(\xi+u)+z_1(\xi-u))}
\notag\\
&&{}\times
\bar u(p')\left[\gamma_+\,H(u,\xi,t)+\frac{i\sigma^{+\nu}\Delta_\nu}{2m}
E(u,\xi,t)\right]u(p)\,,
\label{GPD1}
\end{eqnarray}
where
\begin{align}
P=\frac{p+p'}{2}\,, && \Delta=p'-p\,,&& t=\Delta^2\,,&& \xi = \frac{p_+-p'_+}{p_++p'_+}\,,
\end{align}
$m$ is the nucleon mass and $u(p)$ is the nucleon spinor.
As always, $P_+=P_\mu n^\mu$ etc.
The GPDs $H(u,\xi,t)$ and $E(u,\xi,t)$ depend on Bjorken variable which we call here $u$
in order to distinguish from the quark coordinate, skewedness parameter $\xi$ and the
invariant momentum transfer $t$.

Gegenbauer moments of the GPDs can be written in terms of reduced matrix elements of
local conformal operators
\begin{eqnarray}
\int du \, C_N^{3/2}\left(\frac{u}\xi\right)\,H(u,\xi,t)
&=&
\sum_{k,\text{even}}^N (-1)^k (2\xi)^{k-N} \big[\widetilde{H}_{Nk}(t)+E_{Nk}(t)\big] \,,
\notag\\
\int du \, C_N^{3/2}\left(\frac{u}\xi\right)\,E(u,\xi,t)
&=&
\sum_{k,\text{even}}^{N+1}(-1)^{k+1}(2\xi)^{k-N} E_{Nk}(t)
\end{eqnarray}
where
\begin{eqnarray}\label{iO}
(-i)^N\vev{p'|\mathcal{O}_N^V(n)|p}&=&
\bar u(p')\slashed{n}u(p)\, \widetilde{H}_{N}(n;p,p')
+\frac{\bar u(p')u(p)}{m}\,E_{N}(n;p,p')\,
\end{eqnarray}
and
\begin{eqnarray}
\widetilde{H}_{N}(n;p,p') &=& \sum_{k,\text{even}}^N\widetilde{H}_{Nk}(t)\Delta_+^k P_+^{N-k}\,,
\nonumber\\
E_{N}(n;p,p') &=& \sum_{k,\text{even}}^{N+1} E_{Nk}(t)\Delta_+^k P_+^{N+1-k}\,.
\end{eqnarray}
Note that we define the generalized form factors $E_{Nk}(t)$ as coefficients of the scalar
structure $\bar u(p')u(p)$ instead of a more usual $\bar u(p')i\sigma^{+\nu}\Delta_\nu u(p)$
which leads to a redefinition of the $\tilde{H}_{Nk}(t)$ form factors; hence a ``tilde''
in the notation.
The description in terms of the generalized form factors of conformal operators
has an advantage compared to the standard parametrization in terms of simple moments,
$A_{nk}(t), B_{nk}(t), C_{n}(t)$~\cite{Diehl:2003ny},
in that $E_{Nk}(t)$ and $\widetilde{H}_{Nk}(t)$ have autonomous scale dependence to one-loop
accuracy. This property is convenient in applications~\cite{Mueller:2005ed,Kirch:2005tt}.

The distributions (\ref{GPD1}) are defined as matrix elements of the operators directly on the light cone,
and they enter the OPE through the leading twist projection of the quark-antiquark operators
at generic (non-light-like) separations:
\begin{eqnarray}\label{iO1}
(-i)^N\vev{p'|[\mathcal{O}_N^V(x)]_{lt}|p}&=&
\bar u(p')\slashed{n}u(p)\, \widetilde{H}_{N}(x;p,p')
+\frac{\bar u(p')u(p)}{m}\,E_{N}(x;p,p')\,.
\end{eqnarray}
Using the explicit expression for the projection operator $\Pi(x,\partial_n)$ we obtain,
to the $O(x^2)$ accuracy
\begin{eqnarray}\label{815}
\widetilde{H}_N(x;p,p')&=&\sum_{k,\text{even}}^N\widetilde{H}_{Nk}(t)(\Delta x)^k (P x)^{N-k}-\frac{x^2}{4(N+1)}\sum_{k,\text{even}}^{N-2}
(\Delta x)^k(Px)^{N-k-2}
\notag\\
&&{}
\times\biggl[(k+1)(k+2)
t\,\widetilde{H}_{Nk+2}(t)+(N-k)(N-k-1)\left(m^2-\frac{t}4\right)\widetilde{H}_{Nk}(t)\biggr]\,,
\notag\\
E_N(x;p,p')&=&\sum_{k,\text{even}}^{N+1}E_{Nk}(t)(\Delta x)^k (P x)^{N+1-k}-\frac{x^2}{4(N+1)}\sum_{k,\text{even}}^{N-1}
(\Delta x)^k(Px)^{N-k-1}
\notag\\
&&{}
\times\biggl[(k+1)(k+2)
t\,E_{Nk+2}(t)\!+\!(N+1-k)(N-k)\left(m^2-\frac{t}4\right)E_{Nk}(t)
\notag\\
&&{}\phantom{\Biggl[\Big[} +2(N-k)m^2 \widetilde{H}_{Nk}(t)\biggr].
\end{eqnarray}
The terms $O(x^2)$ in these equations give rise to a finite-$t$ and target mass corrections to hard exclusive scattering
reactions, which is analogous to the Nachtmann correction in deep-inelastic scattering. For the particular case of DVCS
such corrections were studied in
\cite{Radyushkin:2000ap,Belitsky:2001hz,Belitsky:2000vx,Belitsky:2010jw,%
Geyer:2004bx,Blumlein:2006ia,Blumlein:2008di}.
The new contribution of this work is the calculation of corrections due to the ``kinematic''
twist-four operator $(\partial \mathcal{O})_N^V$  which must be taken into account
alongside the Nachtmann correction to maintain Lorentz and electromagnetic gauge invariance.

The relevant matrix elements can be defined by an expression similar to Eq.(\ref{iO1}):
\begin{align}\label{}
(-i)^{N-1}\vev{p'|(\partial\mathcal{O})_N^V(x)|p}=\bar u(p')\slashed{x}u(p)\, {\widetilde H}^{(4)}_N(x)+
\frac1{m}\bar u(p')u(p)\, E^{(4)}_N(x)\,.
\end{align}
Taking into account that
\begin{equation}
\vev{p'|(\partial{\mathcal{O}})_N^V(x)|p}
=\frac1{N+1}i\Delta^\mu\frac{\partial}{\partial x^\mu}\vev{p'|\mathcal{O}_N^V(x)|p}
\end{equation}
we obtain from Eq.~(\ref{815})
\begin{eqnarray}
{\widetilde H}^{(4)}_N(x)&=&\frac{1}{2(N+1)^2}\sum_{k,\text{odd}}^{N-1}(\Delta x)^{k}(Px)^{N-k-1}
\biggl[(k+1)(2N+2-k)\, t \widetilde{H}_{Nk+1}(t)
\notag\\
&&{}
-(N-k+1)(N-k+2)\Big(m^2-\frac{t}{4}\Big)\widetilde{H}_{Nk-1}(t)\biggr]\,,
\notag\\
{E}^{(4)}_N(x)&=&\frac{1}{2(N+1)^2}\sum_{k,\text{odd}}^{N}(\Delta x)^{k}(Px)^{N-k}
\biggl[(k+1)(2N+2-k)\, t E_{Nk+1}(t)
\notag\\
&&{}
-(N-k+2)\Big[(N-k+3)\Big(m^2-\frac{t}{4}\Big)E_{Nk-1}(t)+
2m^2\widetilde{H}_{Nk-1}(t)\Big]\biggr]\,,
\end{eqnarray}
These expressions have similar level of complexity compared to Nachtmann-type corrections
in Eq.~(\ref{815}), but a somewhat different structure, which means that they may result
in a different dependence on the quark momentum fraction.

Applications to concrete processes require dedicated studies and this
task goes far beyond the scope of our paper.
Main question that has to be addressed is whether QCD factorization
is applicable for a given process for kinematic twist-four contributions.

\section{Summary and Conclusions}
Operator product expansion belongs to the most important tools in the
analysis of hard reactions in QCD. Whereas OPE for forward scattering
(i.e. neglecting operators that include total derivatives) was thoroughly
studied in connection with deep-inelastic
lepton-nucleon scattering, the general case of off-forward kinematics has
never been analyzed systematically beyond the leading twist. A new
phenomenon which is specific for off-forward kinematics is appearance
of higher-twist operators that are total derivatives of the operators of
the leading twist. In this work we show that
the structure of such contributions to the leading order in strong
coupling is governed by conformal invariance.
We call these terms kinematic, as they do not involve new
nonperturbative input compared to the leading twist. Our study has been
fuelled by potential applications
to hard exclusive scattering processes that involve generalized parton
distributions, in particular DVCS, transition form factors of the type
$\gamma^*\to M\gamma$, and to calculations of weak decays of $B$-mesons in
the framework of QCD factorization or light-cone sum rules.

As a principal result, we provide a set of projection operators that allow
one to extract the ``kinematic'' part of an arbitrary flavor-nonsinglet
chiral-even twist-four operator in QCD.
We also calculate, to the same accuracy, the kinematic contribution to the
time-ordered product of two electromagnetic currents. This result makes
possible a complete calculation of the finite-$t$ and target mass
corrections in two-photon reactions that are amenable to factorization to
twist-four accuracy.
We stress that the proof of validity of QCD factorization in a given process, at
least for the kinematic contributions considered here, is a separate
question which is beyond the scope of this paper.
We hope that factorization of kinematic twist-four contributions can be
shown at least for some selected reactions, because their structure is
intertwined with the leading twist and strongly constrained by electromagnetic
gauge and Lorentz invariance. Clarification of this issue
is certainly the most important task for future studies.

Apart from this topic, our study can be extended in several directions.
In this work we have only considered chiral-even operators, i.e. ones built of a pair
of left-handed or right-handed quarks.
{}For some applications, e.g. $B$-meson decays, it may be necessary to
include contributions of chiral-odd operators (i.e. made of one left-handed and one
right-handed quark) as well. This extension should be straightforward.
A more difficult question concerns
kinematic projection of flavor-singlet and in particular pure gluon
operators. Such contributions are needed e.g. for the calculation of
finite-$t$ and target mass corrections that involve gluon GPD.
From a theory perspective, it would also be interesting
to reproduce some of our results using a different technique. Our final
formula for the kinematic contribution to the $T$-product of two
electromagnetic currents appears to be considerably
simpler as compared to the intermediate expressions. This suggests that a
more direct derivation of this result might be possible by considering
three-point correlation functions of conformal operators
with the electromagnetic currents to two-loop accuracy in the critical
dimension.



\acknowledgments

The work by A.M. was supported by the DFG, grant BR2021/5-2,
and RFFI, grant 09-01-93108.

\appendix

\section*{Appendices}
\addcontentsline{toc}{section}{Appendices}

\renewcommand{\theequation}{\Alph{section}.\arabic{equation}}
\renewcommand{\thetable}{\Alph{table}}
\setcounter{section}{0}
\setcounter{table}{0}

\section{Evolution Hamiltonian for non-quasipartonic operators}\label{App:renorm}
%
The matrix of anomalous dimensions for a set of local operators $O_i$ is defined as
\begin{align}
\gamma_{ik}=  Z^{-1}_{ij}\mu\frac{\partial}{\partial\mu} Z_{jk} \,,&&[O]_{i}=Z^{-1}_{ik} O^B_{k},
\end{align}
where $[O]_{i}$, $O^B_{i}$ are  renormalized and bare operators, respectively.
The renormalized operators satisfy the matrix RG equation
\begin{align}\label{rg-0}
\left(\mu\frac{\partial }{\partial\mu}+
      \beta(g)\frac{\partial }{\partial g} +\gamma_{ik}(g)\right)\,[O]_{k}=0\,,
\end{align}
which, equivalently, can be cast in the form of an integro-differential equation for the
generating functions (light-ray operators):
\begin{align}\label{rg-1}
\left(\mu\frac{\partial }{\partial\mu}+
      \beta(g)\frac{\partial }{\partial g} +\gamma (g)\right)\,[O](z_1,\ldots,z_N)=0\,.
\end{align}
Here $\gamma$ is an integral operator, which we write as
\begin{align}
\gamma=\frac{\alpha_s}{2\pi}\, {\mathbb H}\,,
\end{align}
where $\alpha_s=g^2/4\pi$.

The operator $\mathbb{H}_{QQ}$ is determined by  one-loop counterterms to the nonlocal operator
and can be written as
\begin{align}
\mathbb{H}_{QQ} = N_c\,\mathbb{H}_{QQ}^{(1)}-\frac1{N_c}\mathbb{H}_{QQ}^{(-1)}-3C_F-2N_c\sigma_g\,,
\end{align}
where $\sigma_g=b_0/4N_c$,  $b_0=\frac{11}3N_c-\frac23 n_f$. The ``Hamiltonians''
$\mathbb{H}^{(\pm 1)}$ are $3\times3$ matrices which are written in terms of $SL(2)$ invariant
integral operators
\begin{align}
\mathbb{H}_{QQ}^{(1)}=&(\widehat{\mathcal{H}}_{12}+\widehat{\mathcal{H}}_{23})\,\mathbb{I}+
\begin{pmatrix}
  2\mathcal{H}_{12}^d-\mathcal{H}^+_{12} & \mathcal{H}^{e,(1)}_{23}&
  z_{12}(\mathcal{H}_{12}^+ +  \widetilde{\mathcal{H}}_{12}^+)
\\[2mm]
        \mathcal{H}_{32}^{e,(1)} &-2\mathcal{H}^+_{12}
               &0
\\[2mm]
z_{12}^{-1}\Pi_0&0&-2(2\mathcal{H}_{12}^++\widetilde{\mathcal{H}}_{12}^+)
\end{pmatrix}\,,
\notag\\[2mm]
\mathbb{H}_{QQ}^{(-1)}=&\widehat{\mathcal{H}}_{13}\,\mathbb{I}+
\begin{pmatrix}
2P_{12}\mathcal{H}_{12}^d-\mathcal{H}^+_{13}&
-P_{23}\mathcal{H}^{e(1)}_{23}&
-2z_{12}\mathcal{H}_{12}^-
\\[2mm]
       -P_{23} \mathcal{H}_{23}^{e(1)}&
2\mathcal{H}^-_{12}+\mathcal{H}_{13}^d+P_{23} \mathcal{H}^{e(2)}_{23}
               &z_{13}\mathcal{H}_{13}^+
\\[2mm]
\frac{2}{z_{12}}\mathcal{H}_{12}^-\Pi_0
&
\frac1{z_{13}}\Pi_0
&
             6\mathcal{H}_{12}^- -2\mathcal{H}_{13}^+  -P_{23} \mathcal{H}^{e(1)}_{23}
\end{pmatrix}\,.
\label{A:HQQ}
\end{align}
Explicit expressions for the two-particle kernels can be found in
Ref.~\cite{Braun:2009vc}.

We want to find a scalar product such that $\mathbb{H}_{QQ}$ is hermitian.
Write, schematically
\begin{eqnarray}
   \mathbb{H}\, \overrightarrow{\Psi} &=&
\begin{pmatrix}
H_{11}& H_{12} & H_{13}\\H_{21}& H_{22} & H_{23} \\ H_{31}& H_{32} & H_{33}
\end{pmatrix}
\begin{pmatrix}
       \Psi_1^{(1,1,1)}\\  \Psi_2^{(1,\frac32,\frac12)} \\ \Psi_3^{(\frac32,\frac32,1)}
\end{pmatrix}\,,
\end{eqnarray}
where the superscripts on the three components of the ``wave function'' indicate the corresponding
$SL(2)$ representations.
Since the diagonal $H_{kk}$ kernels are $SL(2)$ invariant operators in their subsectors,
it is natural to look for the scalar product of the form
\begin{eqnarray}
 \langle\!\langle \Phi,\Psi\rangle\!\rangle = \sum_{k=1}^3 a_k \langle \Phi_k,\Psi_k\rangle
\end{eqnarray}
where $\langle \Phi_k,\Psi_k\rangle$ is the standard $SU(1,1)$ scalar product
for the appropriate representation.
We want to achieve that $H^\dagger = H$, i.e.
\begin{eqnarray}
  \langle\,\langle \Phi, H\Psi\rangle\!\rangle & \equiv&
\langle\!\langle H^\dagger \Phi, \Psi\rangle\!\rangle \,=\,  \langle\!\langle H \Phi, \Psi\rangle\!\rangle
\end{eqnarray}
 for arbitrary $\Phi, \Psi$. In particular we can take
\begin{eqnarray}
 \Phi \,=\,
\begin{pmatrix}
       \phi^{(1,1,1)}\\ 0 \\ 0
\end{pmatrix}\,, \qquad
 \Psi \,=\,
\begin{pmatrix}
       0 \\  \psi^{(1,\frac32,\frac12)} \\ 0
\end{pmatrix}\,
\end{eqnarray}
and require that
\begin{eqnarray}
 a_1 \langle \phi, H_{12}\psi\rangle_{(1,1,1)} = a_2 \langle H_{21} \phi,\psi\rangle_{(1,\frac32,\frac12)}
\end{eqnarray}
Since the evolution kernels are given by the sum of two-particle operators acting on a chosen
parton pair, e.g.
\begin{eqnarray}
   H_{12} &=& H_{12}^{(12)} + H_{12}^{(23)} + H_{12}^{(31)}
\end{eqnarray}
it is enough to consider two-particle kernels separately, and due to $SL(2)$ invariance
it is enough to consider their action on the corresponding eigenfunctions.
{}For example, take $H_{12}^{(23)}$ and $\phi=\psi= (z_2-z_3)^n$. Then
\begin{eqnarray}
   H_{12}^{(23)} z_{23}^n = h_{12}(n) z_{23}^n \,, \qquad  H_{21}^{(23)} z_{23}^n = h_{21}(n) z_{23}^n
\end{eqnarray}
and we have to require that
\begin{eqnarray}
  a_1\, h_{12}(n)\, ||z^n_{23}||^2_{(1,1)} &=& a_2\, h_{21}(n)\, ||z^n_{23}||^2_{(\frac32,\frac12)}
\end{eqnarray}
which gives the relation between $a_1$ and $a_2$.
Note that this expression has to hold for arbitrary $n$.
Explicit expressions for $h_{ik}$ can be read of Ref.~\cite{Braun:2009vc}
(see the results in invariant representation) after taking the color traces,
or using the expressions in Eq.~(\ref{A:HQQ}).

Completing this calculation for all cases one arrives at the result given in Eq.~(\ref{eq:<<>>}).

\section{Derivation of Eq.~(\ref{AO})}\label{App:B}

To save space, in this Appendix we will use the following notations:
\begin{eqnarray}\label{DQdef}
\mathcal{F}(z_1,z_2,z_3)&=&ig\bar\psi_+(z_1)f_{+\alpha}(z_{2})\psi^\alpha(z_3)\,,
\notag\\
\bar{\mathcal{F}}(z_1,z_2,z_3)&=&ig\bar\psi^{\dot\alpha}(z_1)\bar f_{\dot\alpha+}(z_{2})\psi_+(z_3)\,,
\notag\\
\mathcal{D}(z_1,z_2,z_3)&=&ig[D_{\alpha+}\bar\psi_+](z_1)f_{++}(z_{2})\psi^\alpha(z_3)\,,
\notag\\
\bar{\mathcal{D}}(z_1,z_2,z_3)&=&ig\bar\psi^{\dot\alpha}(z_1)\bar f_{++}(z_{2})[D_{+\dot\alpha}\psi_+](z_3)\,.
\end{eqnarray}

We have to collect contributions to Eq.~(\ref{NN}) from the operators containing
two derivatives acting on the quark or the antiquark, and the contributions from
quark-antiquark-gluon operators as indicated schematically by the second, third and fourth
terms  in  Eq.~(\ref{set0}).
The contribution of the operator $[D^2\bar \psi](z_1n)\bar n \psi(z_2n)$
can be written as
\begin{multline}\label{}
\Big[(z_1\partial_{z_1}+z_2\partial_{z_2}+2)z_1 -\frac12(\partial_{z_1}+\partial_{z_2})  z_1^2
\Big][D^2\bar \psi](z_1n)\bar n \psi(z_2n)
\\
=\left(S_{12}^+-\frac12\partial_{z_2} z_{12}^2\right)[D^2\bar \psi](z_1n)\bar n \psi(z_2n)\,.
\end{multline}
As explained in the text, all terms in $S_{12}^+$ can be omitted since they vanish upon
integration~(\ref{whO++}). Adding the similar contribution with the derivatives acting
on the quark field, one obtains for the sum of the second and the third terms in~(\ref{set0})
\begin{align}\label{B1}
I_1=-\frac12\partial_{z_2} z_{12}^2[D^2\bar \psi](z_1n)\bar n \psi(z_2n)-
\frac12\partial_{z_1} z_{12}^2\bar \psi(z_1n)\bar n [D^2\psi](z_2n)\,.
\end{align}
Using EOM these operators can be rewritten in the form
\begin{eqnarray}
\bar \psi(z_1n)\bar n [D^2\psi](z_2n)&=&-2ig\bar\psi_+(z_1 n) f_{+\alpha}(z_2n)\psi^\alpha(z_2n)=
-2\mathcal{F}(z_1,z_{2},z_2)\,,
\notag\\
{}[D^2\bar \psi](z_1n)\bar n \psi(z_2n)&=&\phantom{-}2ig\bar\psi^{\dot\alpha}(z_1 n) \bar f_{\dot\alpha+}(z_1n)\psi_+(z_2n)
=\phantom{-}2\bar{\mathcal{F}}(z_1,z_{1},z_2)
\end{eqnarray}
so that one gets for~(\ref{B1})
\begin{align}\label{bound}
I_1=-\partial_{z_2} z_{12}^2\bar{\mathcal{F}}(z_1,z_{1},z_2)
+\partial_{z_1} z_{12}^2{\mathcal{F}}(z_1,z_{2},z_2)\,.
\end{align}

Quark-antiquark-gluon operators indicated schematically as $\sim ig \bar\psi F \psi$
(the fourth term in Eq.~(\ref{set0})) actually come in two variations: with and without an
extra covariant derivative acting on the quarks.  The contribution of operators without
extra derivatives, call it $I_2$, reads
%
\begin{eqnarray}\label{sigmamu}
I_2&=&\int_0^1 du\Big[
(\partial_{z_1} z_1+\partial_{z_2} z_2)z_{12}-(\partial_{z_1}+\partial_{z_2})z_{12}z_{21}^u
\Big]\bar\psi(z_1)\bar \sigma^\mu F_{\mu\nu}(z_{21}^u)n^\nu\psi(z_2)
\notag\\
&=&\int_0^1 du\,[\bar u \partial_{z_1}-u\partial_{z_2}]z_{12}^2\bar\psi(z_1)\bar \sigma^\mu F_{\mu\nu}(z_{21}^u)n^\nu\psi(z_2)\,,
\end{eqnarray}
where we can substitute
\begin{align}\label{}
\bar\psi(z_1)\bar \sigma^\mu F_{\mu\nu}(z_{21}^u)n^\nu\psi(z_2)=
-\mathcal{F}(z_1,z_{21}^u,z_2)-\bar{\mathcal{F}}(z_1,z_{21}^u,z_2)\,.
\end{align}
The contribution of the operators with extra derivatives is (omitting the term in $S^+_{12}$)
\begin{equation}\label{LT}
I_3=-\int_0^1 du\Big\{
\bar u \partial_{z_1} z_{12}^3 \bar\psi(z_1)\bar n F_{\mu+}(z_{21}^u)[D^\mu\psi](z_2)
+u\partial_{z_2} z_{12}^3 [D^\mu\bar\psi](z_1)\bar n F_{\mu+}(z_{21}^u)\psi(z_2)
\Big\}.
\end{equation}
Rewriting the gluon strength tensor in terms of $f,\bar f$ one obtains
\begin{eqnarray}\label{DS2}
\bar\psi(z_1)\bar n F_{\mu+}(z_{21}^u)[D^\mu\psi](z_2)&=&-
\bar\psi_+(z_1)\Big(f_{+\alpha}(z_{21}^u) [{D^{\alpha}}_+\psi_+](z_2)
+ \bar f_{+\dot\alpha}(z_{21}^u)[{D_+}^{\dot\alpha}\psi_+](z_2)
\Big)
\notag\\
&=&\frac12\bar {\mathcal{D}}(z_1,z_{21}^u,z_2)
-\mathcal{F}^{(3)}(z_1,z_{21}^u,z_2)-\bar{\mathcal{F}}^{(3)}(z_1,z_{21}^u,z_2)\,
\end{eqnarray}
and similar
\begin{align}\label{DS1}
[D^\mu\bar\psi](z_1)\bar n F_{\mu+}(z_{21}^u)\psi(z_2)=
\frac12{\mathcal{D}}(z_1,z_{21}^u,z_2)
-\mathcal{F}^{(1)}(z_1,z_{21}^u,z_2)-\bar{\mathcal{F}}^{(1)}(z_1,z_{21}^u,z_2)\,,
\end{align}
where we used a notation $\mathcal{F}^{(k)}$ for the derivative over the corresponding light-cone
coordinate:
$$
\mathcal{F}^{(3)}(z_1,z_{21}^u,z_2)=\partial_z\mathcal{F}](z_1,z_{21}^u, z)\Big|_{z=z_2}\,,
\qquad
\mathcal{F}^{(1)}(z_1,z_{21}^u,z_2)=\partial_z\mathcal{F}](z,z_{21}^u,z_2)\Big|_{z=z_1}\,
$$
and similar for $\bar{\mathcal{F}}$.

 Next, we write
\begin{eqnarray}\label{F31}
\mathcal{F}^{(3)}(z_1,z_{21}^u,z_2)&=&
\left(\partial_{z_2}-\frac1{z_{12}}\bar u\frac{d}{du}\right)\mathcal{F}(z_1,z_{21}^u,z_2)\,,
\notag\\
\mathcal{F}^{(1)}(z_1,z_{21}^u,z_2)&=&
\left(\partial_{z_2}-\frac1{z_{12}} u\frac{d}{du}\right)\mathcal{F}(z_1,z_{21}^u,z_2)\,,
\end{eqnarray}
and similar for $\bar{\mathcal{F}}$. Inserting~(\ref{F31}), (\ref{DS2}), (\ref{DS1})
in Eq.~(\ref{LT}) and integrating by parts one arrives at the following expression for the
sum $I_1+I_2+I_3$:
\begin{eqnarray}\label{tO}
(\partial O)_{++}(z_1,z_2)&=&
\partial_{z_1} z_{12}^2\bar{\mathcal{F}}(z_1,z_2,z_2)-\partial_{z_2} z_{12}^2\mathcal{F}(z_1,z_1,z_2)
\notag\\
&&{}
+\int_0^1du\,\biggl\{\partial_{z_1}\partial_{z_2} z_{12}^3
\Big[\mathcal{F}(z_1,z_{21}^u,z_2)+\bar{\mathcal{F}}(z_1,z_{21}^u,z_2)\Big]
\notag\\
&&{}\hspace*{1.0cm}-\frac12u\Big[
\partial_{z_2}z_{12}^3 \mathcal{D}(z_1,z_{21}^u,z_2)+\partial_{z_1} z_{12}^3 \bar{\mathcal{D}}(z_1,z_{12}^u,z_2)\Big]
\biggr\}\,+ \ldots
\end{eqnarray}
The ellipses stand for EOM, contributions of quasipartonic operators and terms
proportional to $S_{12}^+$ which do not contribute to the projection (\ref{whO++})
for the conformal operator.

The last step, we have to rewrite the answer in terms of the operators from the
conformal basis (\ref{eq:Q}), (\ref{eq:barQ}):
\begin{eqnarray}\label{QF-relation}
(\mu\lambda)(\bar\lambda\bar\mu) \mathcal{F} &=& ig [Q_2(z_1,z_2,z_3)-Q_1(z_1,z_2,z_3)]\,,
\nonumber\\
(\mu\lambda)(\bar\lambda\bar\mu) \mathcal{D} &=& 2ig[Q^{(1)}_2(z_1,z_2,z_3)- Q_3(z_1,z_2,z_3)]\,,
\nonumber\\
(\mu\lambda)(\bar\lambda\bar\mu)\bar{\mathcal{F}} &=& ig[\bar Q_2(z_1,z_2,z_3)-\bar Q_1(z_1,z_2,z_3)]\,,
\nonumber\\
(\mu\lambda)(\bar\lambda\bar\mu)\bar{\mathcal{D}} &=& 2ig[\bar Q^{(3)}_2(z_1,z_2,z_3)- \bar Q_3(z_1,z_2,z_3)]\,,
\label{app:calnoncal}
\end{eqnarray}
where, as above, the superscript in $Q^{(1)}_2$, $\bar Q^{(3)}_2$ stands for the
derivative in the corresponding argument. Using these expressions, rewriting the
derivatives in $Q^{(1)}_2$, $\bar Q^{(3)}_2$ in the form (\ref{F31}) and integrating by
parts we obtain the final answer given in the text, Eq.~(\ref{AO}).

\section{Leading twist projector}\label{app:projector}
Here we derive Eq.~(\ref{PIVO}) which establishes a relation between the projectors in
spinor and vector representations. By definition, the action of the projector~(\ref{PI}) on
the function $f(x^\mu)$ is given by the expression
\begin{align}\label{PifPaf}
[\Pi f](x^\mu)=\sum_{k=0}^\infty\frac{(\bar\partial\bar x\partial)^{k}}{[k!]^2}
f\left(n^\mu=\frac12\lambda\sigma^\mu\bar\lambda\right)\Big|_{\lambda,\bar\lambda=0}\,.
\end{align}
By a direct calculation one easily finds
\begin{align*}
{(\bar\partial \bar x\partial)} f(n)=&
\frac12 (\bar\partial \bar x\sigma^\mu \bar\lambda)\frac{\partial}{\partial n^\mu}f(n)=
(x\partial_n)F(n)+\frac14(\lambda \sigma^\nu \bar x\sigma^\mu \bar\lambda)\partial_\mu\partial_\nu f(n)
=x^\mu \widetilde K_\mu f(n)\,,
\end{align*}
where
\begin{align}
\widetilde K_\mu=(n\partial)\partial_\mu+\partial_\mu-\frac12 n_\mu \partial^2\,,
\end{align}
and the derivatives are taken with respect to $n$.
Next
\begin{align}
(\bar\partial \bar x\partial)^k f\left(\frac12\lambda\sigma^\mu\bar\lambda\right)\Big|_{\lambda=0}
=(x \widetilde K)^k\, f(n)\Big|_{n=0}=
f(\partial_n)\, \left(\frac12 x K\right)^k\Big|_{n=0}\,,
\end{align}
where
\begin{align}
K_\mu=2n_\mu (n\partial)- n^2\partial_\mu+2n_\mu\,
\end{align}
is nothing but the generator of special conformal transformations for a scalar field
$\varphi$ with scaling dimension $\dim[\varphi]=1$ ($\varphi(x)\to x^{-2} \varphi(x/x^2)$). Thus
\begin{eqnarray}
f(\partial_n) \left(\frac12x K\right)^k\Big|_{n=0}&=&
f(\partial_n)\left(\frac{\partial}{\partial a}\right)^k\exp\left[\frac12ax^\mu K_\mu\right]\cdot 1\Big|_{n=0,a=0}
\notag\\
&=&
f(\partial_n)\left(\frac{\partial}{\partial a}\right)^k\frac1{1-(xn)a+a^2x^2n^2/4}\Big|_{n=0,a=0}\,.
\end{eqnarray}
Taking into account that the last factor in the above expression is the generating function
for the Gegenbauer polynomials $C_k^{(1)}$
\begin{align}
\frac1{1-(xn)a+a^2x^2n^2/4}=\sum_{k=0}^\infty C_k^{(1)}\left(\frac{(xn)}{\sqrt{x^2n^2}}\right) \left(\frac14a^2x^2 n^2\right)^{k/2}
\end{align}
one obtains
\begin{equation}\label{fK}
f(\partial_n) \left(\frac12x K\right)^k\Big|_{n=0}=
\frac{k!}{2^k}f(\partial_n)\,(x^2 n^2)^{k/2} C_k^{(1)}\left(\frac{(xn)}{\sqrt{x^2n^2}}\right)\Big|_{n=0}\,.
\end{equation}
Finally, inserting~(\ref{fK}) in Eq.~(\ref{PifPaf}) one ends up with the expression in Eq.~(\ref{PIVO}):
\begin{equation}\label{PCGE}
[\Pi f](x)=\sum_{k=0}^\infty
\frac{1}{2^kk!}\,(x^2 \partial^2)^{k/2}
C_k^{(1)}\left(\frac{(x\partial)}{\sqrt{x^2\partial^2}}\right)f(n)\Big|_{n=0}\,.
\end{equation}

\section{Derivation  of Eq.~(\ref{ABC1})}\label{app:ABC1}

Collecting the expressions derived in Sec.~\ref{subsec:twist4} and Sec.~\ref{subsec:ge}
we obtain
\begin{eqnarray}\label{A}
A(n;z_1,z_2)&=&-\frac12\int_0^1du\, u^2(1+\ln u)\int_{z_2}^{z_1} dw\,
w\,\big[\mathcal{F}+\bar{\mathcal{F}}\big](z_1u,wu,z_2u)
\notag\\
&&{}+\frac14\int_0^1du \, u^2\ln u
\biggr\{
z_1z_2\left[i\mathbf{P}^\mu \left[i\mathbf{P}_\mu, O_{++}\right]\right](z_1u,z_2u)
\notag\\
&&{}-2 z_{12}\,
\Big(z_2\,\mathcal{F}(z_1u,z_2u,z_2u)+z_1\,\bar{\mathcal{F}}(z_1u,z_1u,z_2u)\Big)
\notag\\
&&{}+ \int_{z_2}^{z_1}dw \Big\{ z_1(w-z_2)
\Big[u \mathcal{D}(z_1u,wu,z_2u)-2\partial_{z_1}\big[\mathcal{F}+\bar{\mathcal{F}}\big](z_1u,wu,z_2u)\Big]
\notag\\
&&{}+z_2(w-z_1) \Big[u\bar{\mathcal{D}}(z_1u,wu,z_2u)-2\partial_{z_2}
\big[\mathcal{F}+\bar{\mathcal{F}}\big](z_1u,wu,z_2u)\Big]
\Big\}\biggr\}\,,
\end{eqnarray}
\begin{eqnarray}\label{B}
B(n;z_1,z_2)&=&\frac18\int_0^1du\,
(1-u^2+u^2\ln u)
\biggl\{
z_1z_2\left[i\mathbf{P}^\mu \left[i\mathbf{P}_\mu, O_{++}\right]\right](z_1u,z_2u)
\nonumber\\
&&{}\hspace*{3cm}
-2 z_{12}\,
\Big[z_2\,\mathcal{F}(z_1u,z_2u,z_2u)+z_1\,\bar{\mathcal{F}}(z_1u,z_1u,z_2u)\Big]\biggr\}
\nonumber\\
&+&\frac18\int_0^1\!du\int_{z_2}^{z_1}\!\!dw\biggl\{-2u^2\ln u\,
 w\, \big[\mathcal{F}+\bar{\mathcal{F}}\big](z_1u,w u,z_2u)
\nonumber\\
&&+ (1-u^2+u^2\ln u)\biggl[ z_1(w-z_2)
\big[u\mathcal{D}-2\partial_{z_1}\mathcal{F}-2\partial_{z_1}\bar{\mathcal{F}}\big](z_1u,wu,z_2u)
\nonumber\\
&&{}\hspace*{3cm}+z_2(w-z_1) \big[u\bar{\mathcal{D}}
-2\partial_{z_2}\mathcal{F}-2\partial_{z_2}\bar{\mathcal{F}}\big](z_1u,wu,z_2u)
\biggr]
\nonumber\\
&&
+2u^2\Big[(w-z_2)\mathcal{F}+(w-z_1)\bar{\mathcal{F}}\Big](uz_1,uw,uz_2)
\nonumber\\
&&-(1-u^2)
\Big[(w-z_2)w\partial_w\mathcal{F}+(w-z_1)w\partial_w\bar{\mathcal{F}}\Big](uz_1,uw,uz_2)
\nonumber\\
&&
-\frac12(1-u^2) u\Big[
z_1 \mathcal{D}+z_2 \bar{\mathcal{D}}
\Big](uz_1,uw,uz_2)
\biggr\}\,,
\end{eqnarray}
\begin{eqnarray}\label{C}
C(n;z_1,z_2)&=&\frac1{8}\int_0^1du\,(1-u^2)\int_{z_2}^{z_1}dw\,(z_2-w)\biggl\{
\frac12
z_1u\,\Big[\mathcal{D}(uz_1,uw,uz_2)
+\bar{\mathcal{D}}(uz_2,uw,uz_1)\Big]
\notag\\
&&{}
+(w\partial_w+2)\big[\mathcal{F}(uz_1,uw,uz_2)+\bar{\mathcal{F}}(uz_2,uw,uz_1)\big]
\biggr\}\,.
\end{eqnarray}
The next step is to rewrite the results in terms of the conformal operator basis, Eq.~(\ref{QF-relation}),
e.g.
\begin{eqnarray}
C(n;z_1,z_2)&=&\frac{ig}{16(n\tilde n)}\int_0^1du\,(1-u^2)\int_{z_2}^{z_1}dw\,(z_2-w)
\nonumber\\&&{}\times\biggl\{
\Big[z_1\partial_{z_1}Q_2 -uz_1Q_3
+(w\partial_w+2)[Q_2-Q_1]\Big](uz_1,uw,uz_2)
\nonumber\\&&{}
\hspace*{0.35cm}
+ \Big[z_1\partial_{z_1}\bar Q_2 -uz_1 \bar Q_3
+(w\partial_w+2)[\bar Q_2-\bar Q_1]\Big](uz_2,uw,uz_1)
\biggr\}\,.
\label{eqc1}
\end{eqnarray}
This expression can be simplified further by observing that in the present context we are not
interested in contributions that are related to descendants of geometric twist-three operators,
cf. Sec.~2.5.
Consider the following operator identities that are easy to derive by explicit calculation:
\begin{eqnarray}
 \frac12[i\mathbf{P}_{-+},\bar\psi_+(z_1) f_{++}(z_2)\psi_+(z_3)]
 &=& Q_3(z_1,z_2,z_3) + \partial_{z_3} Q_2(z_1,z_2,z_3) + \partial_{z_2} Q_1(z_1,z_2,z_3)
\nonumber\\
 \frac12[i\mathbf{P}_{+-},\bar\psi_+(z_1) \bar f_{++}(z_2)\psi_+(z_3)]
 &=& \bar Q_3(z_1,z_2,z_3) + \partial_{z_1} \bar Q_2(z_1,z_2,z_3) + \partial_{z_2} \bar Q_1(z_1,z_2,z_3)
\nonumber\\
\label{eq:Pt3}
\end{eqnarray}
The expressions on the l.h.s. are total transverse derivatives of twist-three
quasipartonic operators. Hence they have autonomous scale dependence and cannot get mixed with
derivatives of twist-two operators that we are interested in. In other words, the matrix elements
of these operators can self-consistently be put to zero so that Eqs.~(\ref{eq:Pt3})
allows one to eliminate all occurrences of $Q_3$ ($\bar Q_3$) in Eq.~(\ref{eqc1})
in favor of $Q_{1,2}$ ($\bar Q_{1,2}$).
In the present case an additional simplification occurs: the operators $Q_1,\bar Q_1$ drop out
from the resulting expression and the derivatives acting on $Q_2,\bar Q_2$ get collected into
a combination $z_1\partial_{z_1}+w\partial_w+z_2\partial_{z_2}+3$ which can be replaced  by
$u\partial_u+3$. Integrating by parts in $u$ one arrives at:
\begin{equation}\label{Ce0}
C(n;z_1,z_2)=-\frac{ig}{8(n\tilde n)}\int_0^1du\int_{z_2}^{z_1}dw\,(w-z_2)
\Big[Q_2(uz_1,uw,uz_2) +\bar Q_2(uz_2,uw,uz_1)\Big]\,.
\end{equation}
Using the notations in Eq.~(\ref{PhibarPhi}) this expression becomes
the one in the last line in Eq.~(\ref{ABC1}).

In the similar way we obtain after some algebra
\begin{eqnarray}
4\, A(n;z_1,z_2)
&=& z_1 z_2 \!\int_0^1\!\!\!du\,u^2 \ln u\,\Big[i\mathbf{P}^\mu,\big[i\mathbf{P}_\mu,
     O_{++}(uz_1,uz_2)\big]\Big]
\nonumber\\
&&{} +\frac{ig}{(n \tilde n)}\int_0^1 \!\!du\, u^2\biggl\{
\int_{z_2}^{z_1}\!\!dw\Big[z_2 Q_1-wQ_2\Big](uz_1,uw,uz_2)
\notag\\
&&{}\hspace*{2cm}- \ln u \,(z_2\partial_{z_2}) z_{21} \int_{z_2}^{z_1}\!\!dw\Big[ Q_2-Q_1\Big](uz_1,uw,uz_2)
\biggr\}
\notag\\
&&{}
+\frac{ig}{(n \tilde n)}\int_0^1 \!\!du\, u^2\biggl\{
\int_{z_2}^{z_1}\!\!dw\Big[z_1 \bar Q_1-w\bar Q_2\Big](uz_1,uw,uz_2)
\notag\\
&&{}\hspace*{2cm}-
\ln u \,(z_1\partial_{z_1})z_{12}\int_{z_2}^{z_1}\!\!dw \Big[ \bar Q_2-\bar Q_1\Big](uz_1,uw,uz_2)
\biggr\}
\end{eqnarray}
and
\begin{eqnarray}
8\,{B}(n;z_1,z_2) &=&
 z_1 z_2 \!\int_0^1\!\!\!du\,(1\!-\!u^2\!+\!u^2\ln u)\,\Big[i\mathbf{P}^\mu,\big[i\mathbf{P}_\mu,
     O_{++}(uz_1,uz_2)\big]\Big]
\nonumber\\&&{}+
  \frac{ig}{(n \tilde n)}
 \int_0^1\!\!\!du\,\biggl\{ (1\!-\!u^2\!+\!u^2\ln u)
 (z_2\partial_{z_2})z_{21} \int_{z_2}^{z_1}\!\!\!\! dw\, [Q_1-Q_2](uz_1,uw,uz_2)
\nonumber\\&&{} \hspace*{2cm}
- (1-u^2)\int_{z_2}^{z_1}\!\!\!dw\,[z_2Q_1- w Q_2](uz_1,uw,uz_2)\biggr\}
\nonumber\\&&{}+
  \frac{ig}{(n \tilde n)}
 \int_0^1\!\!\!du\,\biggl\{ (1\!-\!u^2\!+\!u^2\ln u)
 (z_1\partial_{z_1})z_{12} \int_{z_2}^{z_1}\!\!\!\! dw\, [\bar Q_1-\bar Q_2](uz_1,uw,uz_2)
\nonumber\\&&{} \hspace*{2cm}
- (1-u^2)\int_{z_2}^{z_1}\!\!\!dw\,[z_1\bar Q_1- w \bar Q_2](uz_1,uw,uz_2)\biggr\}.
\end{eqnarray}
In this case the operator $Q_1$ does contribute, but can be dispensed off using the
relation (which is equivalent to Eq.~(\ref{eq:killQ1}))
\begin{align}
Q_1(z_1,w,z_2)=\partial_{z_2} z_{12}\int_{w}^{z_1}d\eta\frac{z_1-\eta}{(z_1-w)^2}Q_2(z_1,\eta,z_2)\,.
\end{align}
With this substitution, after some algebra one arrives at the expressions given in Eq.~(\ref{ABC1}).

\section{Identities for three-particle wave functions  $\Psi_{Nk}(z_1,z_2,z_3)$}\label{app:Id}

The ``wave functions''  $\Psi_{Nk}(z_1,z_2,z_3)$ (\ref{def:psiNk}) of three-particle nonquasipartonic twist-four
operators have a rather complicated algebraic structure. In practice one usually has to deals with integrals
of these functions over the light-cone position of one of the fields (usually gluon) with simple polynomial
weights which follow from the structure of Feynman diagrams. Such integrals can be much simpler compared to
the wave functions themselves, see Eq.~(\ref{SS20}) for an example.
Although this equation can easily be verified by a direct calculation, it makes sense to show that
it is due to conformal invariance. Apart of being more elegant, this derivation allows one to understand
why a simplification has been possible for this particular case, and why it does not happen for some other
integrals.

Consider
\begin{equation}
  \Phi_N(z_1,z_2) = \int_0^1 du\, u\, \Psi^{(2)}_N(z_1,z_{21}^u,z_2)
= \frac{1}{z_{12}^2}\int_{z_2}^{z_1} \!\!\!dw\, (w-z_2)\Psi^{(2)}_N(z_1,w,z_2)\,,
\label{E11}
\end{equation}
where $\Psi^{(2)}_N(z_1,w,z_2)$ transforms according to the tensor product of the
representations $T^{j_1=1}\otimes T^{j_2=3/2}\otimes T^{j_3=1/2}$ under $SL(2)$ transformations~(\ref{SL2R}).
If we prove that $\Phi_N(z_1,z_2)$ transforms according to $T^{j_1=2}\otimes T^{j_2=1}$, i.e. the
integral operator in (\ref{E11}) is an intertwining operator between these representations, then
\begin{equation}
    (S_+^{(2,1)})^k \Phi_N(z_1,z_2) = \int_0^1 du\, u\, \Big[(S_+^{(1,\frac32,\frac12)})^k\Psi^{(2)}_N\Big](z_1,z_{21}^u,z_2)
\end{equation}
which is equivalent to Eq.~(\ref{SS20}) since $z_{12}^2 (S_+^{(2,1)})^k = z_{12}^2 (S_+^{(1,0)})^k$.

Behavior under conformal transformations can conveniently be analyzed using the reproducing
kernel $\mathcal{K}$. Using the definition in Eq.(\ref{repr}) we can write
\begin{equation}
\Phi_N(z_1,z_2)=\int_0^1 du \,u \iiint\prod_{k=1}^3[\mathcal{D} \xi_k]^{(1,\frac32,\frac12)}
\,\mathcal{K}_{1}(z_1,\xi_1)\mathcal{K}_{\frac32}(z_{21}^u,\xi_2)\mathcal{K}_{\frac12}(z_2,\xi_3)
\,\Psi^{(2)}_N(\xi_1,\xi_2,\xi_3).
\end{equation}
Now the integral over $u$ can be taken, producing a product of two reproducing kernels
with different spins, cf. Eq.~(\ref{feynman1})
\begin{equation}
  \int_0^1 du \,u\, \mathcal{K}_{\frac32}(z_{21}^u,\xi_2) = \frac12 \mathcal{K}_{1}(z_1,\xi_2)\mathcal{K}_{\frac12}(z_1,\xi_2)
\end{equation}
which is nothing but a Feynman representation for the product of reproducing kernels,
fully analogous to the product of propagators.

\begin{figure}[ht]
\begin{center}
 \includegraphics[width=3.5cm]{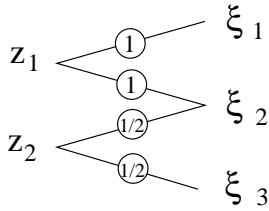}
\caption{Diagrammatic representation of Eq.~(\ref{E12})}
\label{fig:2}
\end{center}
\end{figure}

Thus
\begin{equation}
\Phi_N(z_1,z_2)=\frac12 \iiint\prod_{k=1}^3[\mathcal{D} \xi_k]^{(1,\frac32,\frac12)}
\,\mathcal{K}_{1}(z_1,\xi_1)
\mathcal{K}_{1}(z_1,\xi_2)\mathcal{K}_{\frac12}(z_1,\xi_2)
\mathcal{K}_{\frac12}(z_2,\xi_3)
\,\Psi^{(2)}_N(\xi_1,\xi_2,\xi_3).
\label{E12}
\end{equation}
Following the analogy between reproducing kernels and propagators~\cite{hep-th/0210216,nlin/0512047}, this equation can
be represented as a ``Feynman diagram'' shown in Fig.~\ref{fig:2}, where the numbers in
the circles specify the conformal spins. Using $SL(2)$ transformation properties
of the reproducing kernels (\ref{kernelstrafo}) it is easy to show that
transformation properties in $z$-vertices are determined by the sum of the conformal
spins of the propagators (reproducing kernels) attached to the vertex, in our case
$T^{j=2}$ for $z_1$ and $T^{j=1}$ for $z_2$, which is the stated result.

Thus, it is seen that the possibility to get a simple representation is tightened to
the possibility to make use of the Feynman parametrization of the type in Eq.~(\ref{feynman1}).
Repeating the above arguments, one can easily derive
\begin{eqnarray}
 \int_0^1 du\, u\, \Big[(S_+^{(1,\frac32,\frac12)})^k\Psi^{(2)}_N\Big](z_1,z_{21}^u,z_2)
&=& (S_+^{(2,1)})^k   \int_0^1 du\, u\,\Psi^{(2)}_N(z_1,z_{21}^u,z_2)\,,
\nonumber\\
  \int_0^1 du\, \bar u\, \Big[(S_+^{(1,\frac32,\frac12)})^k\Psi^{(2)}_N\Big](z_1,z_{21}^u,z_2)
&=& (S_+^{(\frac32,\frac32)})^k  \int_0^1 du\, \bar u\,\Psi^{(2)}_N(z_1,z_{21}^u,z_2)\,,
\nonumber\\
  \int_0^1 du\, \Big[(S_+^{(1,1,1)})^k\Psi^{(1)}_N\Big](z_1,z_{21}^u,z_2)
&=& (S_+^{(\frac32,\frac32)})^k  \int_0^1 du\,\Psi^{(1)}_N(z_1,z_{21}^u,z_2)\,,
\end{eqnarray}
however, e.g.
$$
  \int_0^1 du\,u\, \Big[(S_+^{(1,1,1)})^k\Psi^{(1)}_N\Big](z_1,z_{21}^u,z_2)
$$
cannot be simplified in the similar manner. In our calculation of the OPE for the product of
two electromagnetic currents this last integral appears in the contributions of the
handbag and gluon emission diagrams, but cancels in their sum. This cancellation proves to be
crucial in order to obtain a relatively simple final result, e.g. in the form (\ref{pobeda}).


%

\end{document}